\documentclass[a4paper,11pt]{article}

\usepackage{jheppub} 
\usepackage{lineno}
\usepackage{float}

\title{Holographic homogeneous superfluid on the sphere
}

\author[a,b]{Meng Gao,}
\emailAdd{gm@mail.bnu.edu.cn}

\affiliation[a]{School of Physics and Astronomy, Beijing Normal University, Beijing 100875,
China}
\affiliation[b]{Key Laboratory of Multiscale Spin Physics, Ministry of Education, Beijing Normal University, Beijing 100875, China}

\author[c,d,e]{Zhuan Ning,}
\emailAdd{ningzhuan17@mails.ucas.ac.cn}

\affiliation[c]{School of Fundamental Physics and Mathematical Sciences, Hangzhou Institute for Advanced Study, University of Chinese Academy of Sciences, Hangzhou 310024, China}
\affiliation[d]{CAS Key Laboratory of Theoretical Physics, Institute of Theoretical Physics, Chinese Academy of Sciences, Beijing 100190, China}
\affiliation[e]{School of Physical Sciences, University of Chinese Academy of Sciences, Beijing 100049, China}

\author[e,f]{Yu Tian,}
\emailAdd{ytian@ucas.ac.cn}

\affiliation[f]{Institute of Theoretical Physics, Chinese Academy of Sciences, Beijing
100190, China}

\author[a,b]{Hongbao Zhang}
\emailAdd{hongbaozhang@bnu.edu.cn}

\abstract{In this paper, we extend the study of holographic superfluids from planar topology to spherical topology, inspired by recent studies on Bose-Einstein condensation (BEC) on shell-shaped geometry. We investigate the superfluid phase transition from normal fluid and its Quasi-Normal Modes (QNMs) on the sphere. 
It turns out that the critical temperature for the superfluid phase transition on the sphere is higher than that in the planar case. We investigated four different solutions in the backgrounds of large and small black holes. The calculation of free energy selects the most stable solution. Finally, after calculating the quasi-normal modes and their dynamic behavior, we obtained three different channels similar to the planar superfluid case, along with the ``first" hydrodynamic excitation mode.}

\begin{document}

\maketitle

\section{Introduction and motivation}
The exploration of superfluidity \cite{ONSAGER,Transportphenomena}, a state of matter characterized by zero viscosity and perfect fluidity without dissipating energy, has been a fascinating subject in the realm of condensed matter physics. Traditionally, superfluids have been studied in the context of Bose-Einstein condensates (BECs), where the phenomenon is observed at extremely low temperatures \cite{London,Anderson,Davis,GP,BEC-Superfluidity,NonlinearPhenomena}. Since then, the theoretical framework and experimental techniques for studying superfluids have been thriving, leading to novel insights and applications in various fields, including quantum information processing \cite{Quantumcopying,Bosemetal,QuantumSimulation}, superconductivity \cite{Superconductivity,Superconductivity1}, and astrophysics \cite{NeutronStar,NeutronStar1}.
In recent years, there has been a surge of interest in studying superfluids on shell-shaped systems \cite{Shell-shapedAtomic,Bereta,BEC-sphere}, inspired by the unique properties that arise from the curvature and topology of confining geometry \cite{Observation,microgravity,microgravity1,orbit}. In the field of condensed matter, Ref.\cite{Bereta} theoretically studied the critical temperature of Bose-Einstein condensation in a spherical potential well and discussed the effect of dimensional changes of the spherical potential well, from 3D to 2D, on the critical temperature. However, they cannot rigorously obtain a finite critical temperature for Bose Einstein condensation on the 2D sphere using the semi-classical approximation method. Moreover, as the external radius $b=a+\delta$ goes to the internal radius $a$, $\delta \rightarrow 0$, Taylor expansion approximation about Bessel function gives the conclusion that it need an infinite amount of energy to excite the radial degree of freedom. Therefore, the so-called 2-dimensional sphere they refer to is not a strictly 2-dimensional spherical system. In contrast, the holographic gravity model \cite{S.A,C.P} in this paper can easily and analytically construct a mathematical 2-dimensional spherical system to study the phase transition of Bose-Einstein condensation, where the thickness of its shell is strictly zero. Ref.\cite{BEC-sphere} investigated Bose-Einstein condensation in spherical systems, which includes two cases: noninteracting Bose gas and interacting Bose gas. In the non-interacting Bose gas system, they found that the smaller the radius of the spherical surface, the higher the critical temperature for Bose-Einstein condensation. Furthermore, the critical temperature for the phase transition on the spherical surface is always higher than that in the planar system, which is mirrored in our paper. In addition to theoretical research, there have also been some experimental advancements. A spherical potential well can be constructed experimentally \cite{PhysRevLett.86.1195,PhysRevA.69.023605,Garraway_2016} to confine the atoms in
this kind of bubble trap. Recently, with the construction of the NASA Cold Atom Lab (CAL) inside the International Space Station \cite{Aveline2020}, experiments involving the cold atoms system in a bubble trap under microgravity conditions have become possible \cite{Lundblad_2019}. Furthermore, ultracold bubbles between two-component cold atomic gases has been observed in microgravity aboard CAL \cite{Observation}, which confines the Bose-Einstein condensation formed by one component, restricting its exclusion to the surface shell of the solid sphere of the Bose-Einstein condensate formed by the other component. Ultracold bubbles are precisely spherical BEC systems, and the spherical holographic superfluid model has the potential to describe ultracold bubbles.
The two-dimensional surface will exhibit richer topological structures and curvature behaviors and offer a rich playground for investigating the interplay between superfluid properties and the geometry structure, potentially leading to a new picture of phase transition and collective excitations \cite{Hollow,Staticanddynamic,Topologicalsuperfluid} that are not present in the planar case. Therefore, the study of spherical superfluids is of great significance, and the results are expected to be experimentally validated.

Inspired by the progress in these experiments and the theoretical research about shell-shaped Bose Einstein condensations in the realm of condensed matter, we have begun to use the holographic gravity approach to study the Bose-Einstein condensation of cold atoms in a two-dimensional spherical system to explore the new phenomena that may be induced by the spatial topology and curvature within it. One of our results shows that the superfluid phase transition temperature on the 2-dimensional unit sphere is indeed higher than that in the planar topology case, which is consistent with the result in Ref.\cite{BEC-sphere}. As for the case of interacting Bose gases in Ref.\cite{BEC-sphere}, it can be realized by adding interaction terms for the scalar field in a holographic gravity model, but this is not the focus of this article.

The holographic principle, or AdS/CFT correspondence, has emerged as a powerful tool in theoretical physics, allowing the study of strongly coupled systems through gravity duals in higher-dimensional spacetime \cite{Susskind,Maldacena,Witten}. The study of superfluidity in holographic systems has garnered significant interest due to its implications for understanding strongly correlated systems and quantum phase transitions, such as holographic superfluid \cite{key-9,key-10,key-11,key-12,Guo,gaomeng} and superconductors \cite{Nishioka}, where the phase transitions and properties of these systems can be mapped to black hole solutions in Anti-de Sitter (AdS) spacetime. Despite considerable advances in the comprehension of holographic superfluids within planar topology, the extension to spherical topology remains notably unexplored. This paper seeks to address this gap by exploring the phase transitions and collective modes of holographic superfluids on the sphere, and plans to understand how the spherical geometry affects the superfluid phase transition as well as the dynamics of collective excitation modes.
 
Furthermore, The discussion in \cite{Kaku} includes holographic superconducting phase and linear perturbation analysis under spherical topology; however, it primarily addresses the scenario of small black holes, which are inherently thermodynamically unstable spacetime structures. Moreover, they just focus on the case of high frequency $\omega$, ignoring other intriguing collective modes for the case of low frequency. In our study, we both focus on the background geometry from large black hole and small black hole, and investigate collective elementary excitation coming from small $\omega$ and compare them with the flat case. There exist four different solutions: hairy large black hole, bald large black hole, hairy small black hole, and bald small black hole. For the four distinct solutions, we analyzed their phase-transition behavior and calculated their free energy. On the other hand, we calculated the quasi-normal modes as well as its dynamic behavior associated with temperature, chemical potential, and angular quantum number three parameters. Moreover, we obtain the ``first" hydrodynamic excitation mode, which is reflected in the literature \cite{Topologicalsuperfluid}. Meanwhile, we observed three kinds of quasi-normal modes coming from different channels.

The paper is organized as follows. In the next section, we introduce the holographic superfluid setup under a spherical topology background. In Sec.$\textup{III}$, we study the phase transition from normal state to superfluid state. In Sec.$\textup{IV}$, we perform a linear analysis to explore collective excitation modes. In Sec.$\textup{V}$, we draw our conclusions and some outlook.

\section{Holographic setup}

The simplest holographic superfluid model is described as the Abelian-Higgs model coupled to Einstein's gravity in the asymptotically AdS spacetime, and the corresponding action is given by \cite{S.A,C.P}
\begin{equation}
I=\frac{1}{16\pi G}\int_{\mathcal{M}}d^{4}x\sqrt{-g}\left[R-2\Lambda+\frac{1}{e^{2}}\mathcal{L}_{matter}\right],\label{eq:HHH model}
\end{equation}
where the Lagrangian for
matter fields reads
\begin{equation}
\mathcal{L}_{matter}=-\frac{1}{4}F_{ab}F^{ab}-\left|D\Psi\right|^{2}-m^{2}\left|\Psi\right|^{2}.\label{eq:Lagrangian}
\end{equation}
Here $G$ is Newton's gravitational constant, $\Lambda$ is the negative cosmological constant and related to the AdS radius as $L^2=-3/\Lambda$.
$D_{a}=\nabla_{a}-iA_{a}$, with $\nabla_{a}$ the covariant
derivative compatible to the metric. $\Psi$ is a complex scalar field coupled to the gauge potential $A_{a}$, with mass $m$ and charge $e$. In what follows, we shall work with the probe limit, namely the back-reaction of the matter fields to the background metric can be neglected, which can be achieved by taking the limit of $e\rightarrow \infty$. As such, for our purpose, we take the spherically symmetric Schwarzschild-AdS$_{4}$ spacetime
\begin{equation}
ds^{2}=\frac{L^2}{z^{2}}\left[-f(z)dt^{2}+\frac{dz^{2}}{f(z)}+L^2(d\theta^{2}+\sin^2{\theta}d\varphi^{2})\right],\label{eq:metric}
\end{equation}
as our background geometry, where the blackening factor $f\left(z\right)=1+\frac{z^{2}}{L^2}-\left(\frac{z}{z_{h}}\right)^{3}\left(1+\frac{z_{h}^{2}}{L^2}\right)$ with $z_{h}$ the horizon location. Hawking temperature is given by
\begin{equation}
T=\frac{|f^\prime\left(z_h\right)|}{4\pi}=\frac{3+z_h^2/L^2}{4\pi z_h}.
\end{equation}
In the following, we shall work in the units with $L=1$, then the above temperature will be interpreted as the temperature of the dual boundary system living on the unit sphere by holography. Note that, As usual, in this paper, we adopt natural units ($c=1, G=1, \hbar = 1$), where all physical quantities have the dimension of length. Additionally, for convenience, we set the AdS radius to 1 ($L = 1$), so all physical quantities are dimensionless. Obviously, in natural units, at $z_h=\sqrt{3}$, Hawking temperature arrives at its minimal value $T_{min}=\frac{\sqrt{3}}{2\pi}$, below which there is no black hole solution and above which there are two black hole solutions. Moreover, there exists a first order Hawking-Page phase transition at $T_{HP}=\frac{1}{\pi}$, above which the thermodynamically stable state is given by the large black hole and below which the thermodynamically stable state is given by the thermal AdS \cite{HP}. However, the large black hole is dynamically stable at least at the linear level all the way down to the minimal temperature $T_{min}$. Similarly, although the small black hole is thermodynamically unstable, it is also dynamically stable at least at the linear level all the way down to $T_{min}$. So in what follows, we would like to take both large and small black holes as our background geometry, regardless of whether the background geometry is thermodynamically stable or not. 

The dynamics of the matter fields on the aforementioned background is controlled by the following equations of motion
\begin{equation}
\nabla_{a}F^{ab} =J^{b},\quad 
D_{a}D^{a}\Psi-m^{2}\Psi =0,\label{eq:Klein-Gordon equation}
\end{equation}
with $J^{b}=i[\Psi^{*}D^{b}\Psi-\Psi\left(D^{b}\Psi\right)^{*}]$.
Accordingly, the asymptotic behavior for the bulk fields near the AdS boundary
can be obtained as follows
\begin{equation}
A_{\nu}=a_{\nu}+b_{\nu}z+\cdots,\quad 
\Psi=\Psi_{-}z^{\Delta_{-}}+\Psi_{+}z^{\Delta_{+}}+\cdots,\label{eq:psi asymptotic solution}
\end{equation} 
where we have already worked with the axial gauge $A_z=0$ with $\nu$ only taking the boundary coordinates and $\Delta_\pm=\frac{3}{2}\pm\sqrt{\frac{9}{4}+m^2}$. 
According to the holographic dictionary, $b_\nu$ is interpreted as the $U(1)$ conserved current for the boundary system sourced by $a_\nu$. In particular, $a_t$ is interpreted as the chemical potential acting on the boundary system, and $b_t=-\rho$ with $\rho$ the boundary particle number density. On the other hand, 
for simplicity but without loss of generality, below we set $m^{2}=-2$. As a result, we have  $\Delta_{-}=1$ and $\Delta_{+}=2$, where both $\Psi_-$ and $\Psi_+$ can serve as the source, corresponding to 
the standard and alternative quantizations, respectively. In what follows, we shall work exclusively with the standard quantization, where the expectation value of the dual scalar operator is given by
\begin{equation}
    \langle O\rangle=\frac{\delta S_{ren}}{\delta \Psi_-}=\Psi_+^*
\end{equation}
with the renormalized action $S_{ren}=S-\int d^3x\sqrt{-h}|\Psi|^2$ \cite{Guo}. When the scalar source is turned off, $\langle O\rangle\neq 0$, the bulk black hole carries a scalar hair, corresponding to the superfluid state on the boundary with $\langle O\rangle$ interpreted as the superfluid condensate, otherwise it is bald, corresponding to the normal fluid state. 

\section{Holographic superfluid phase transition}
In this section, we shall investigate the phase transition to holographic static and homogeneous superfluid from the normal fluid. As such, the non-vanishing bulk fields can be assumed to be $\Psi(z)\equiv z\psi(z)$ and $A_{t}(z)$ with $\psi(z)$ being also real. Then the corresponding equations of motion reduces to 
\begin{equation}
0=\frac{z^2A_t^2\psi}f+(2-2f+zf^{\prime})\psi+z^2f^{\prime}\psi^{\prime}+z^2f\psi^{\prime\prime},\quad 
0=f A_t^{\prime\prime}-2 A_t \psi^2,\label{eq:At equation}
\end{equation}
where the prime represents the derivative with respect to the radial direction $z$.

\begin{figure}[H]
\includegraphics[scale=0.3]{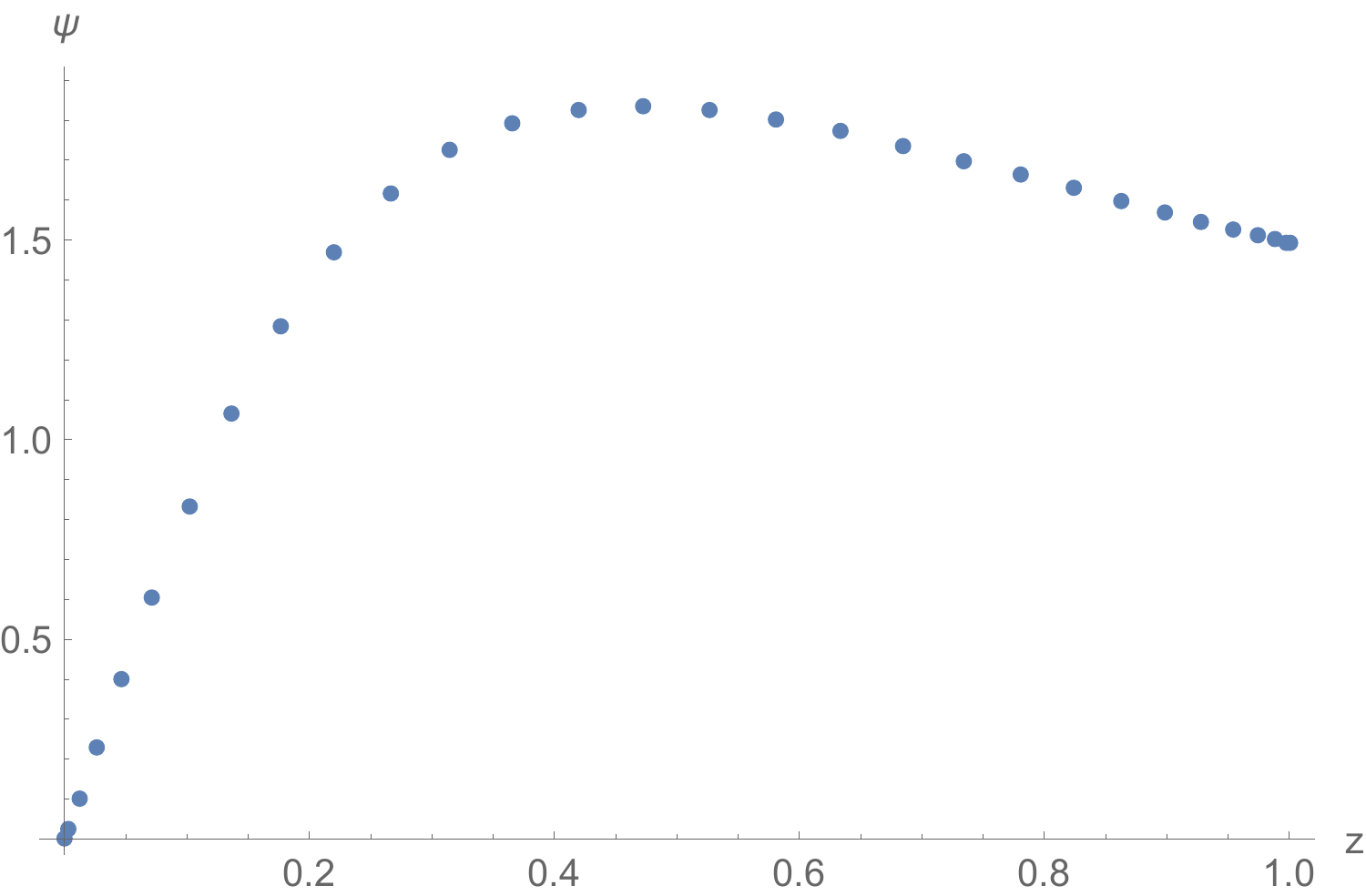}
\includegraphics[scale=0.3]{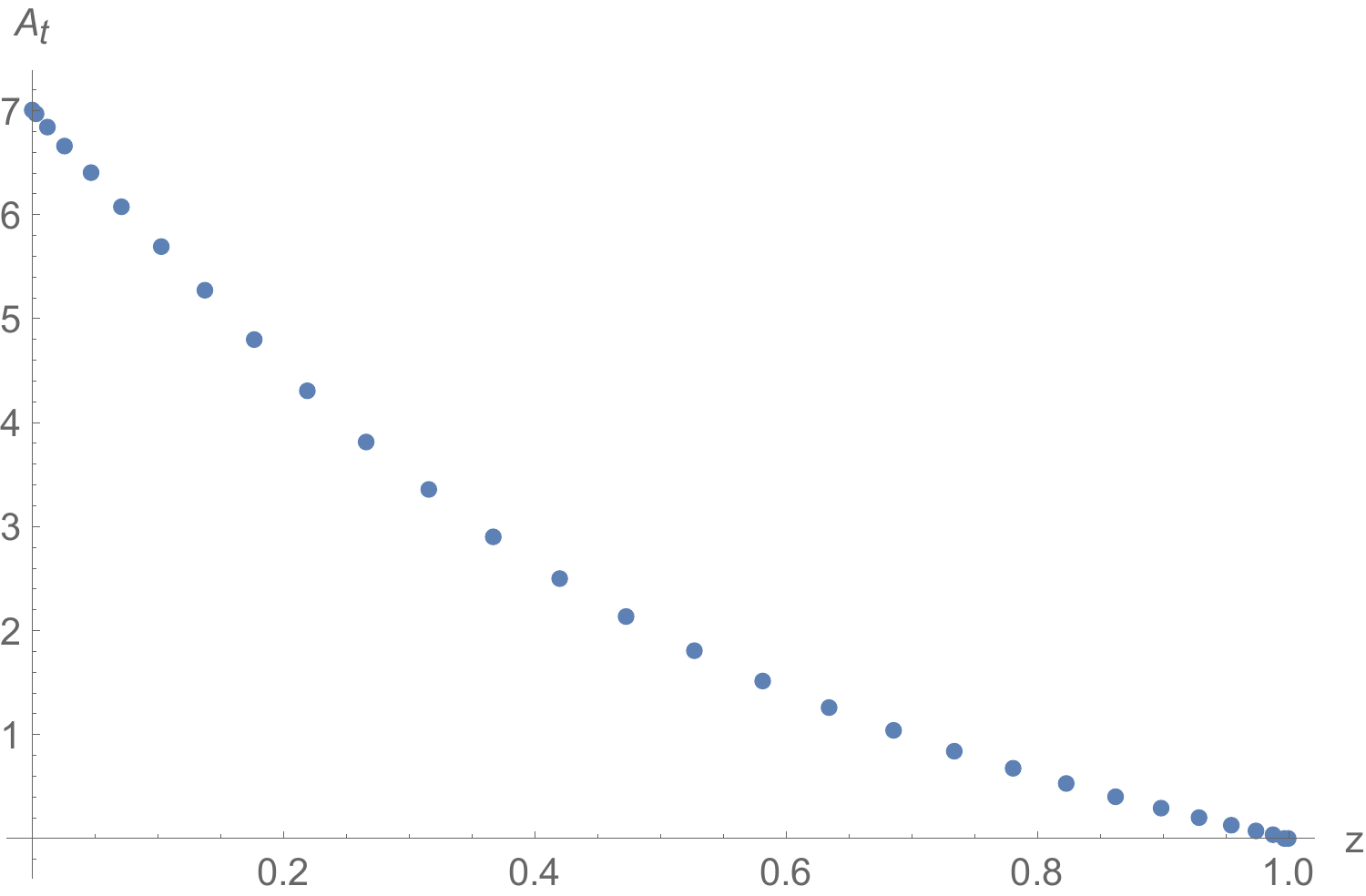}
\caption{The bulk profile for the scalar field $\psi$ and gauge field $A_t$ at $z_h=1$ and $\mu=7$. \label{fig:static configuration}}
\end{figure}

\begin{figure}[H]
\includegraphics[scale=0.3]{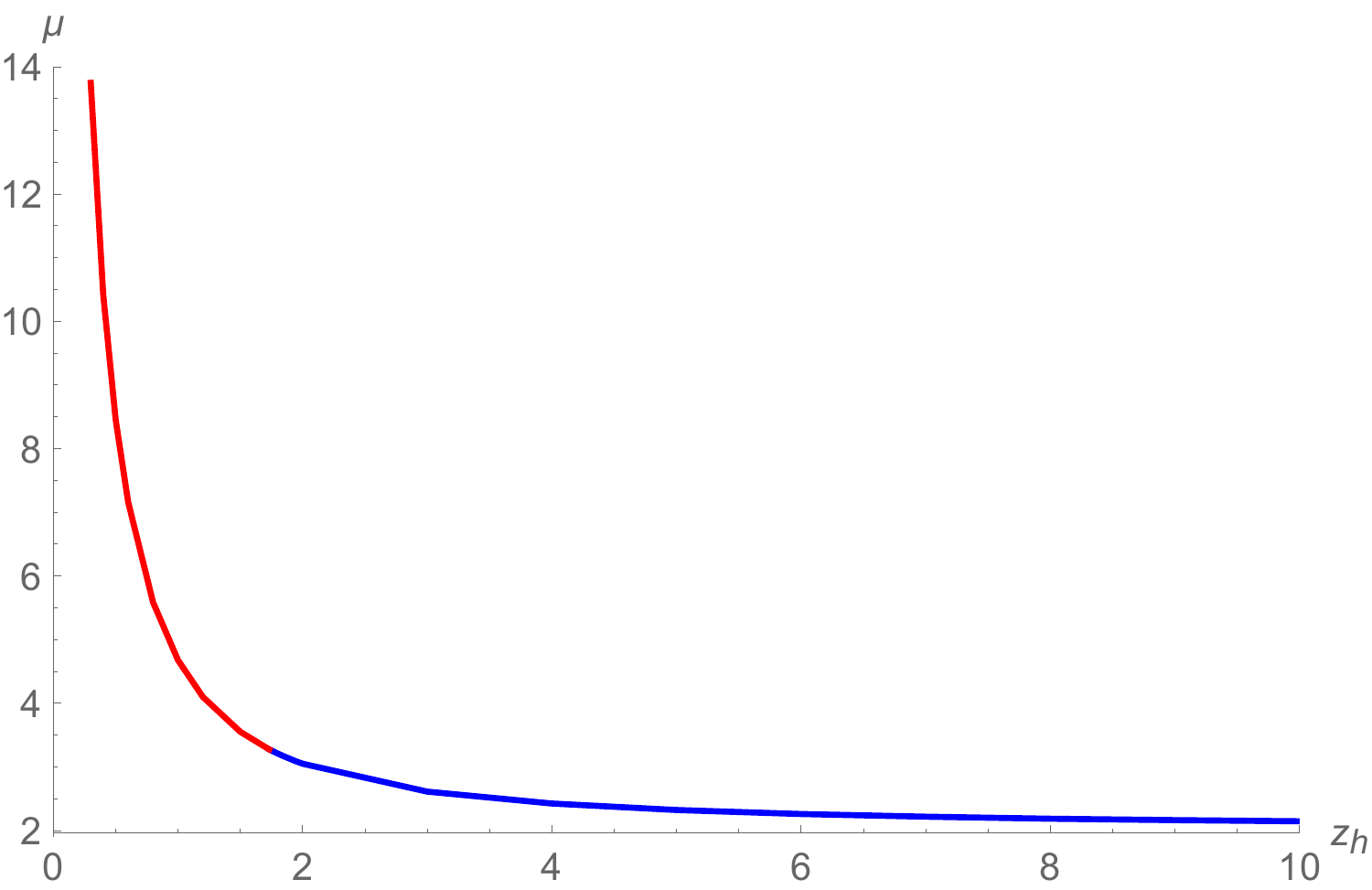}
\includegraphics[scale=0.3]{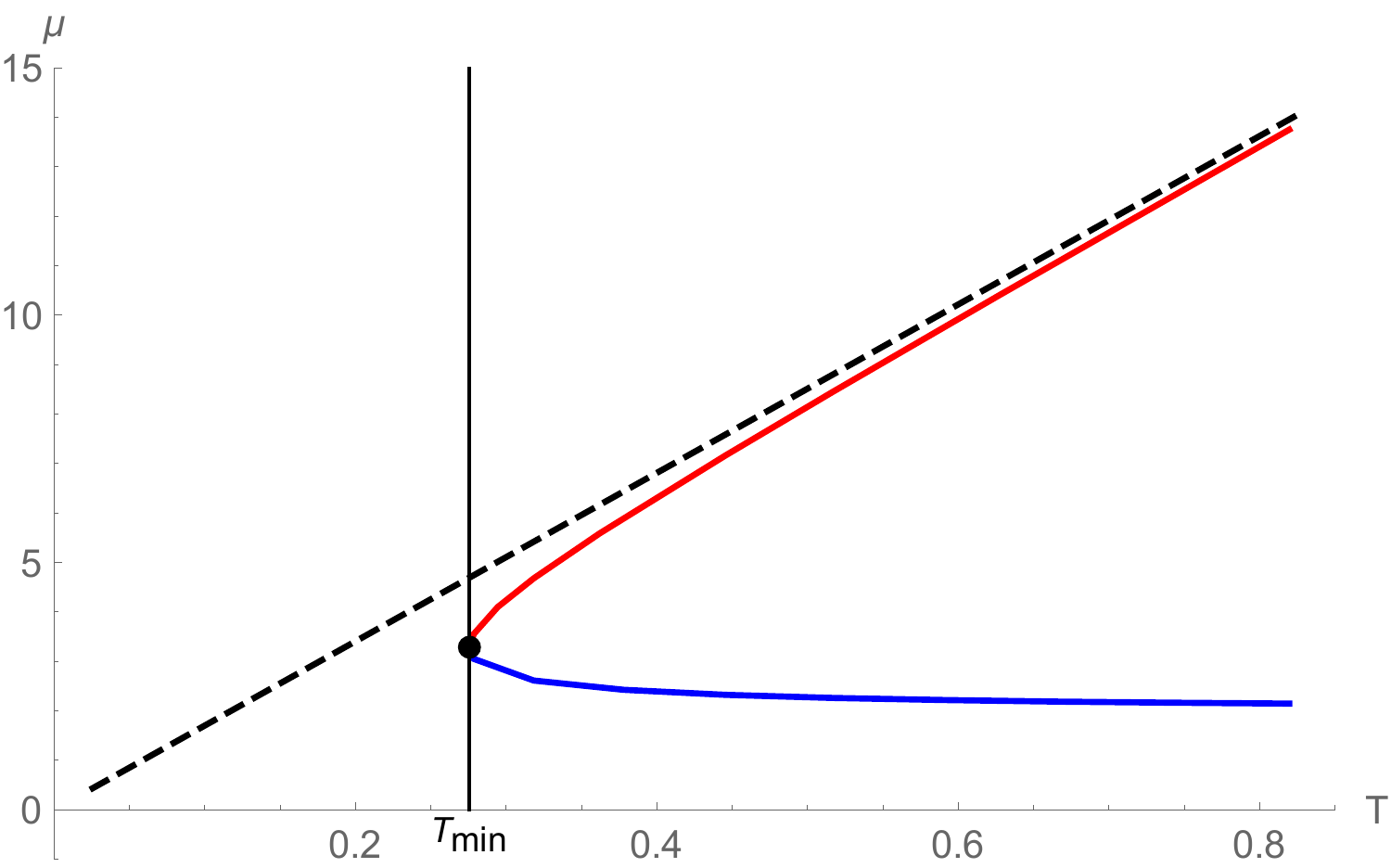}
\caption{The critical line for the phase transition from the normal fluid to superfluid, where the red and blue lines denote the critical line for the phase transition occurring on the large and small black holes, respectively. As a comparison, the dashed straight line as the critical line for the corresponding phase transition occuring on the planar black hole is also depicted in the right plot. The region above the critical line corresponds to the superfluid phase and the region below the critical line corresponds to the normal fluid state. 
\label{fig:muc-zh}}
\end{figure}

The bulk solution dual to the normal fluid state is given analytically by
\begin{equation}
    \psi=0,\quad A_t=\mu(1-\frac{z}{z_h}).
\end{equation}
However, for the superfluid state, we have no dual analytic bulk solution, for which we resort to the pseudo-spectrum method to discretize the spatial direction and Newton iteration method to solve the resulting non-linear algebraic equations. As a demonstration, we plot in Figure \ref{fig:static configuration} the numerical profile for the bulk fields dual to the superfluid state at $z_h=1$ and $\mu=7$.

As further shown in the left plot of Figure \ref{fig:muc-zh}, we find that with the decrease of the background black hole size, the critical chemical potential for the phase transition from the normal state to the superfluid phase decreases. By rephrasing it on the $\mu-T$ plane in the right plot of Figure \ref{fig:muc-zh}, we find that the large and small black holes as the background geometry give rise to distinct behaviors for the critical line. Namely, for the large black hole as the background geometry, the critical chemical potential increases with the temperature. While for the small black hole as the background geometry, the critical chemical potential decreases exotically with the temperature. As a comparison with the phase transition occurring on the planar black hole background, we also depict in the right plot of Figure \ref{fig:muc-zh} the corresponding critical line, which is exactly straight due to the scaling symmetry of the boundary system. As one can see, for a fixed chemical potential, the critical temperature for the phase transition to the superfluid on the unit sphere is always larger than that for the phase transition to the superfluid on the plane. This implies that the normal fluid is easier to spontaneously break to the superfluid phase when confined on the unit sphere. It is noteworthy that this qualitative feature is also shared by the non-interacting Bose-Einstein condensates on the sphere using the conventional method \cite{BEC-sphere}. On the other hand, the critical line for the phase transition occurring on the large black hole asymptotically approaches the critical line for the phase transition occurring on the planar black hole in the large chemical limit. This is supposed to be reasonable in the sense that this limit corresponds to the probe of the ultraviolet scale physics of the boundary system, to which the infrared scale set by the size of the boundary system is supposed to be irrelevant. 

\begin{figure}
\includegraphics[scale=0.3]{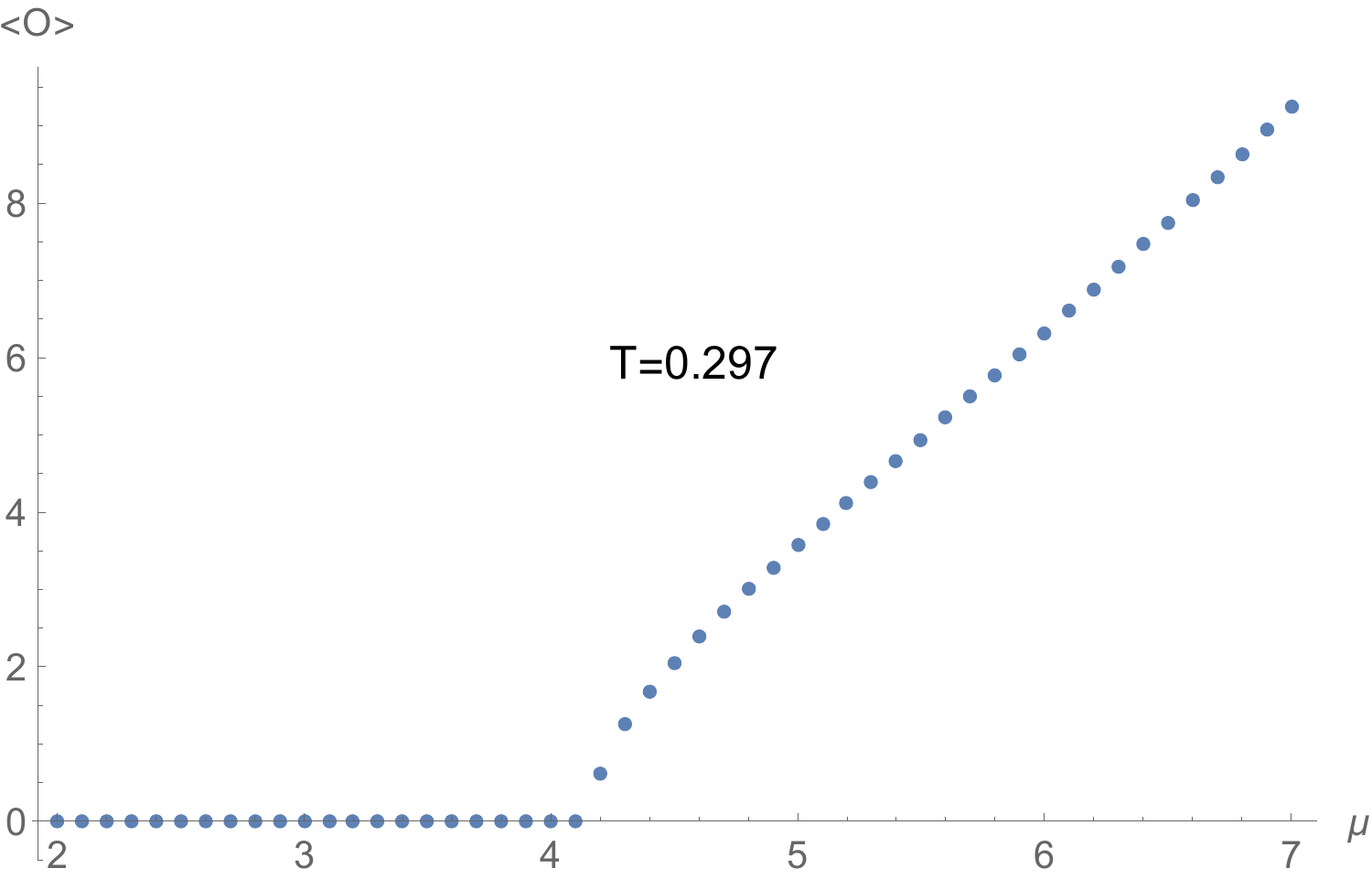}
\includegraphics[scale=0.3]{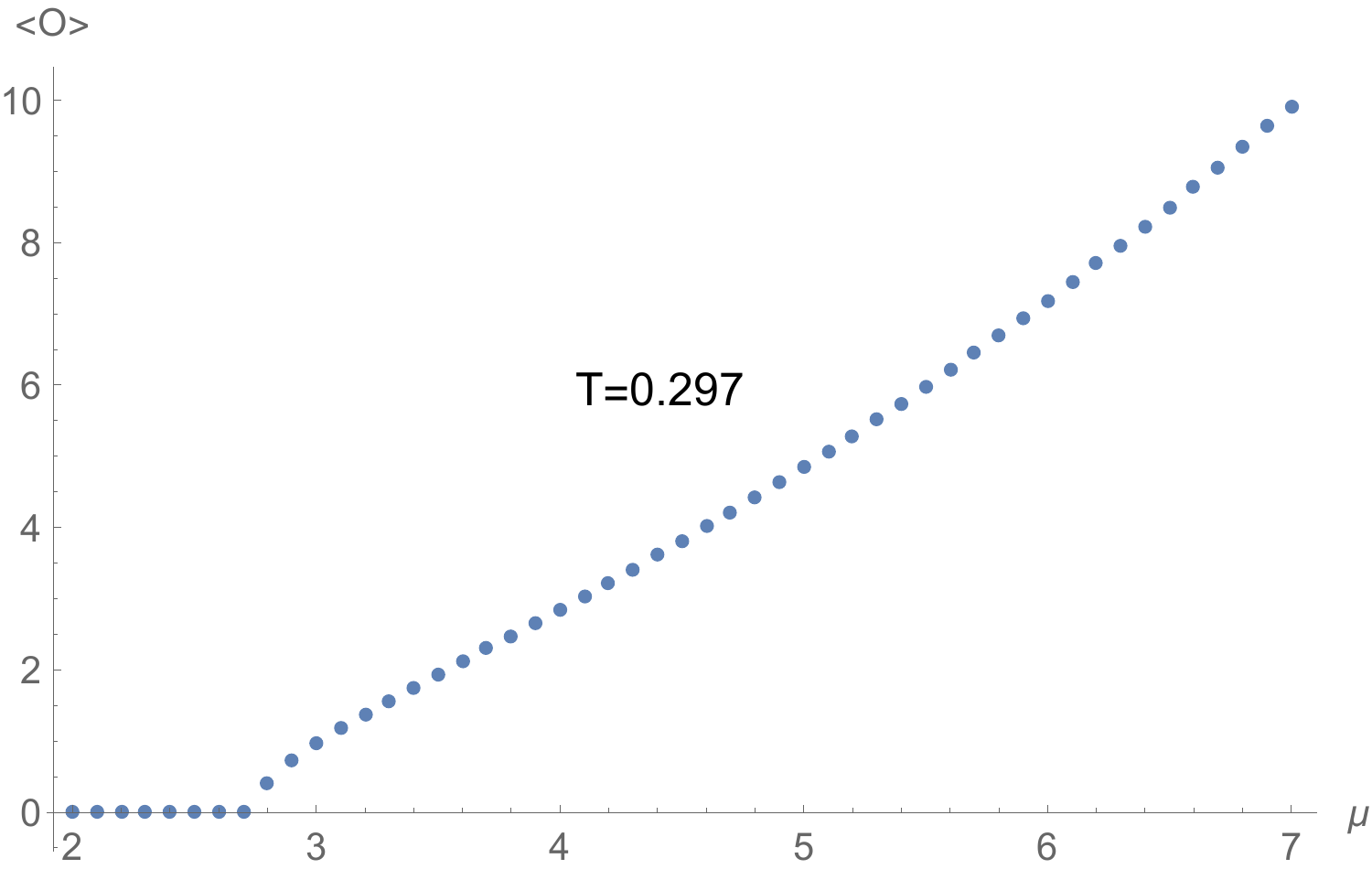}

\includegraphics[scale=0.3]{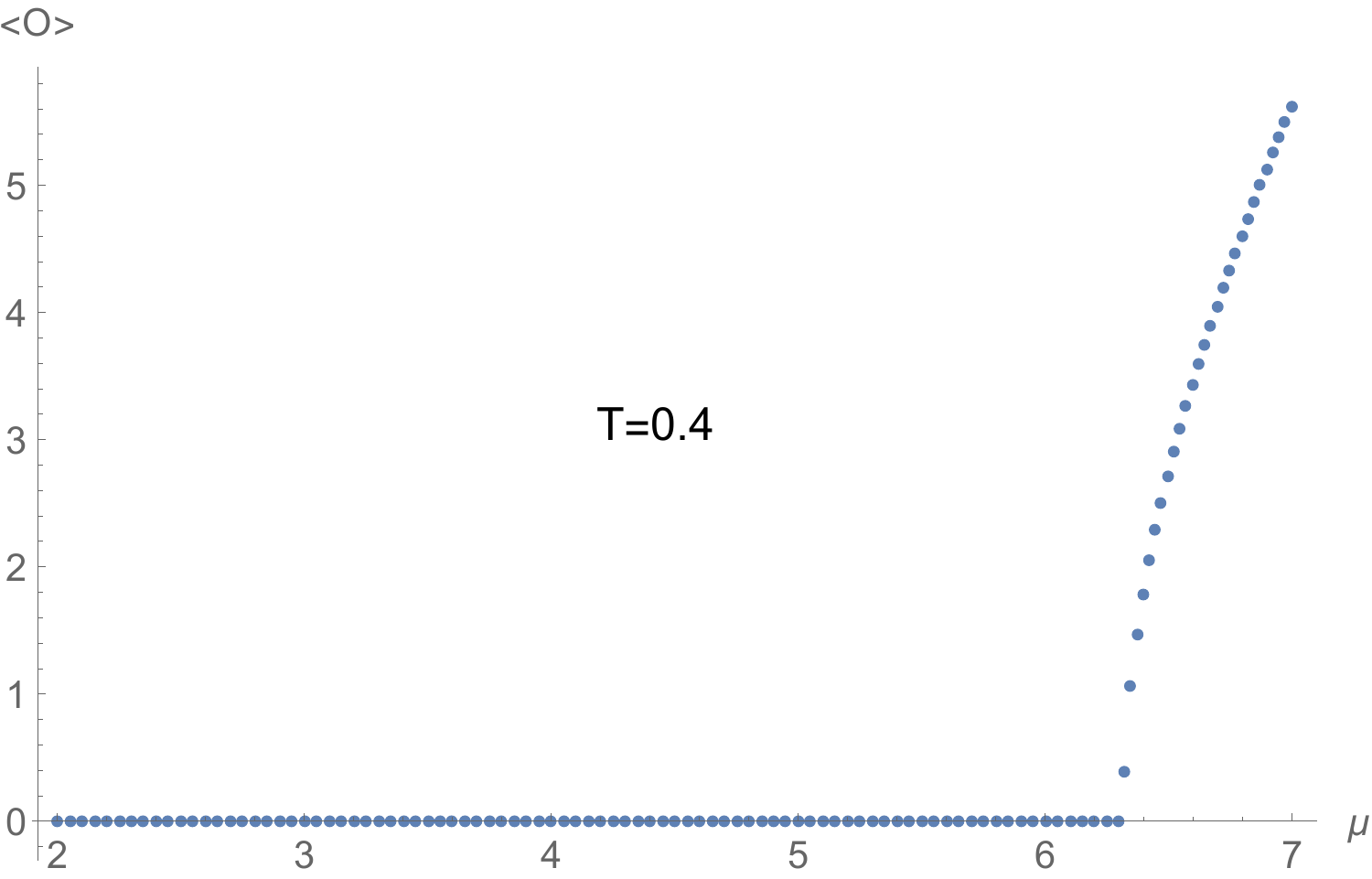}
\includegraphics[scale=0.3]{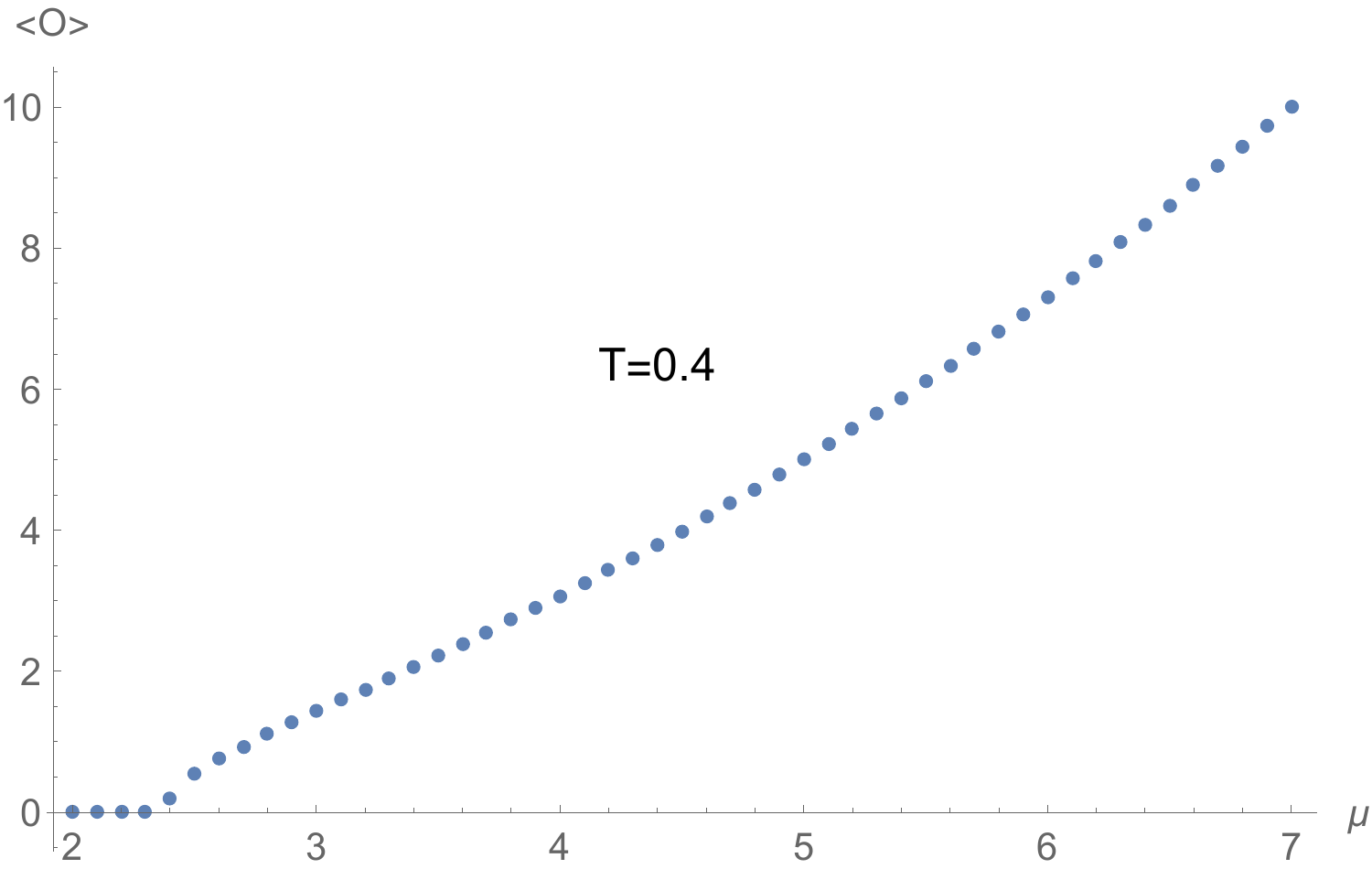}

\caption{The variation of the condensate with respect to the chemical potential at fixed temperature. The temperature is chosen to be lower than $T_{HP}$ for the upper plots and higher than $T_{HP}$ for the lower plots. The left and right plots are for the large and small black holes as the background, respectively.    \label{fig:phase_transition}}
\end{figure}

\begin{figure}
\includegraphics[scale=0.3]{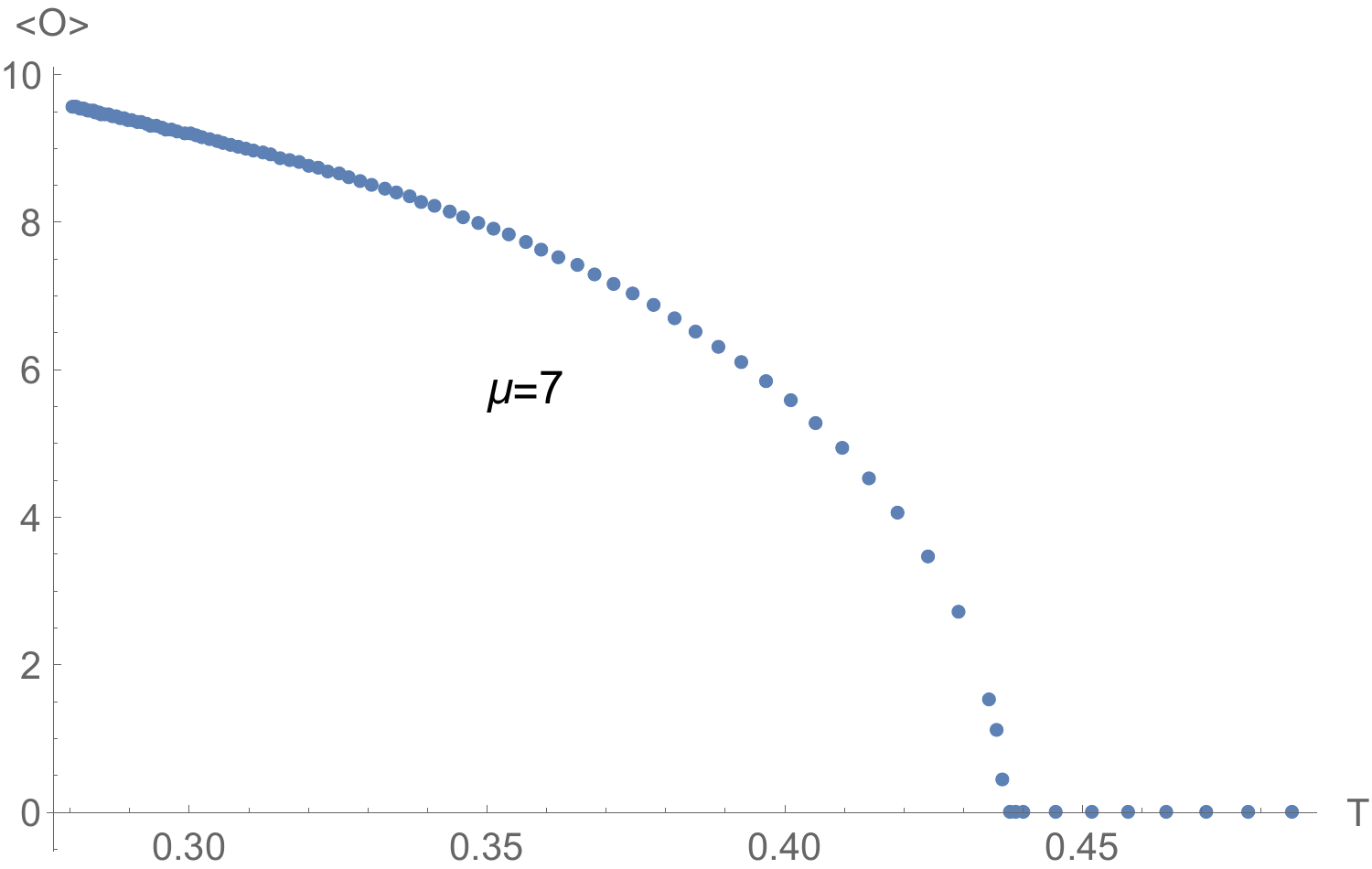}
\includegraphics[scale=0.3]{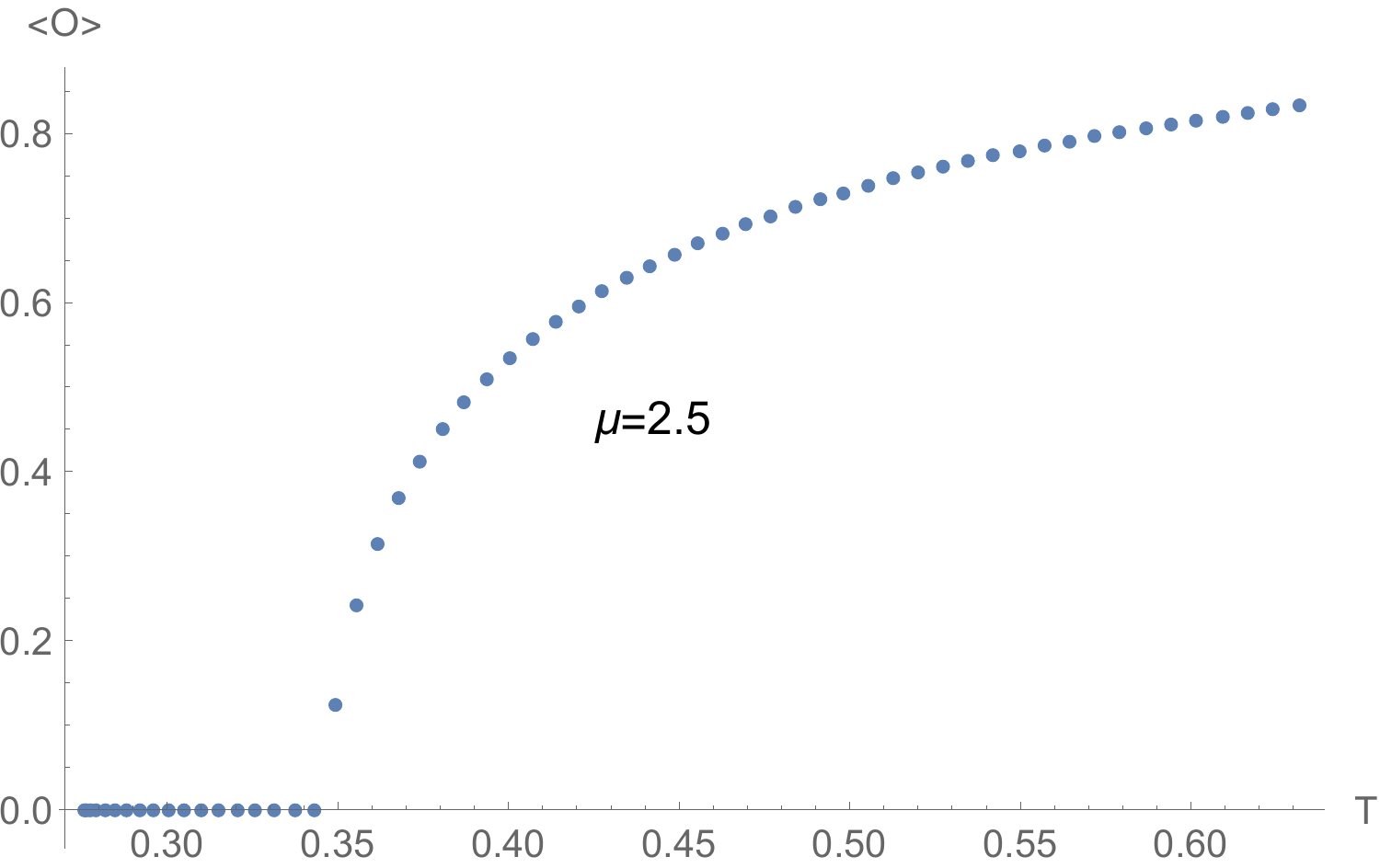}

\caption{The variation of the condensate with respect to the temperature at fixed chemical potential, where the left plot and right plot are for the large black hole and small black hole, respectively.   \label{fig:temperature_transition}}
\end{figure}

Next we present the typical condensation behaviors across the phase transition point in Figure \ref{fig:phase_transition} and Figure \ref{fig:temperature_transition} at fixed temperature and chemical potential, respectively. As shown in Figure \ref{fig:phase_transition}, the condensation displays a normal second order phase transition with respect to the chemical potential at fixed temperature. In particular, this behavior is universal in the sense that neither does it depend on whether the temperature under consideration is higher or lower than the Hawking-Page temperature, nor depends on whether the involved background geometry is given by the large or small black hole. On the other hand, as demonstrated in Figure \ref{fig:temperature_transition}, although the condensation associated with the large black hole as the background geometry still exhibits a normal second phase transition with respect to the temperature at fixed chemical potential, the condensation associated with the small black hole as the background geometry demonstrates an anomalous second-order phase transition in the sense that below some critical temperature, the condensate instead vanishes, while above the critical temperature, the condensate increases with the increase of the temperature.

\begin{figure}
\includegraphics[scale=0.3]{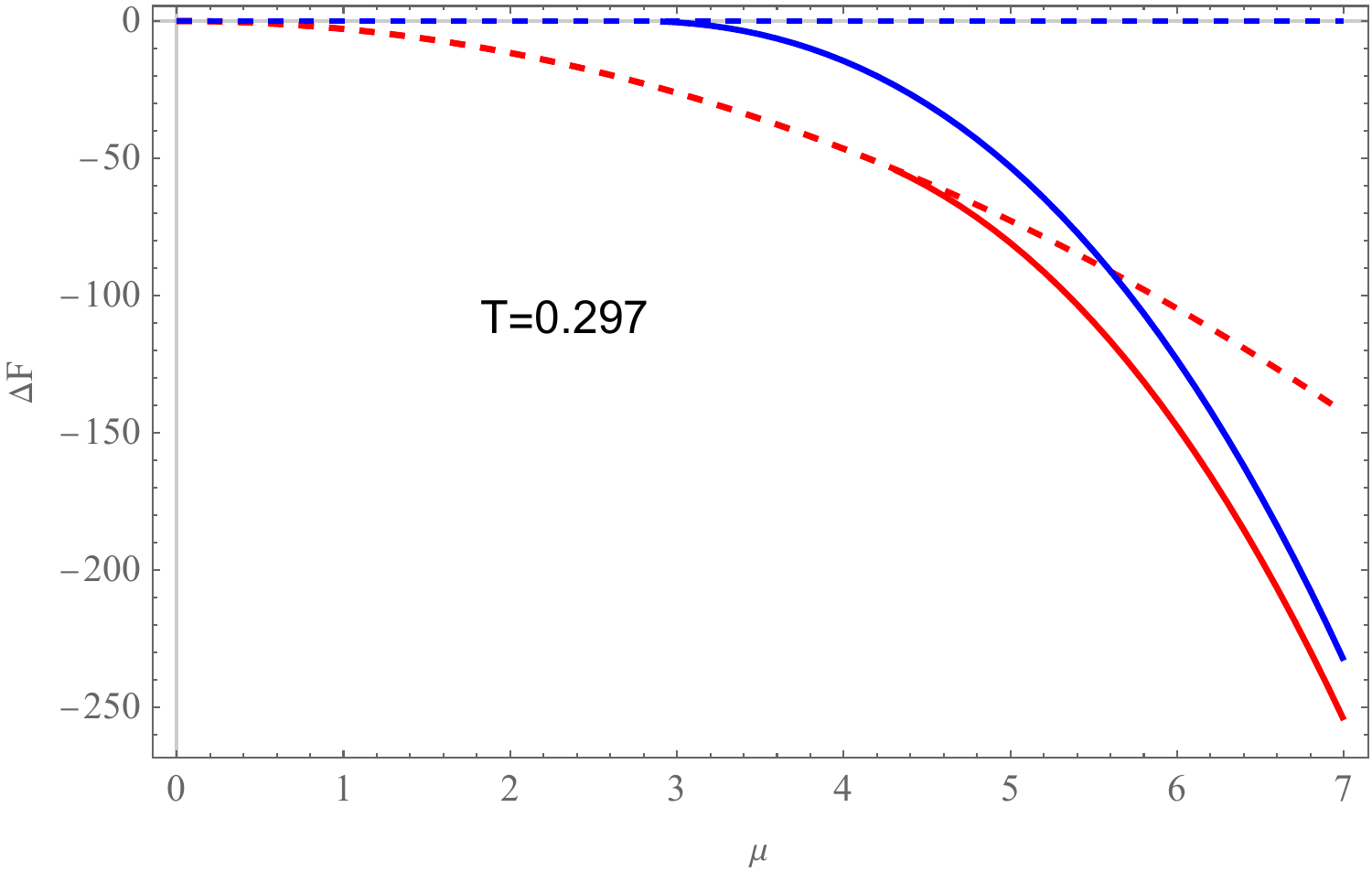}
\includegraphics[scale=0.3]{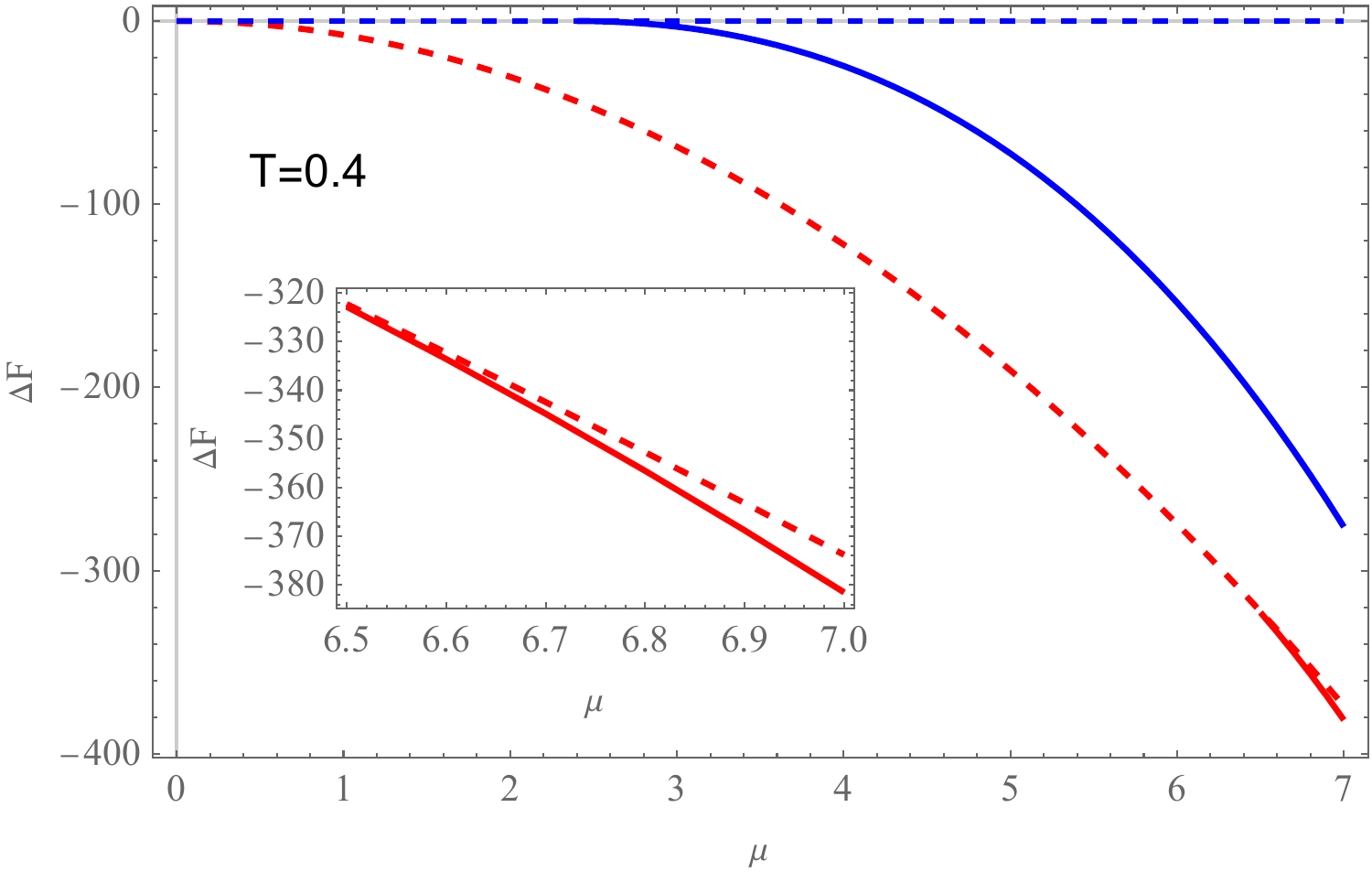}
\includegraphics[scale=0.3]{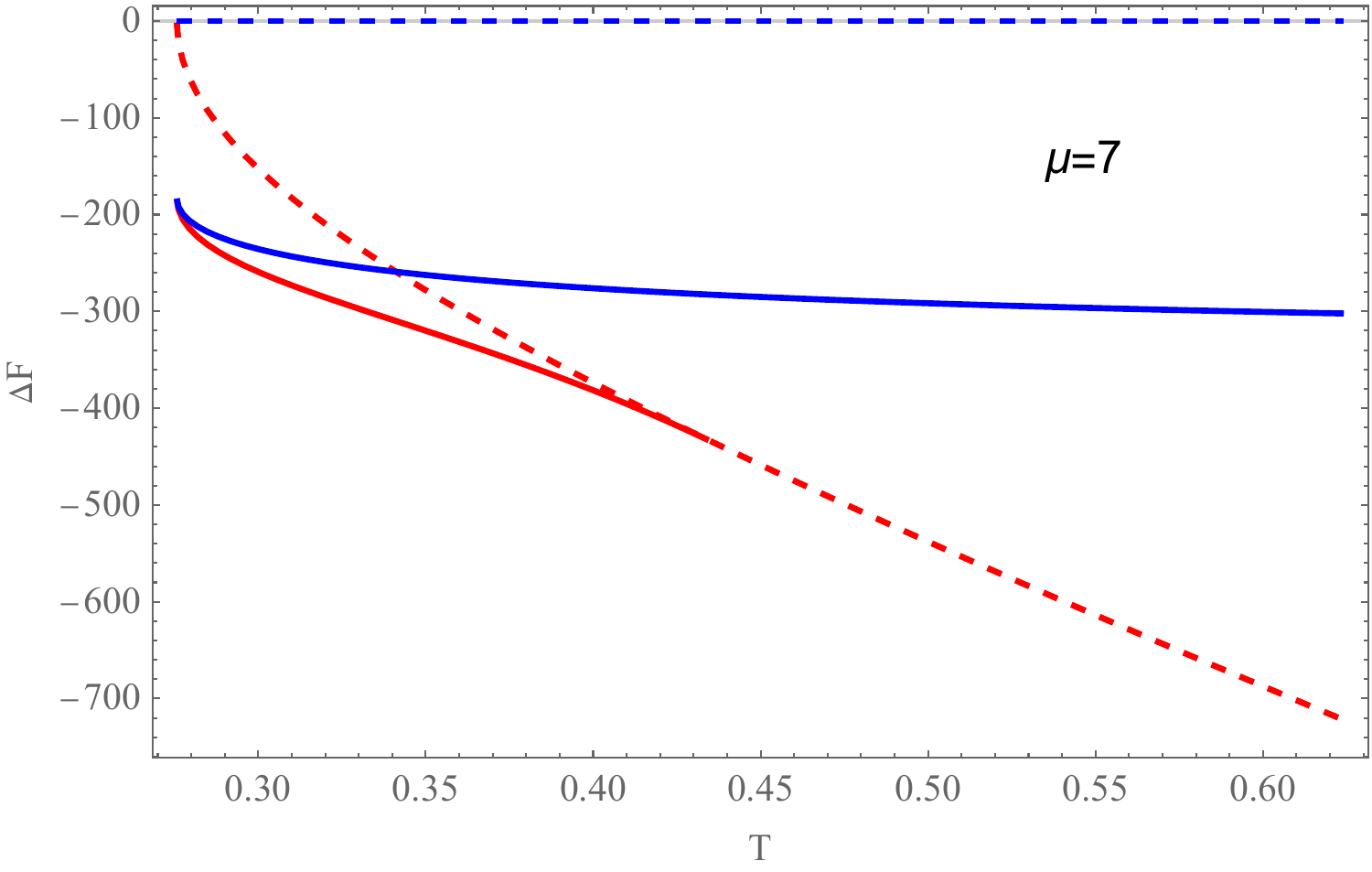}
\includegraphics[scale=0.3]{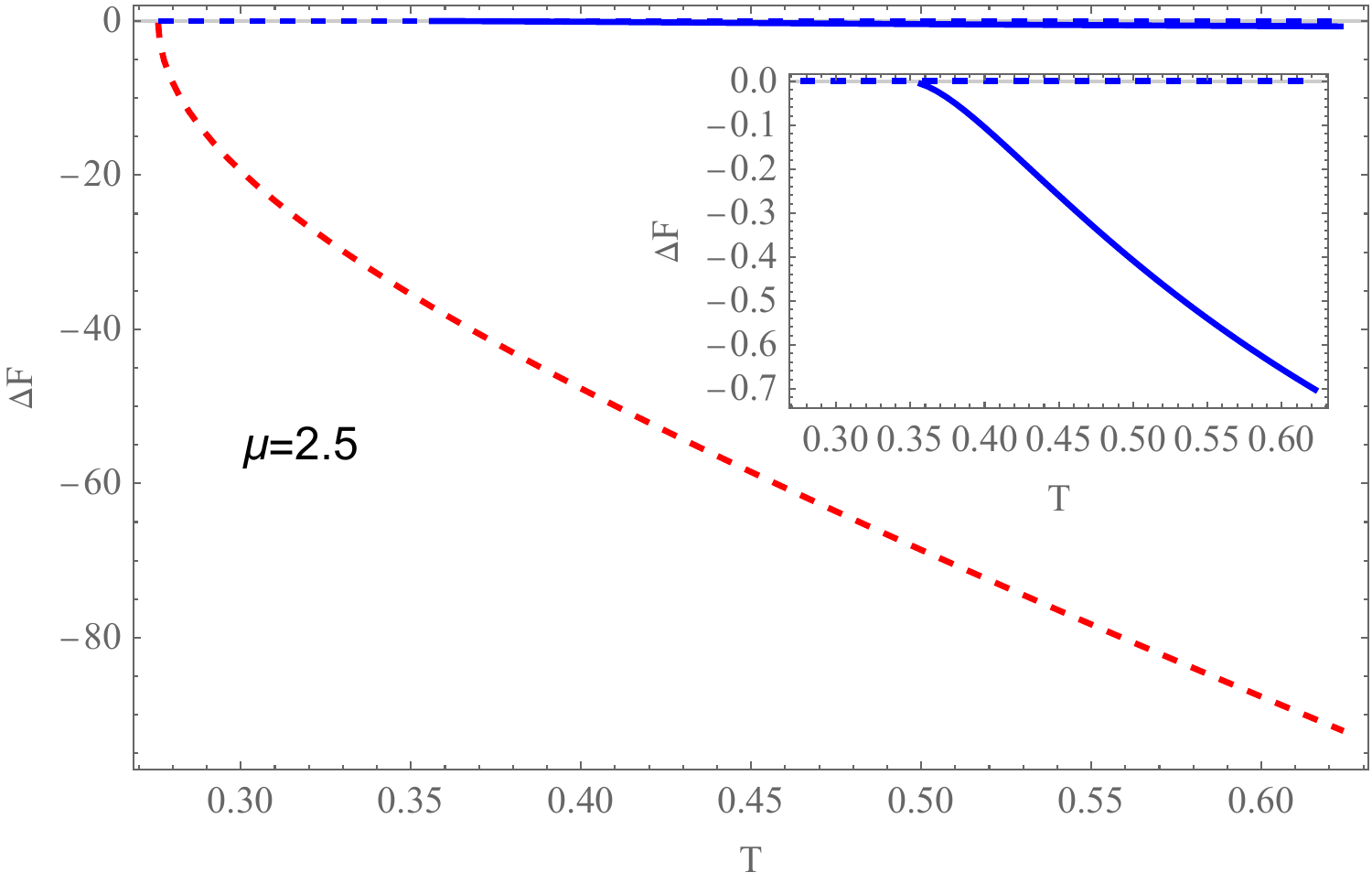}

\caption{The difference in the free energy between all other possible phases and the bald small black hole. The dashed horizontal line denotes the free energy for the bald small black hole, which we set to be zero. The blue, dashed red, and red lines individually denote the free energy of hairy small black hole, bald large black hole, and hairy large black hole with respect to that for the bald small black hole. \label{fig:free_energy}}
\end{figure}

Last, in order to confirm the above phase transition in the phase diagram, we are left with working in the grand canonical ensemble to perform the analysis of the free energy, which can be obtained by the aid of the on-shell condition and boundary condition as
\begin{align}
F
=& -\frac{1}{2}\Big[4\pi\int_{0}^{z_h} z^{-4}j^{\nu}A_{\nu}dz+4\pi\left.\left(z^{-4}F^{z\nu}A_{\nu}\right)\right|_{z=0}\Big] \nonumber\\
=&4\pi\int_0^{z_h}dz\left(\psi^2\frac{A_t^2}f\right)-2\pi\mu\rho.\label{eq:Free energy equation}
\end{align}
With it, we plot in Figure \ref{fig:free_energy} the resulting difference of the free energy of all other possible phases with respect to that for the bald small black hole at given temperature and chemical potential. As is well known, the equilibrium at fixed temperature is unstable for the lower mass of black hole but is locally stable for the higher mass of black hole \cite{HP}. As such the free energy of large black holes is lower than that of small black holes. In this paper, we only calculated the free energy of the matter field in two different black hole backgrounds. The study finds that the free energy of the matter field in the large black hole background is smaller than that in the small black hole background. In conclusion, the total free energy of matter and gravitational fields in the large black hole spacetime background is lower than that in the small black hole case,
which indicates that the physical system in the large black hole background is more thermodynamically stable. On the other hand, in this article, we focus on the probe limit case, therefore, if we consider the contribution of gravitational free energy and matter together, the free energy of the matter field can be neglected, which has no significance for the present work. Here, we use the small bald black hole as a reference, because we can directly compare the thermodynamic stability of small and large black holes, and we are not concerned with their absolute values, only with their relative sizes, so we can determine which one is more stable. The results indicates that we will focus only on the case of large black holes in our future studies. 

As one can see from the free energy difference, the phase transition indicated in Figure \ref{fig:phase_transition} and Figure \ref{fig:temperature_transition} is actually a second-order phase transition. In addition, the hairy large black hole, if such a solution exists, always carries the lowest free energy, thus thermodynamically favorable. Otherwise, the lowest free energy carrier is the bald large black hole. 

\section{Collective modes of holographic superfluid}

In this section, we will perform the linear perturbation analysis of our holographic superfluid and extract its collective modes, which correspond to the quasi-normal modes in the bulk. As such, we like to shift to the ingoing Eddington–Finkelstein coordinates, where the metric reads
\begin{equation}
    ds^2=\frac1{z^2}\left[-f\left(z\right)dt^2-2dzdt+d\theta^2+\sin^2\theta d\varphi^2\right].\label{eq:metric-EF}
\end{equation}
Accordingly, the equations of motion for the bulk matter fields can be written explicitly as follows
\begin{align}
0=& \psi\left(2-z^2A_\theta^2-z^2A_\varphi^2\csc^2\theta-iz^2A_\theta\cot\theta-2f+z\partial_zf-iz^2\csc^2\theta\partial_\varphi A_\varphi-iz^2\partial_\theta A_\theta\right) \nonumber \\
 & +z^2\big(-2iA_\varphi\csc^2\theta \partial_\varphi\psi+\csc^2\theta \partial_\varphi^2\psi-2iA_\theta\partial_\theta\psi+\cot\theta \partial_\theta\psi+\partial_\theta^2\psi+2iA_t\partial_z\psi\big) \nonumber \\
 & +z^2\left(\partial_zf\partial_z\psi+f\partial_z^2\psi-2\partial_t\partial_z\psi \right)+i\psi z^2\partial_zA_t \label{eq:psi},\\ 
0= & -\partial_z^2A_t+\left(\cot\theta\right)\partial_zA_\theta+i\left(\psi^*\partial_z\psi-\psi\partial_z\psi^*\right)+\left(\csc^2\theta\right)\partial_z\partial_\varphi A_\varphi+\partial_z\partial_\theta A_\theta, \label{eq:At}\\
0= & \partial_t\partial_zA_t-f\csc^2\theta\partial_z\partial_\varphi A_\varphi-f\partial_z\partial_\theta A_\theta-\cot\theta\left(\partial_\theta A_t+f\partial_zA_\theta-\partial_tA_\theta\right)+2A_t\psi\psi^*-\partial_\theta^2A_t \nonumber \\
 &-if\left(\psi^*\partial_z\psi-\psi\partial_z\psi^*\right)+i\left(\psi^*\partial_t\psi-\psi\partial_t\psi^*\right)-\csc^2\theta\left(\partial_\varphi^2A_t-\partial_t\partial_\varphi A_\varphi\right)+\partial_t\partial_\theta A_\theta,\label{eq:Az}\\
0= & f\partial_z^2A_\theta+\csc^2\theta\partial_\varphi^2A_\theta-2A_\theta\psi\psi^*-i\left(\psi^*\partial_\theta\psi-\psi\partial_\theta\psi^*\right)-\csc^2\theta\partial_\theta\partial_\varphi A_\varphi+\partial_zf\partial_zA_\theta\nonumber \\
 &+\partial_z\partial_\theta A_t-2\partial_t\partial_zA_\theta, \label{eq:A_theta}\\
0= &-f\partial_z^2A_\varphi+2A_\varphi\psi\psi^*+i\big(\psi^*\partial_\varphi\psi-\psi\partial_\varphi\psi^*\big)-\cot\theta \partial_\varphi A_\theta-\big(\partial_\theta^2A_\varphi-\partial_\theta\partial_\varphi A_\theta\big)\nonumber\\&-\partial_zf\partial_zA_\varphi-\partial_z\partial_\varphi A_t+2\partial_t\partial_zA_\varphi+\cot\theta \partial_\theta A_\varphi, \label{eq:A_varphi}
\end{align}
where the axial gauge $A_z=0$ is also used in the ingoing Eddington-Finkenstein coordinates. The corresponding background solution in this new coordinate system can be obtained readily via the coordinate transformation supplemented by the following gauge transformation 
\begin{equation}
  A\rightarrow A_S+\nabla\beta \quad\psi\rightarrow \psi_{S} e^{i\beta},
\end{equation}
with $\beta=-\int\frac{A_t}fdz$, $A_S$ and $\psi_{S}$ the corresponding background profile in the Schwarzschild coordinates.

To investigate the quasi-normal modes on top of the background in question, we are required to make the following ansatz for the perturbed bulk fields, i.e.,
\begin{align}
\delta\psi= & e^{-i\omega t}q_1\left(z\right)Y_{l,m}\left(\theta,\varphi\right)+e^{i\omega^*t}q_2^*\left(z\right)Y_{l,m}^*\left(\theta,\varphi\right),\label{eq:deltapsi}\\
\delta A_{t}= & e^{-i\omega t}a\left(z\right)Y_{l,m}\left(\theta,\varphi\right)+e^{i\omega^*t}a^*\left(z\right)Y_{l,m}^*\left(\theta,\varphi\right),\label{eq:deltaA_t}\\
\delta A_{\theta}= & e^{-i\omega t}\left[\frac{b\left(z\right)}{\sin\theta}\partial_\varphi Y_{l,m}+c\left(z\right)\partial_\theta Y_{l,m}\right]+e^{i\omega^*t}\left[\frac{b^*\left(z\right)}{\sin\theta}\partial_\varphi Y_{l,m}^*+c^*\left(z\right)\partial_\theta Y_{l,m}^*\right],\label{eq:deltaA_theta}\\
\delta A_{\varphi}= & e^{-i\omega t}\left[-b\left(z\right)\sin\theta\partial_\theta Y_{l,m}+c\left(z\right)\partial_\varphi Y_{l,m}\right]+e^{i\omega^*t}\left[-b^*\left(z\right)\sin\theta\partial_\theta Y_{l,m}^*+c^*\left(z\right)\partial_\varphi Y_{l,m}^*\right],\label{eq:deltaA_varphi}
\end{align}
whereby the linearized perturbation equations read
\begin{align}
0= & z^2f\partial_z^2q_1+\left[2iz^2\left(A_t+\omega\right)+z^2\partial_zf\right]\partial_zq_1+\left[2-2f-\left(l+l^2\right)z^2+iz^2\partial_zA_t+z\partial_zf\right]q_1\nonumber\\&+iz^2\psi\partial_za+2iaz^2\left(\partial_z\psi\right)+ic\left(l+l^2\right)z^2\psi,\label{eq:q1}\\
0= &z^2f\partial_z^2q_2-\left[2iz^2\left(A_t-\omega\right)-z^2\partial_zf\right]\partial_zq_2+\left[2-2f-z^2\left(l+l^2\right)-iz^2\partial_zA_t+z\partial_zf\right]q_2\nonumber\\&-iz^2\psi^*\partial_za-2iaz^2\left(\partial_z\psi^*\right)-ic\left(l+l^2\right)z^2\psi^*,\label{eq:q2}\\
0= &-\partial_z^2a-\left(l^2+l\right)\partial_zc+i\left(\psi^*\partial_zq_1-q_1\partial_z\psi^*-\psi\partial_zq_2+q_2\partial_z\psi\right), \label{eq:a}\\
0= & f\partial_z^2b-2b\psi^*\psi+2i\omega\partial_zb+\left(\partial_zb\right)\left(\partial_zf\right)-bl\left(l+1\right),\label{eq:b}\\
0= & f\partial_z^2c-2c\psi^*\psi+2i\omega\partial_zc+\left(\partial_zc\right)\left(\partial_zf\right)+i\left(q_2\psi-\psi^*q_1\right)+\partial_za,\label{eq:c}\\
0=&2a\psi\psi^*-a\left[-l\left(l+1\right)\right]+2A_t\left(q_1\psi^*+q_2\psi\right)-f\left(\partial_zc\right)\left[-l\left(l+1\right)\right]-i\omega c\left[-l\left(l+1\right)\right]\nonumber\\&-i\omega\partial_z a+\omega\left(q_1\psi^*-q_2\psi\right)+if\left(\psi\partial_zq_2-q_2\partial_z\psi+q_1\partial_z\psi^*-\psi^*\partial_zq_1\right).\label{eq:jz-equation}
\end{align}
It is noteworthy that Eq. (\ref{eq:b}) and Eq. (\ref{eq:c}) should be discarded in the case of $l=0$, where there is indeed essentially no excitation for the $b$ and $c$ modes, which can be seen from Eq. (\ref{eq:A_theta}) and Eq. (\ref{eq:A_varphi}). Eqs. (\ref{eq:q1}-\ref{eq:c}) in the bulk as well as Eq. (\ref{eq:jz-equation}) evaluated at the AdS boundary supplemented with the Dirichlet boundary conditions for all modes on the AdS boundary can be formulated in terms of the generalized eigenvalue problem for the quasi-normal modes, which can be solved readily by our numerical method. Furthermore, note that the $b$ mode decouples from the other modes. So in the case of $l=0$, we have the quasi-normal frequencies from the coupled $q_1$, $q_2$, $a$ modes, while in the case of $l\neq 0$ we have the quasi-normal frequencies separately from the $b$ mode and the coupled $q_1$, $q_2$, $a$, $c$ modes, respectively. 

\begin{figure}
\includegraphics[scale=0.3]{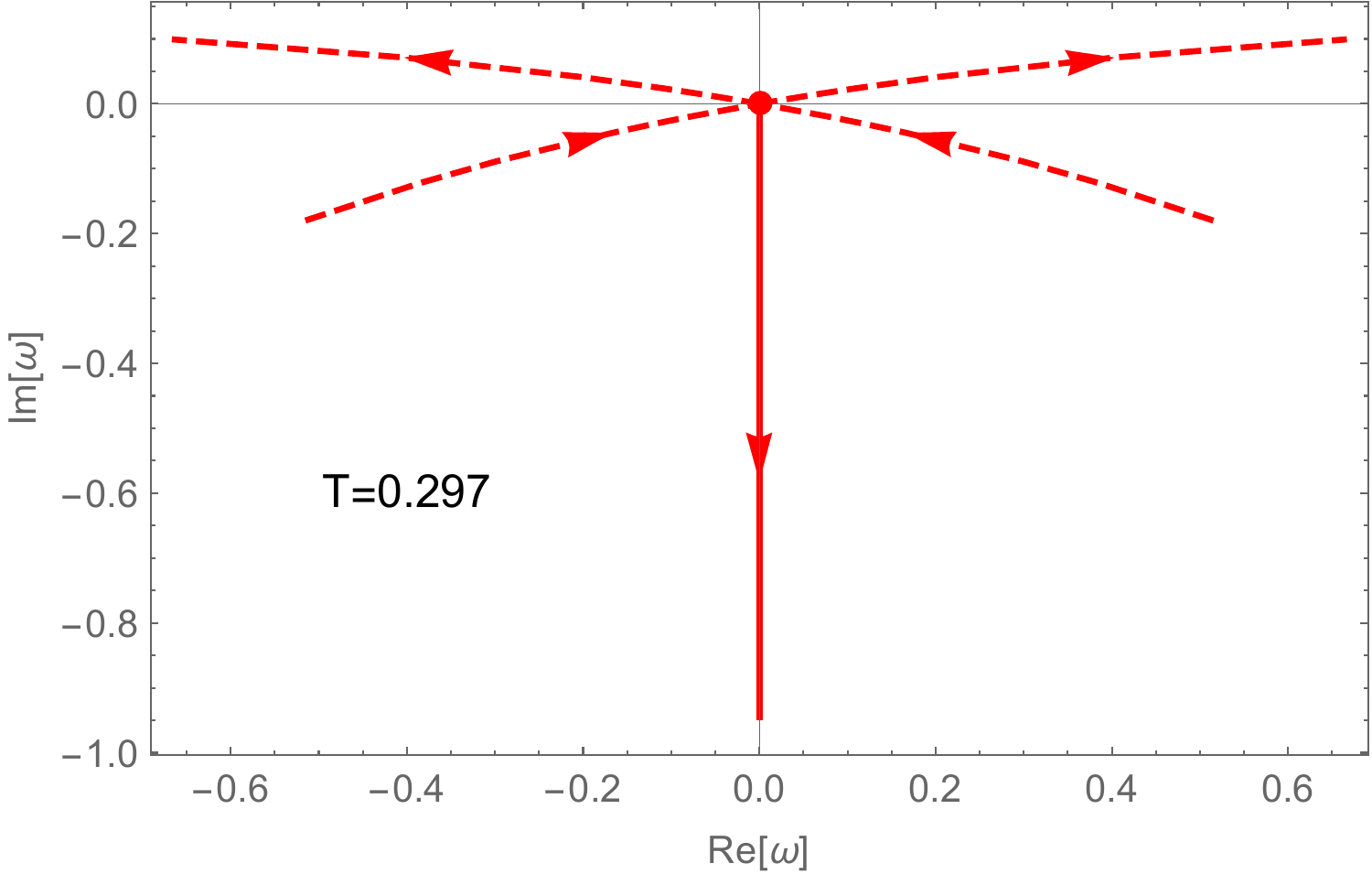}
\includegraphics[scale=0.3]{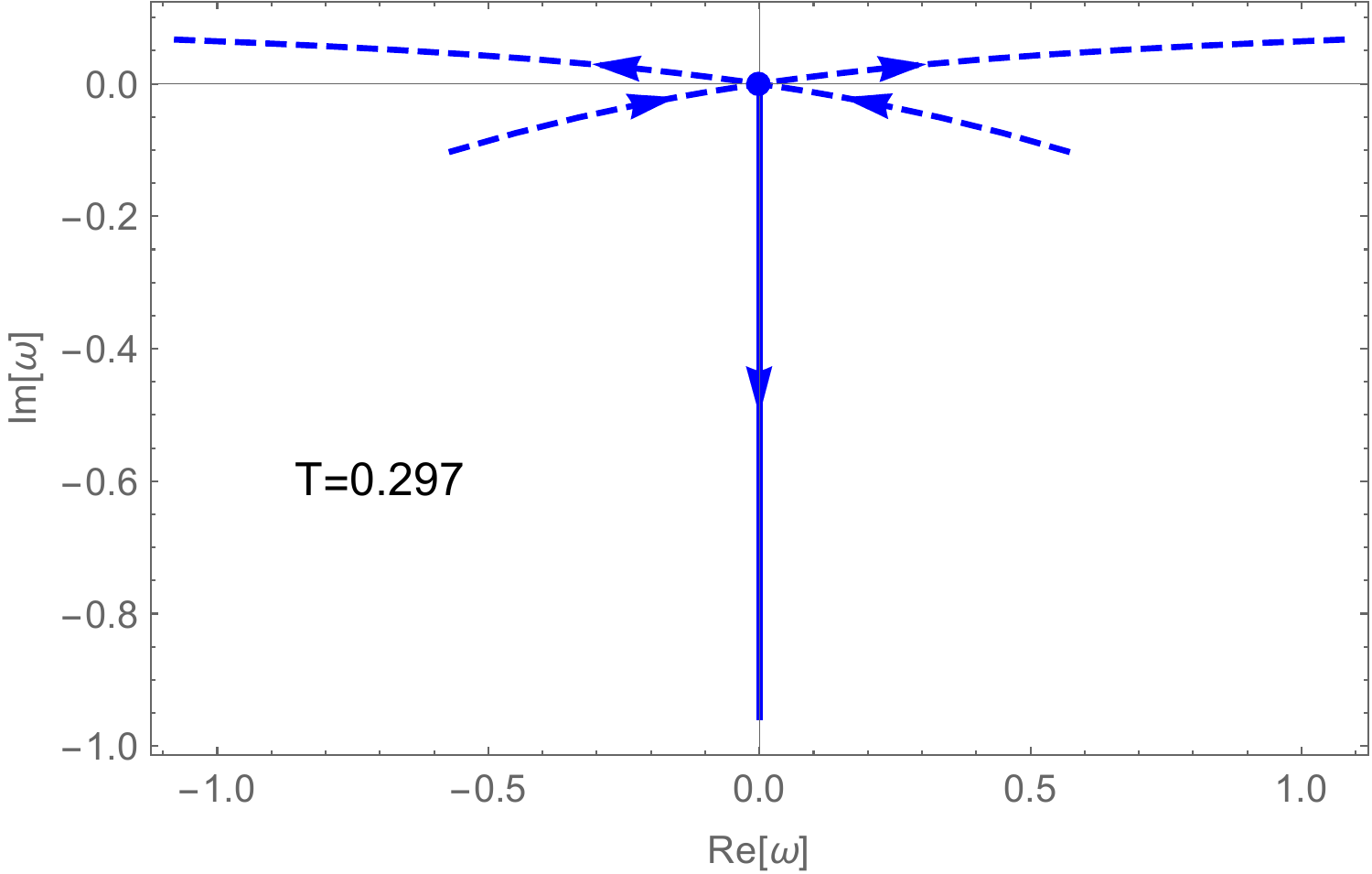}
\includegraphics[scale=0.3]{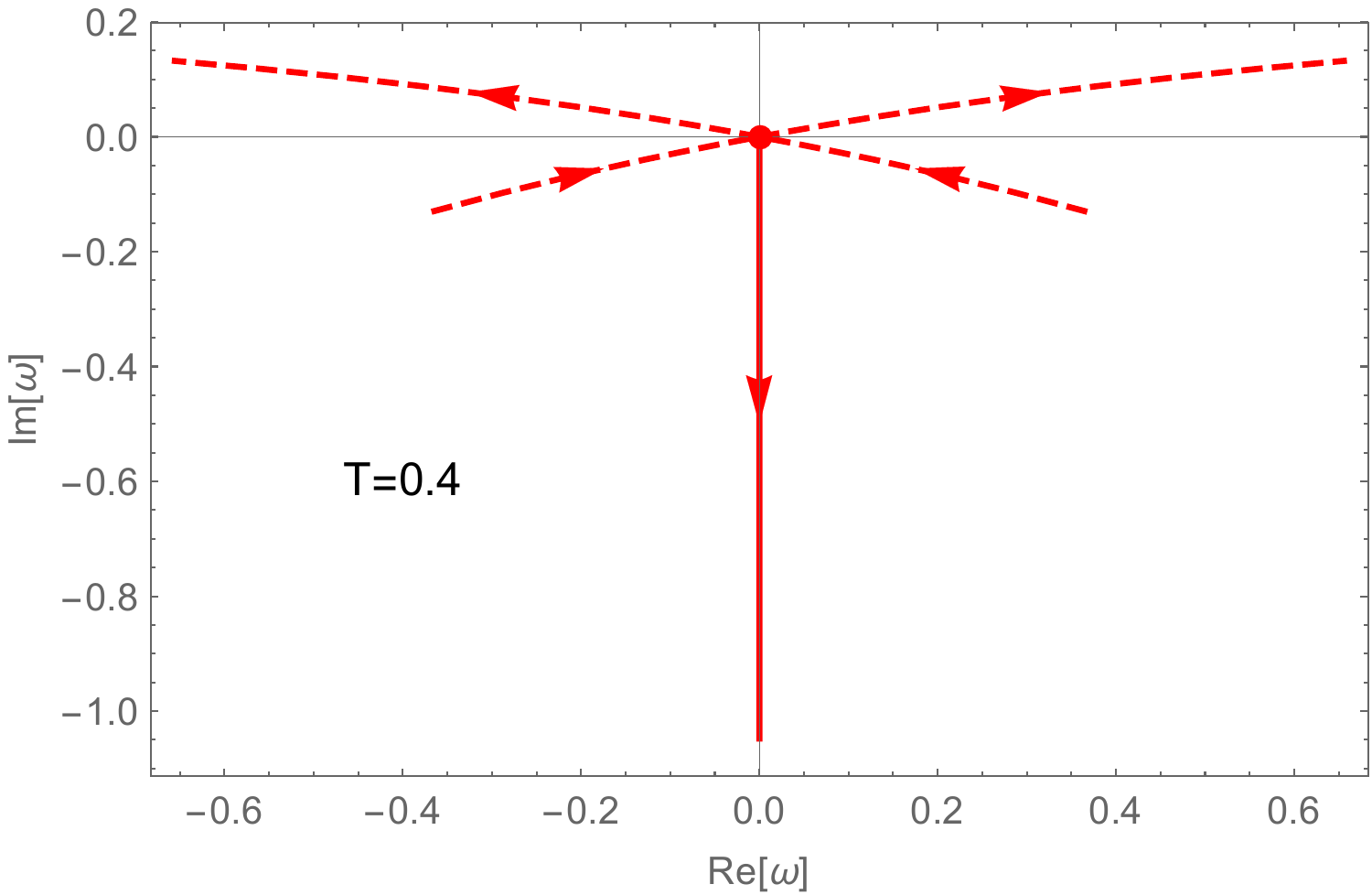}
\includegraphics[scale=0.3]{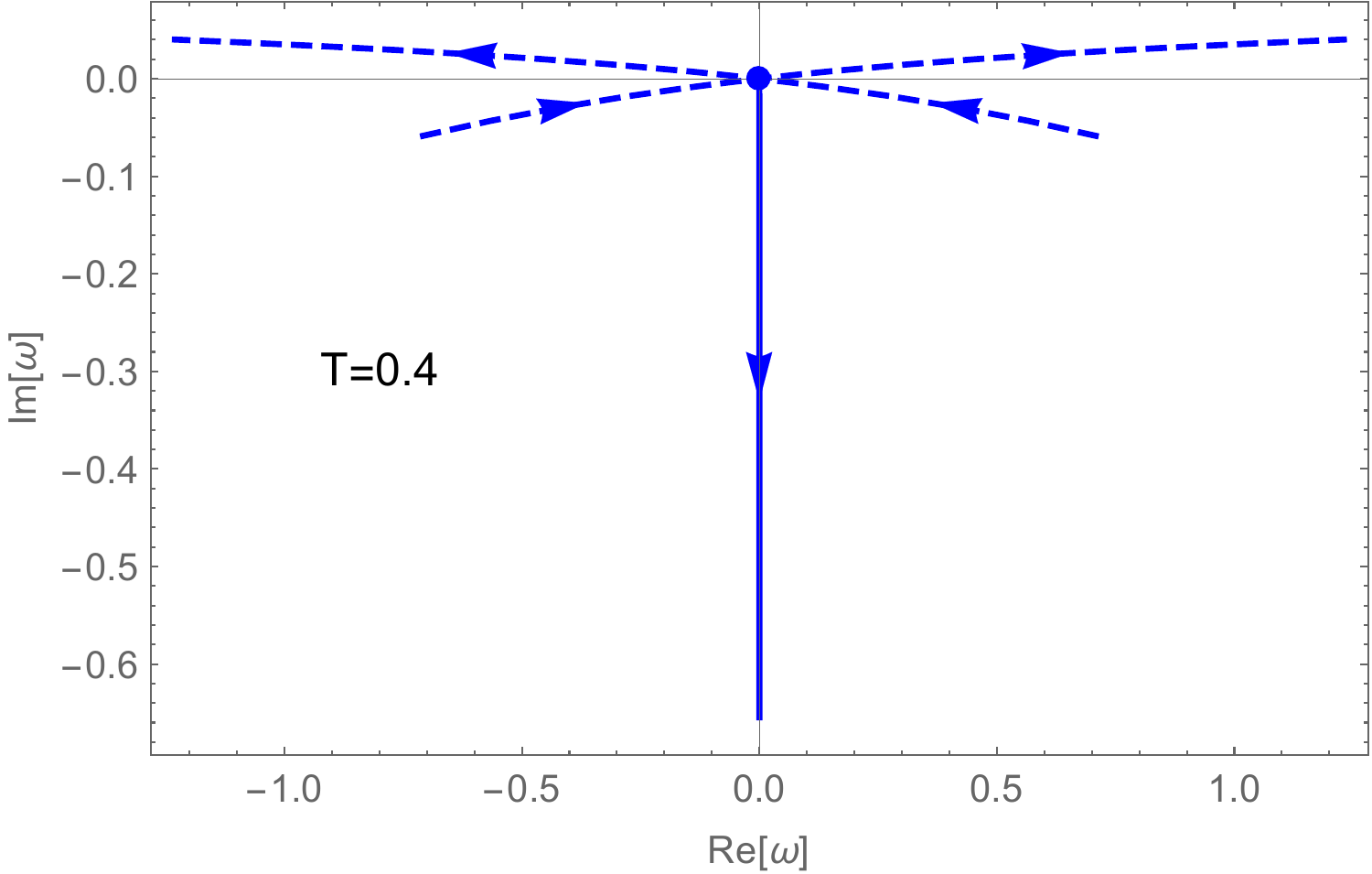}
\caption{The variation of the low-lying quasi-normal modes with the chemical potential at fixed temperature. The dashed red and dashed blue lines are for the bald large and small black holes while the red and blue lines are for the hairy large and small black holes. The chemical potential is varied from 3.0 to 5.5 for the top left panel, from 1.8 to 4.4 for the top right panel, from 5.4 to 7.8 for the bottom left panel, and from  1.4 to 4.0 for the bottom right panel. \label{fig:omega1}}
\end{figure}

\begin{figure}
\includegraphics[scale=0.3]{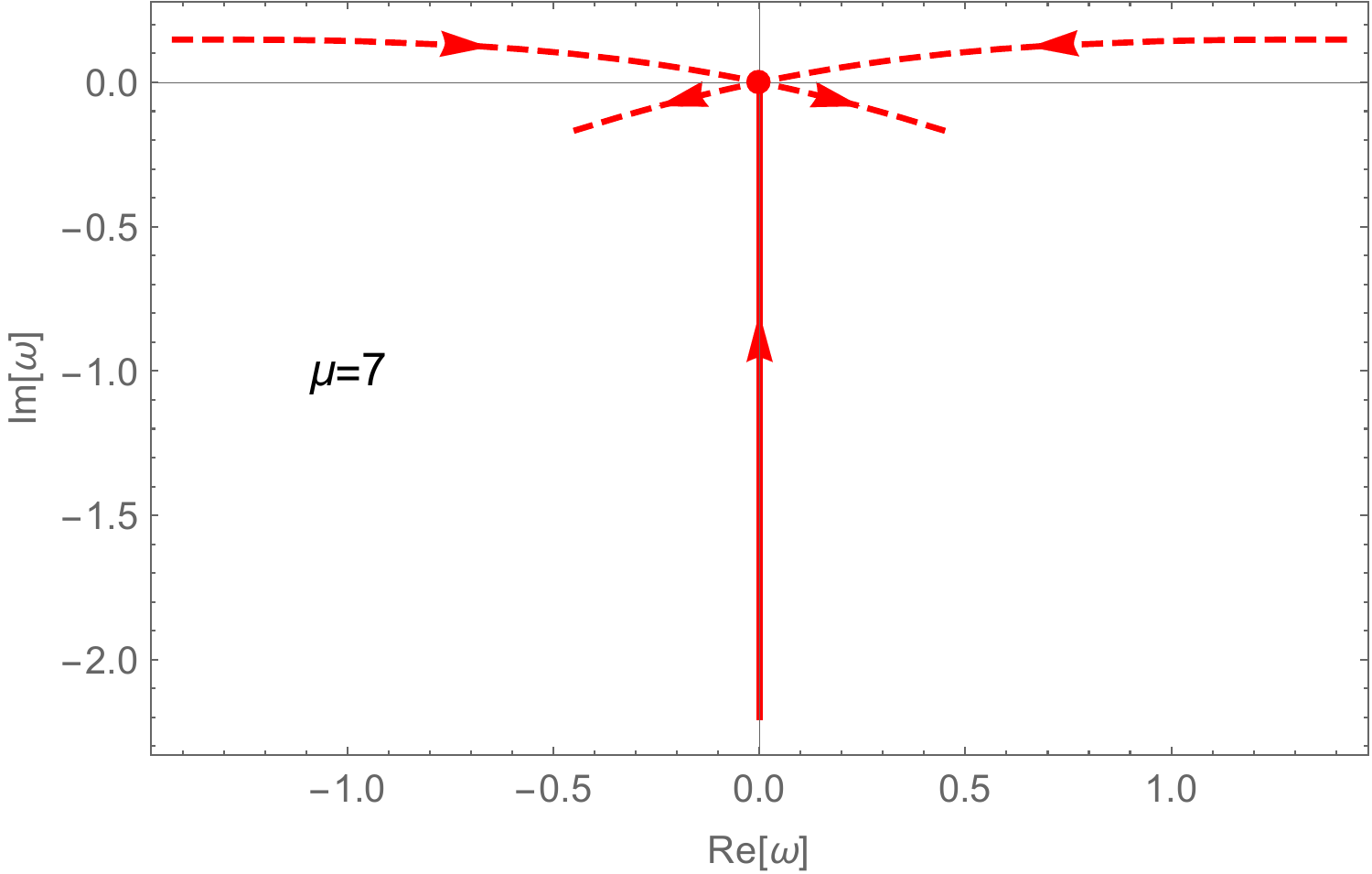}
\includegraphics[scale=0.3]{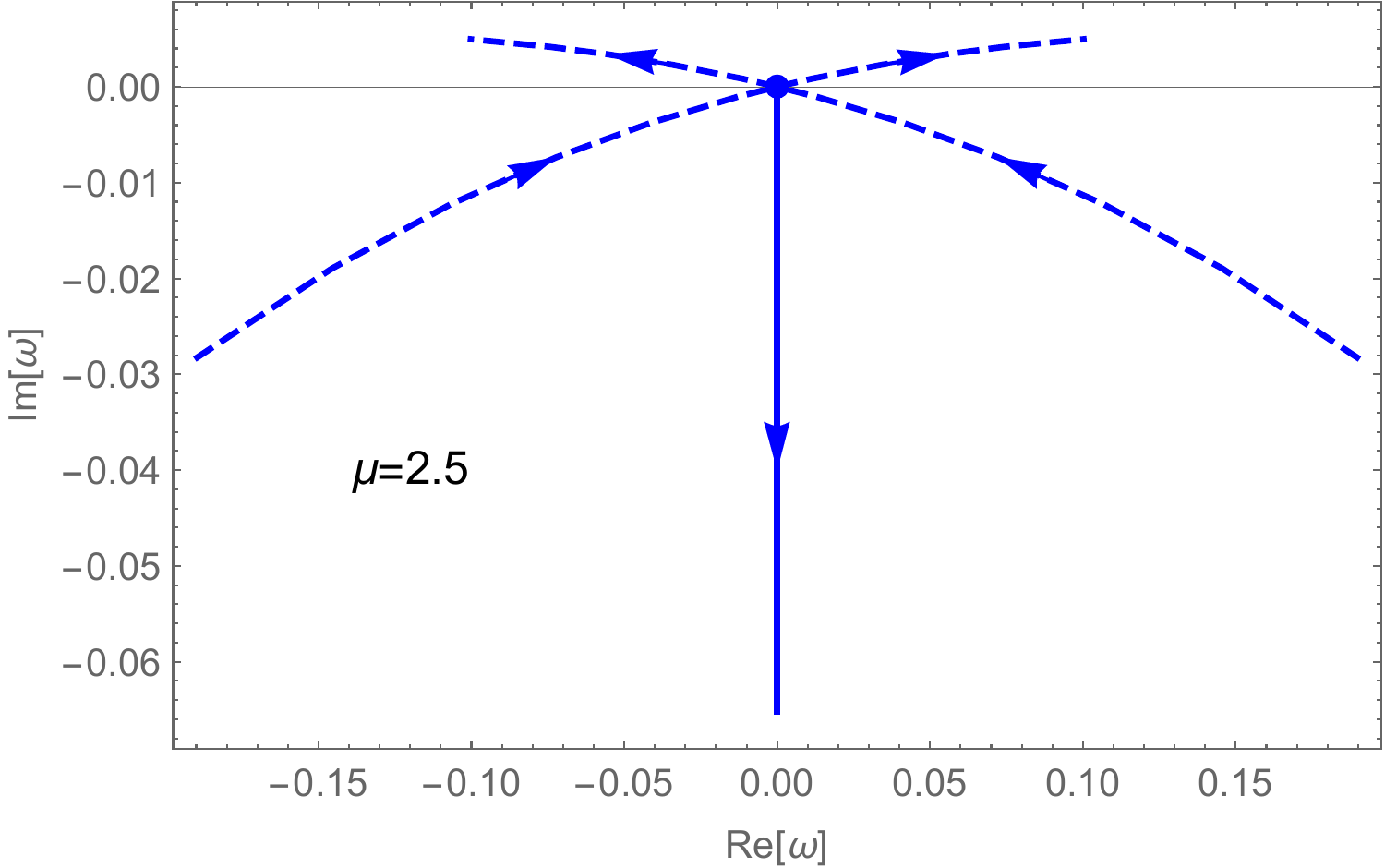}
\caption{The variation of the low-lying quasi-normal modes with the temperature at fixed chemical potential. The dashed red and dashed blue lines are for the bald large and small black holes while the red and blue lines are for the hairy large and small black holes. The temperature is varied from 0.3 to 0.5 for the left panel, and from 0.29 to 0.41 for the right panel.\label{fig:omega2}}
\end{figure}

We first look at the behavior of the $l=0$ quasi-normal modes
near the phase transition, which is depicted in Figure \ref{fig:omega1} and Figure \ref{fig:omega2}. As the normal fluid approaches the critical point, the two low-lying modes, which lie symmetrically in the lower $\omega$ plane, migrate symmetrically towards the origin. Exactly at the critical point, they meet each other at the origin. If the system is kept in the normal fluid state beyond the critical point, one can see that the two low-lying modes will continue to migrate
towards the upper $\omega$ plane, which signals the dynamical instability of the normal fluid state. On the other hand, if the system transits to the superfluid state, one can see that one mode remains at the origin as the Goldstone mode, while the other mode migrates down the imaginary axis as the Higgs mode. Although our system is confined on the finite sphere, the above result indicates the equivalence of the dynamical instability and the thermodynamical instability still holds for our system. 

\begin{figure}
\includegraphics[scale=0.3]{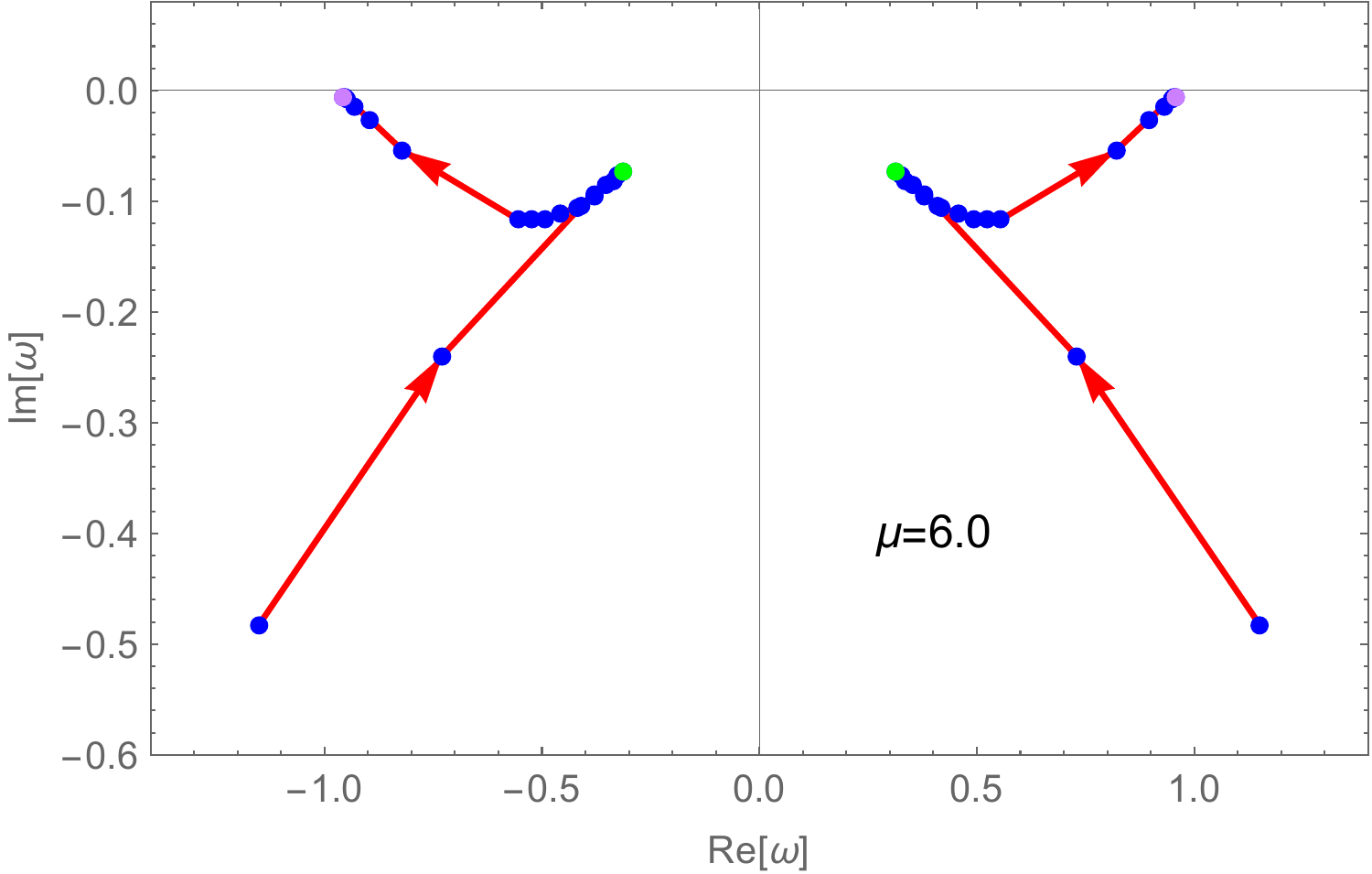}
\includegraphics[scale=0.3]{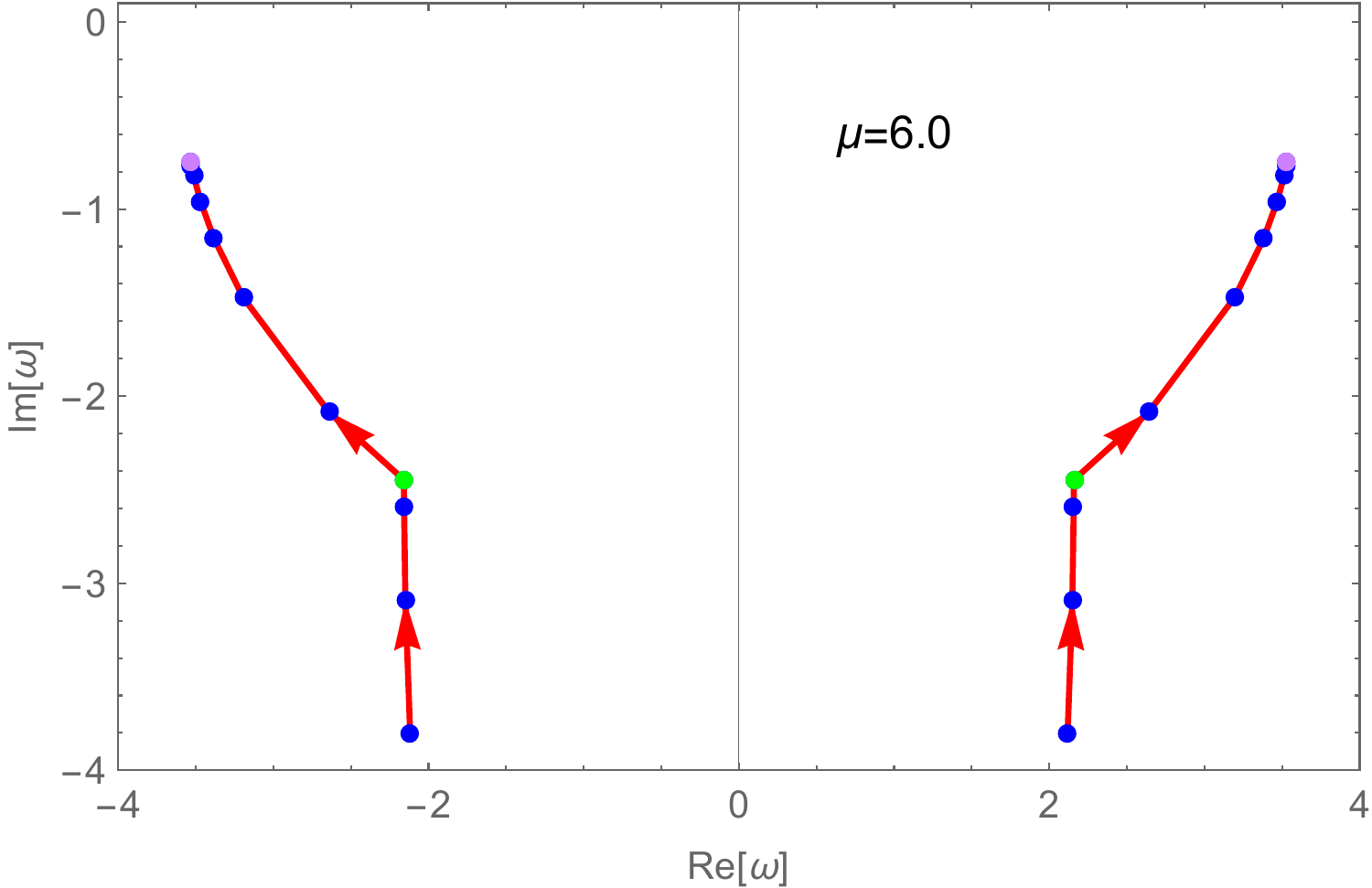}
\includegraphics[scale=0.3]{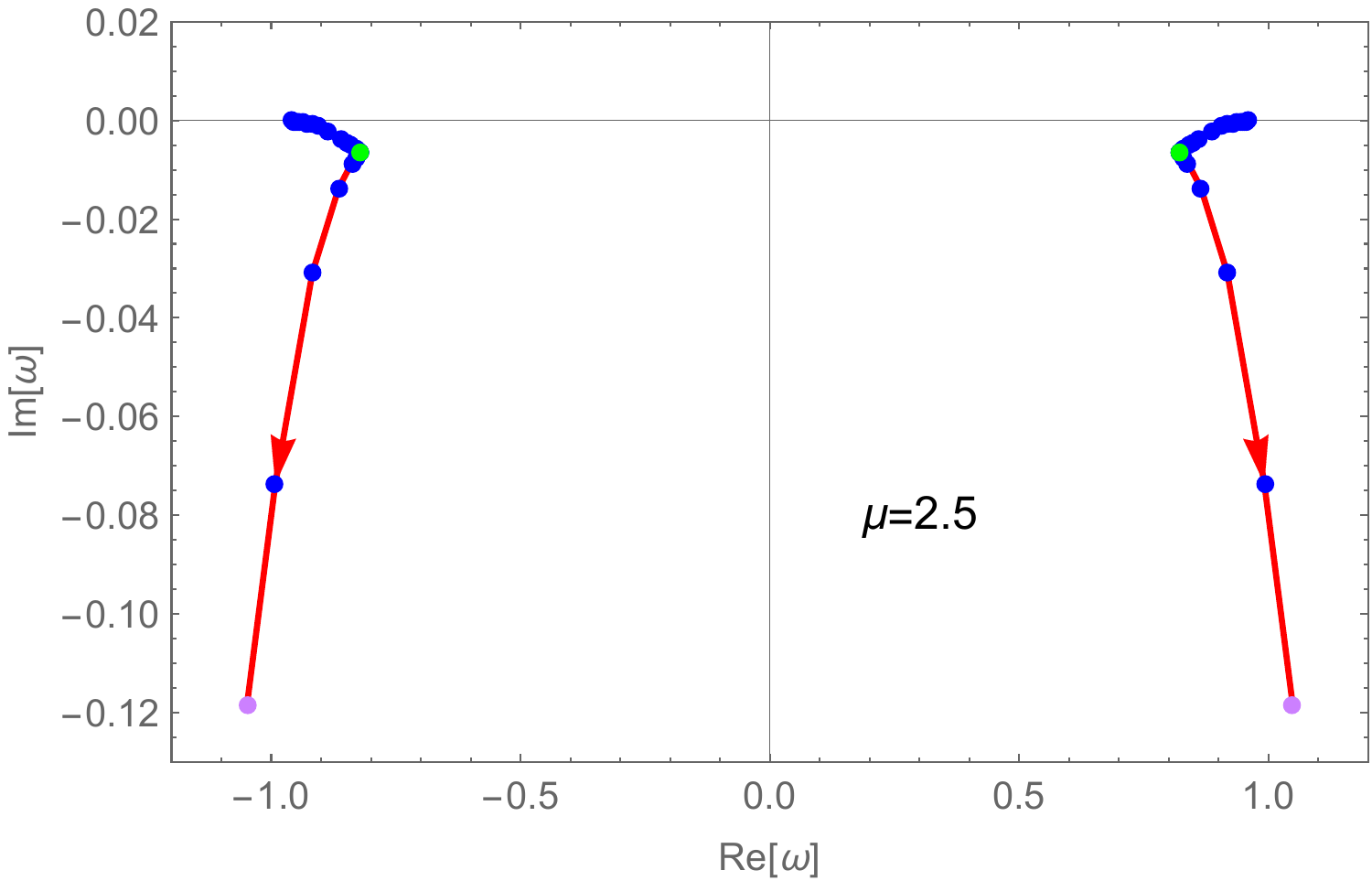}
\includegraphics[scale=0.3]{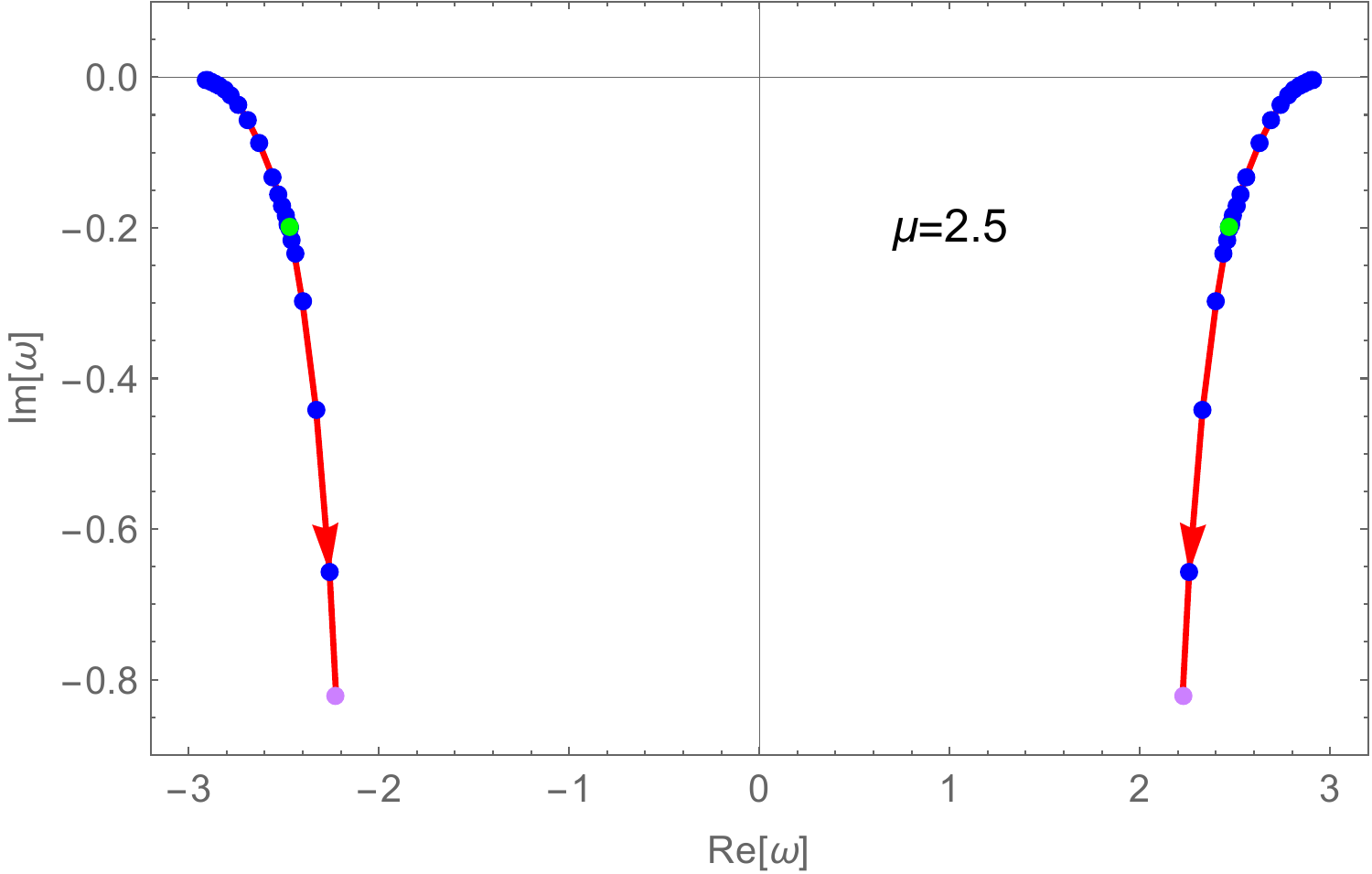}
\caption{The variation of the low-lying quasi-normal modes with the temperature at fixed chemical
potential for large (top plots) and small (bottom plots) black holes, where angular quantum number is set to $l=1$. The temperature is varied from
$T_{min}=\sqrt{3}/2\pi$ to 0.52 for large black hole, from $T_{min}=\sqrt{3}/2\pi$ to 0.743 for small black hole. As the temperature decreases, the two modes move in the direction indicated by the arrow. The left plots are the case of scalar channel and the right plots are from independent equation of $b$ field. The green dot 
represents phase transition from normal state to superfluid state for large black hole, the situation is exactly opposite for small black hole. The purple dot represents the state of the lowest temperature. \label{fig:modes with temperature as L=1}}
\end{figure}

\begin{figure}
\includegraphics[scale=0.3]{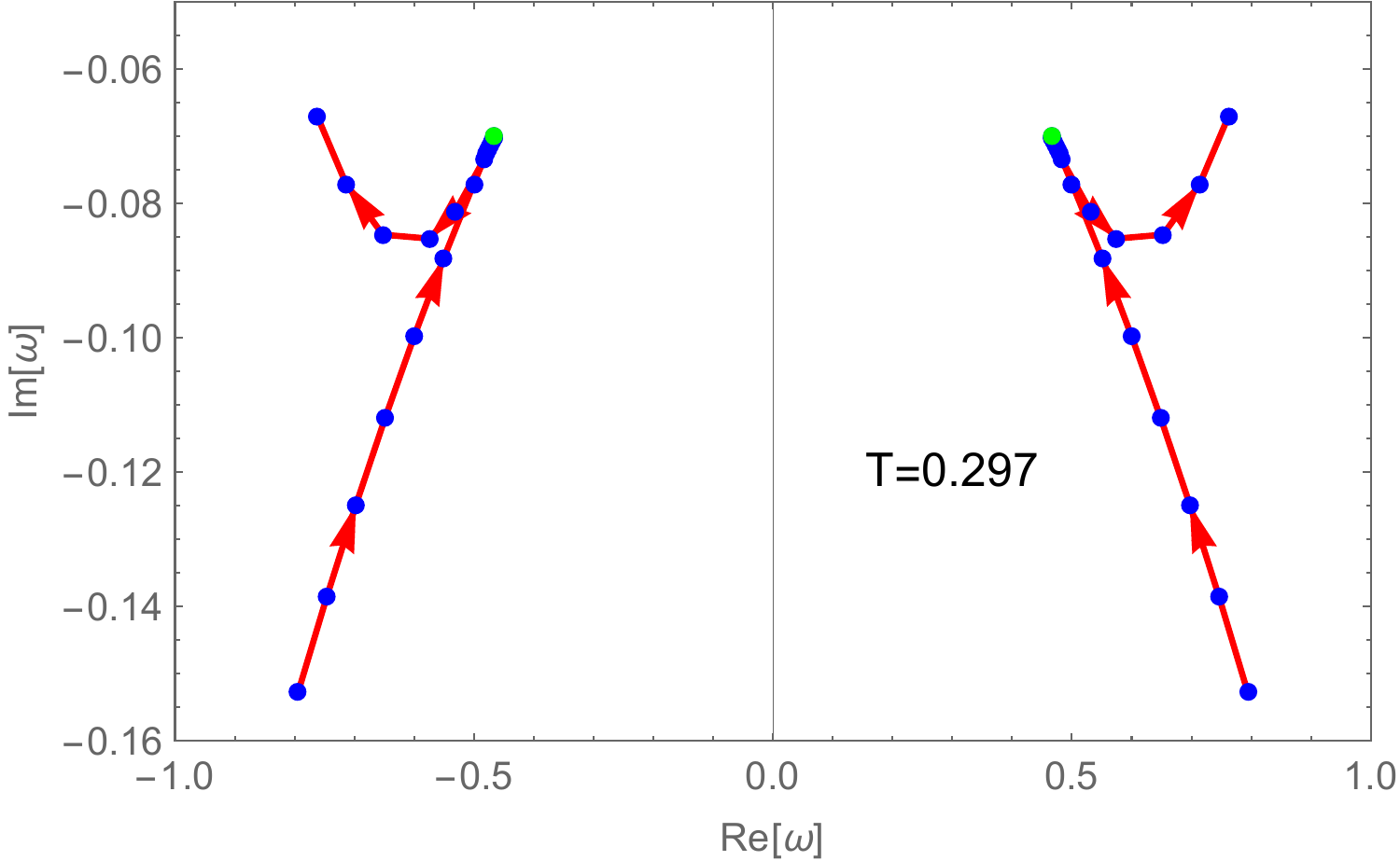}
\includegraphics[scale=0.3]{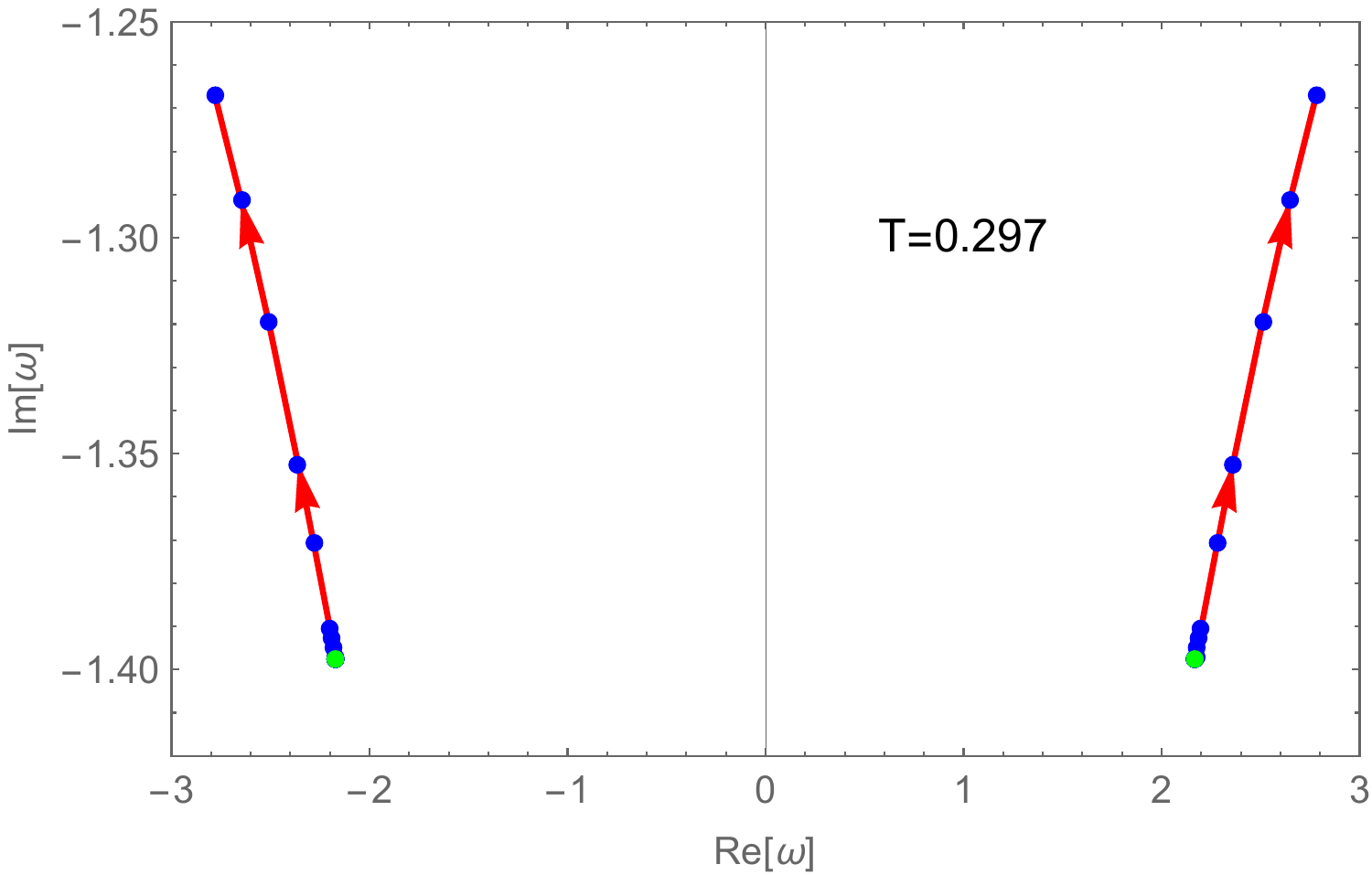}
\includegraphics[scale=0.3]{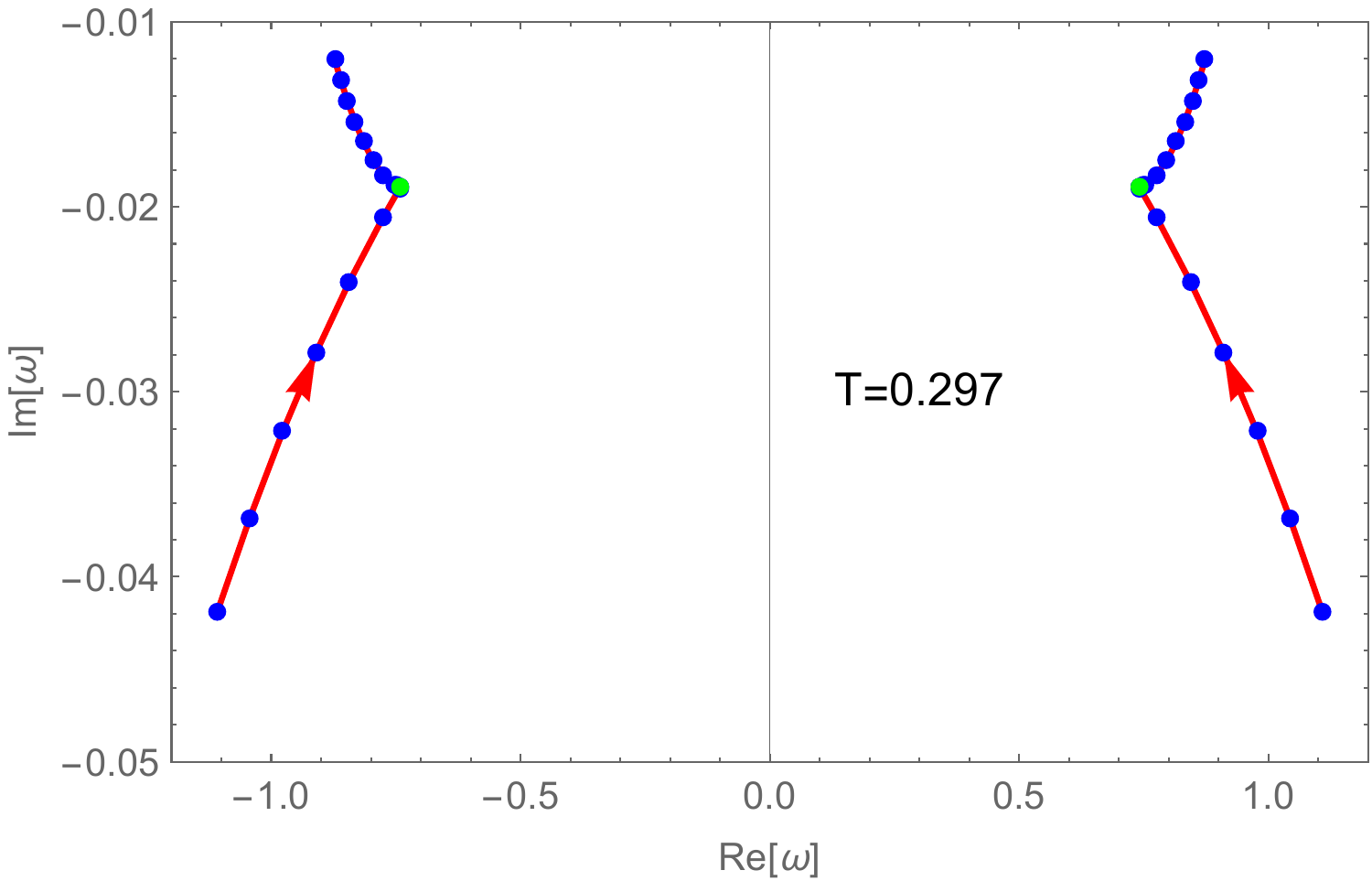}
\includegraphics[scale=0.3]{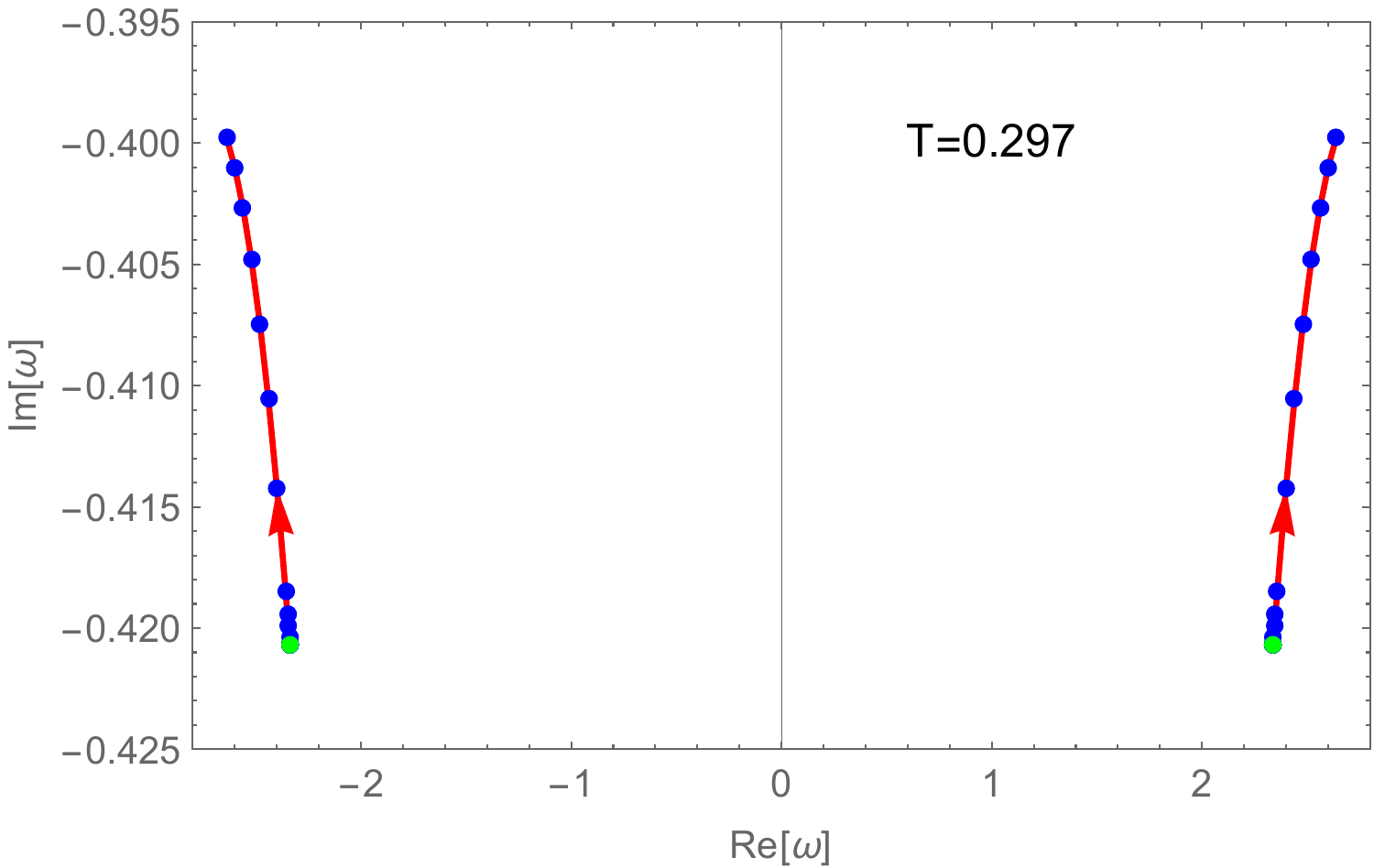}
\caption{The variation of the low-lying quasi-normal modes with chemical
potential at fixed temperature for $l=1$ case under large (top plots) and small (bottom plots) black holes background. Chemical
potential is varied from 3.5 to 5.0 for large black hole, from 2.2 to 3.5 for small black hole. Left plots are the case of scalar channel and the right plots are from independent equation of $b$ field. The green dot 
represents phase transition from normal state to superfluid state for both two types of black holes. Thus, we can identify that the modes from the $b$ field do not move until phase transition occurs.\label{fig:modes with chemical potential as L=1}}
\end{figure}

In what follows, we investigate the behavior of the superfluid system for the case of $l\neq 0$ with large black hole and small black hole being our background geometry, respectively. Initially, we studied the variation of the low-lying quasi-normal modes with the temperature at fixed chemical potential, which is presented in Figure \ref{fig:modes with temperature as L=1}. The two top panels are calculated in the background of the large black hole, the bottom two panels are from small black hole. From the top left panel, two scalar modes move towards the origin with temperature decreasing; however, they change direction after finishing the phase transition represented by the green dot. Until the temperature reaches the lowest value (characterized by purple dot), both modes remain stable all the time. However, the scalar modes in the small black hole (bottom left panel) exhibit anomalies with changes in temperature. When the temperature is lower than the critical point, the system is in a normal flow state, while above the critical temperature the system transits to a superfluid state. The right column is from the independent equation of the $b$ field. It shows an inflection point at the location of the phase transition with large black hole being our background geometry, but no inflection point appears in the case of a small black hole. Meanwhile, we explore the impact associated with chemical potential on the low-lying quasi-normal modes at fixed temperature, which is shown in Figure \ref{fig:modes with chemical potential as L=1}. Similarly, the dynamic behaviors of the low-lying quasi-normal modes undergoes a turning point at the phase transition location. Remarkably, the low-lying quasi-normal modes of $b$ field do not move until the phase transition occurs, which is reasonable according to Equation (\ref{eq:b}).

\begin{figure}
\includegraphics[scale=0.19]{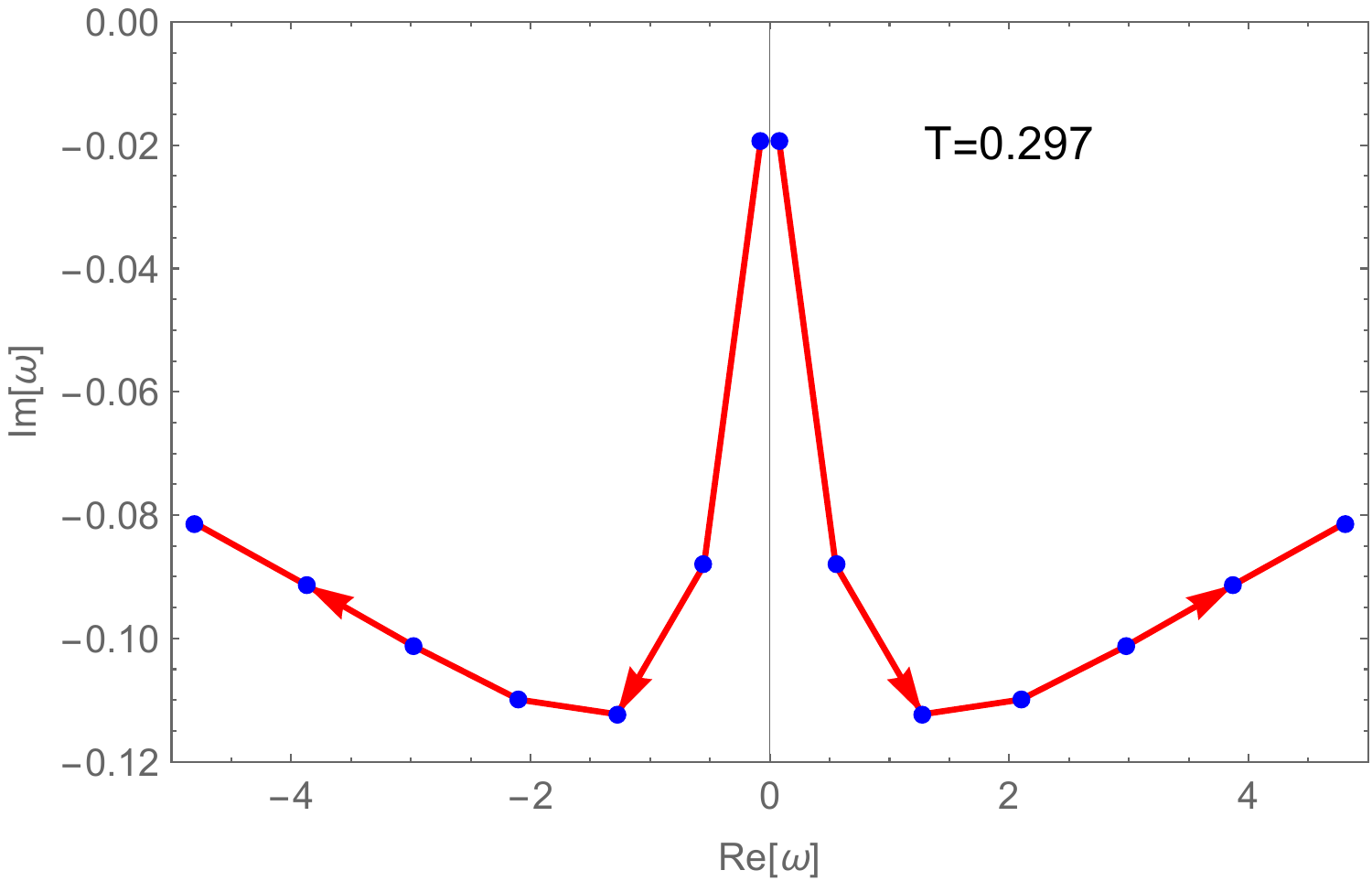}
\includegraphics[scale=0.19]{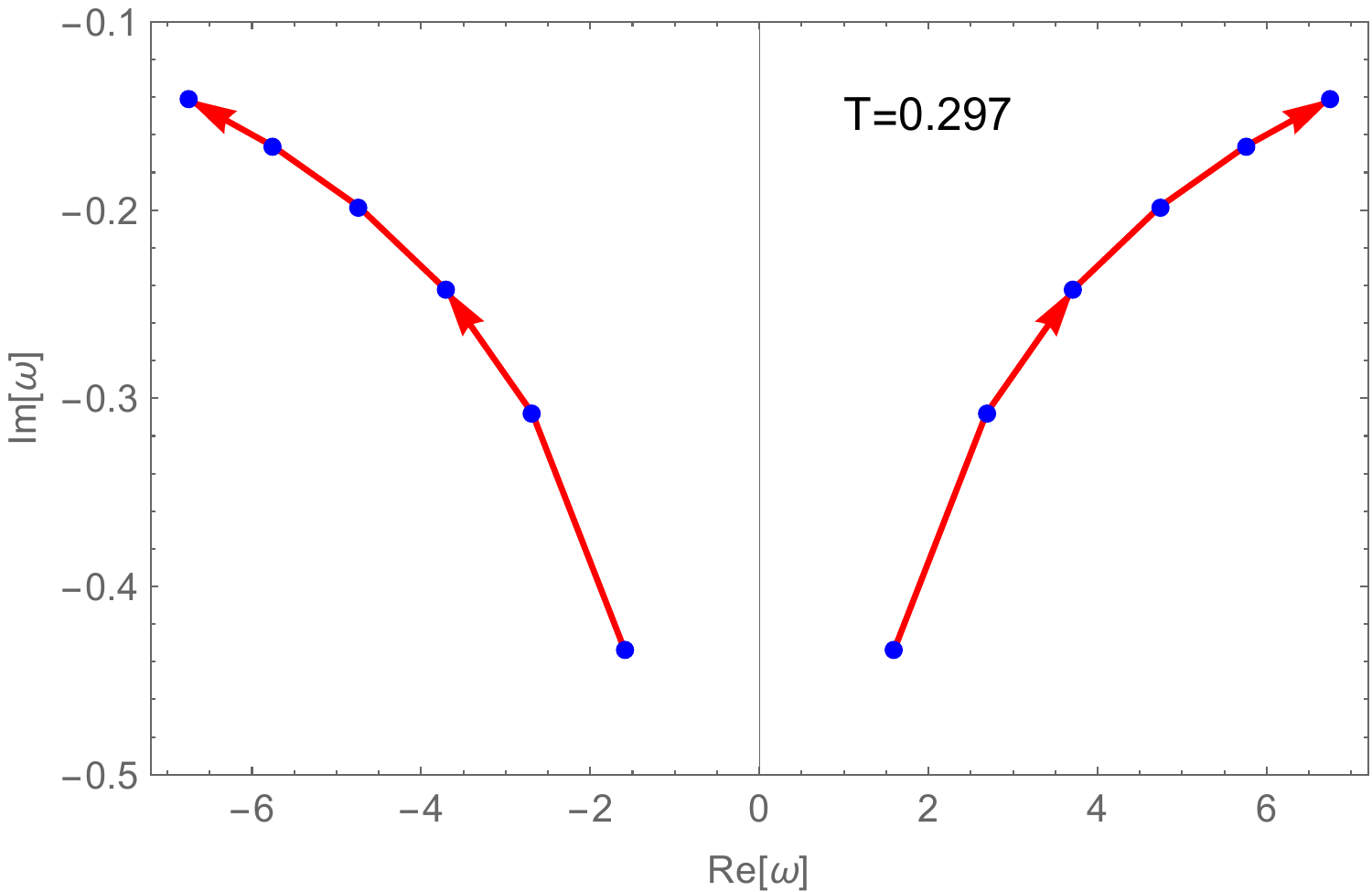}
\includegraphics[scale=0.19]{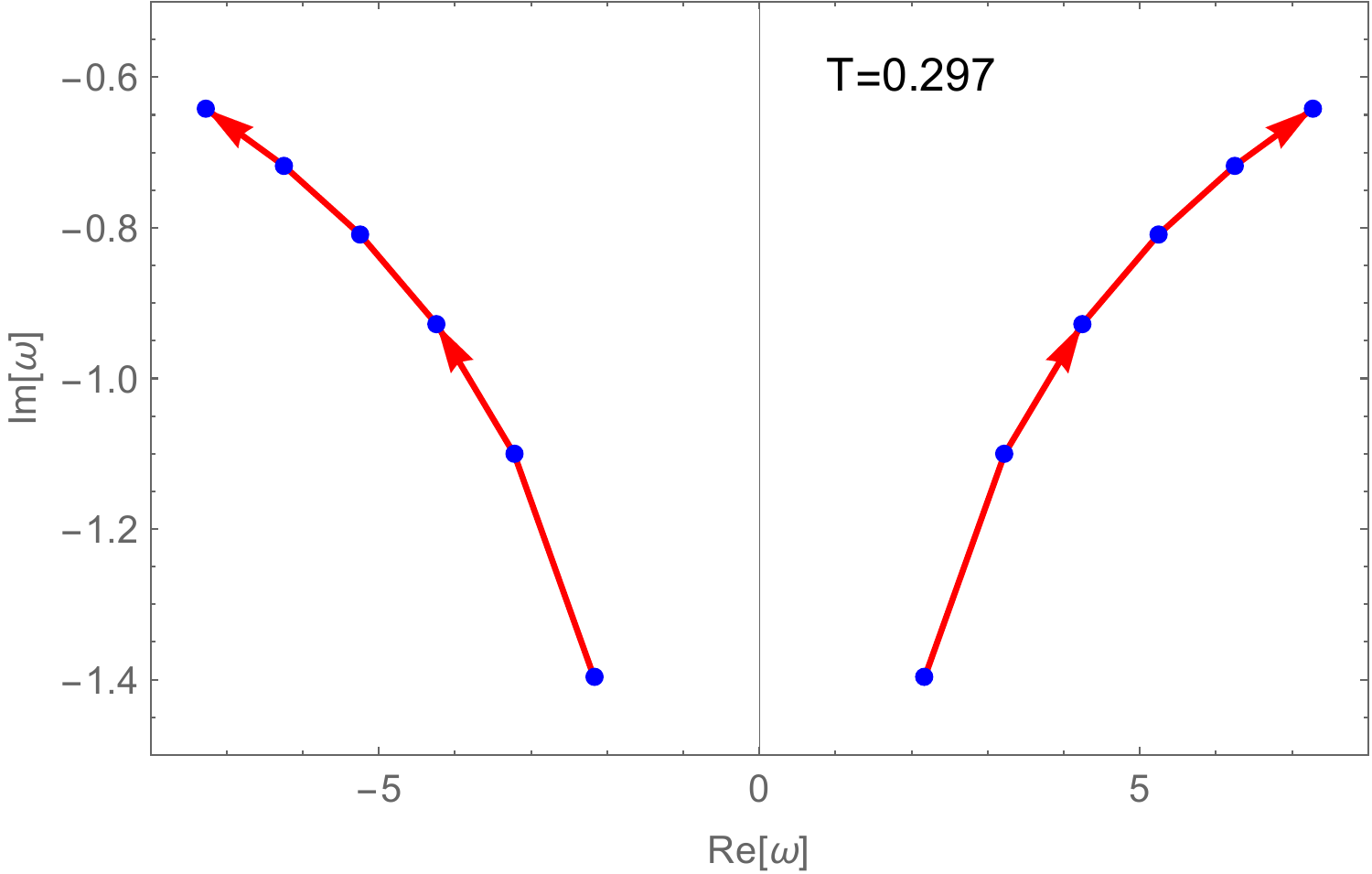}

\includegraphics[scale=0.19]{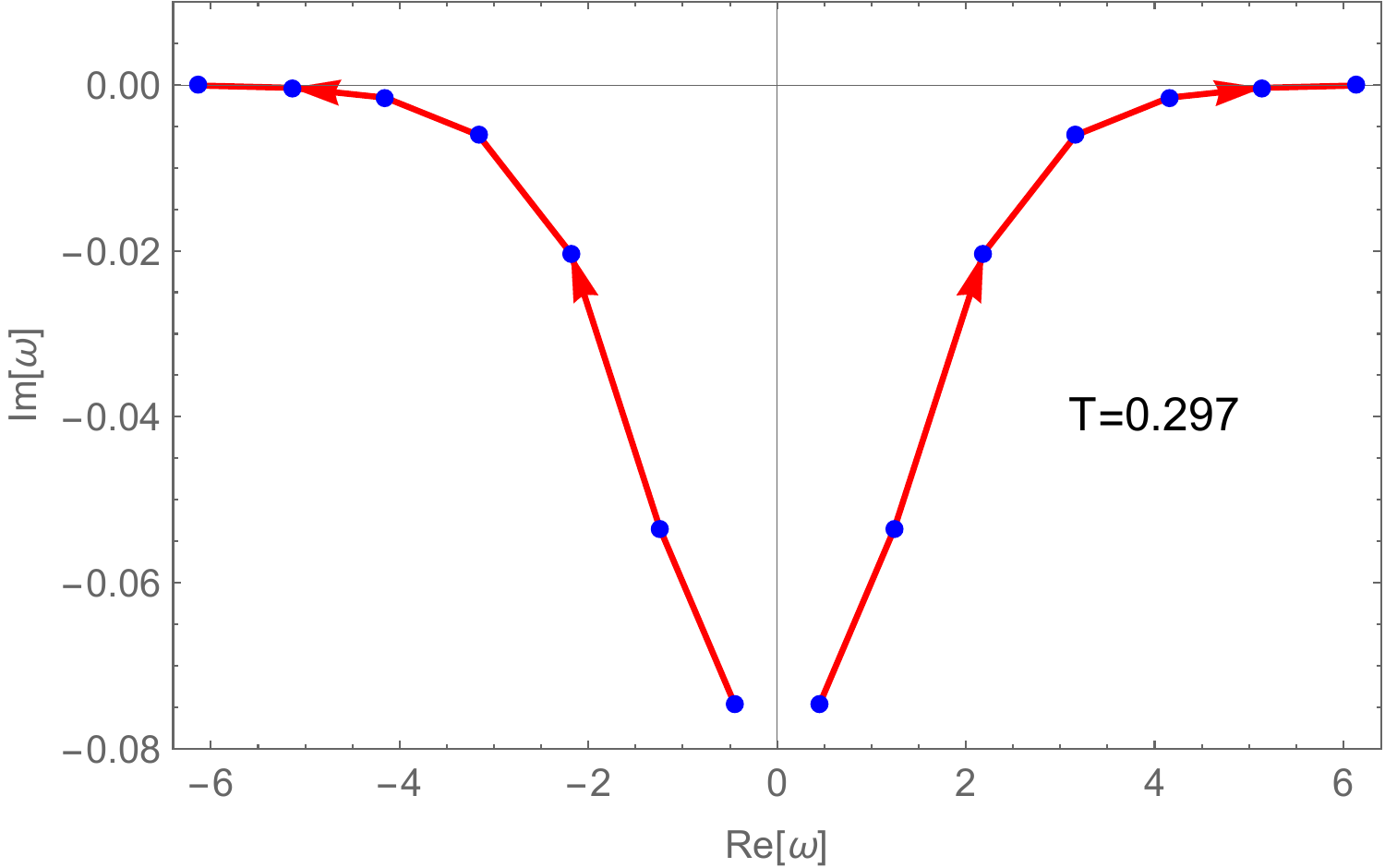}
\includegraphics[scale=0.19]{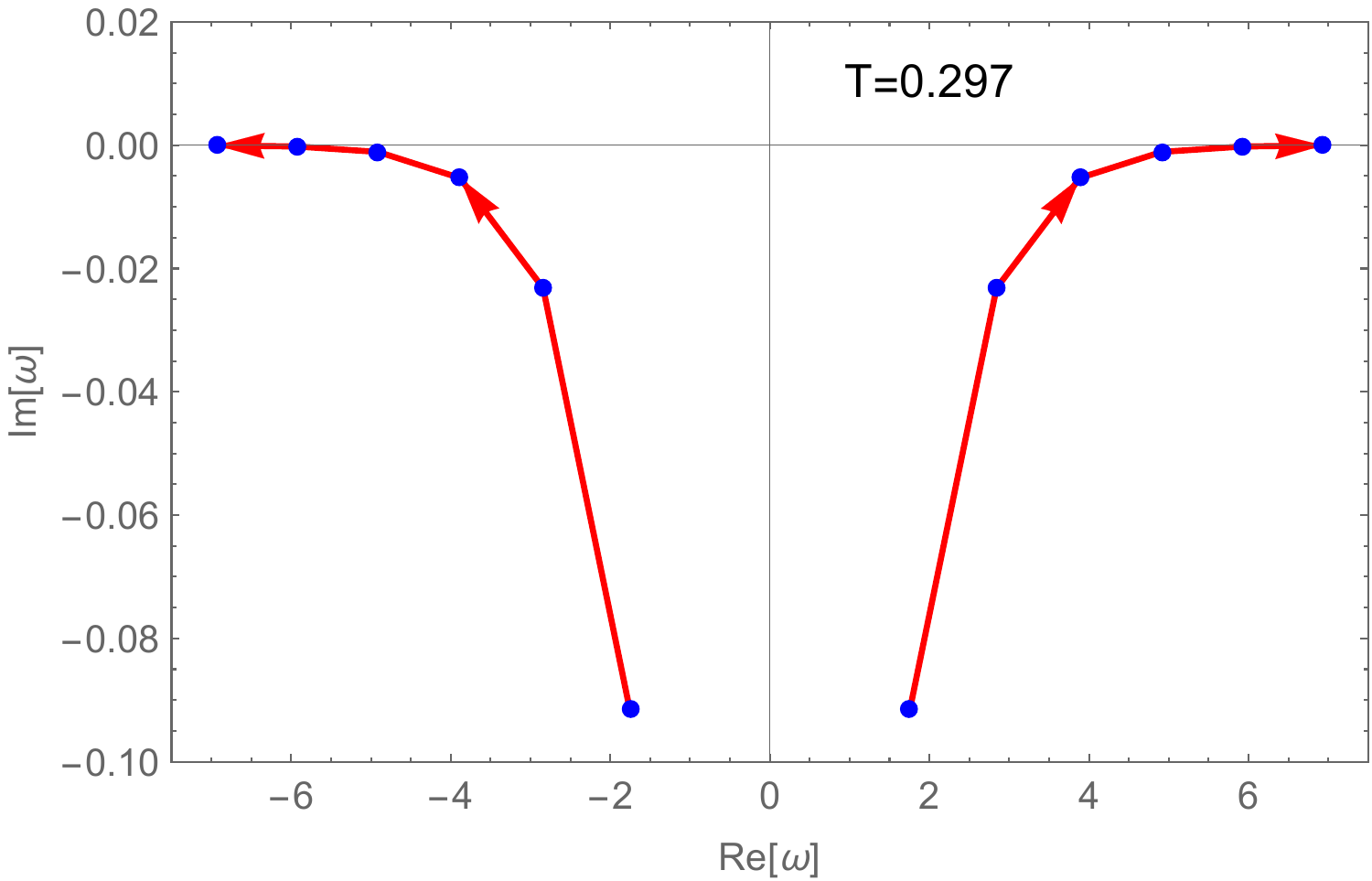}
\includegraphics[scale=0.19]{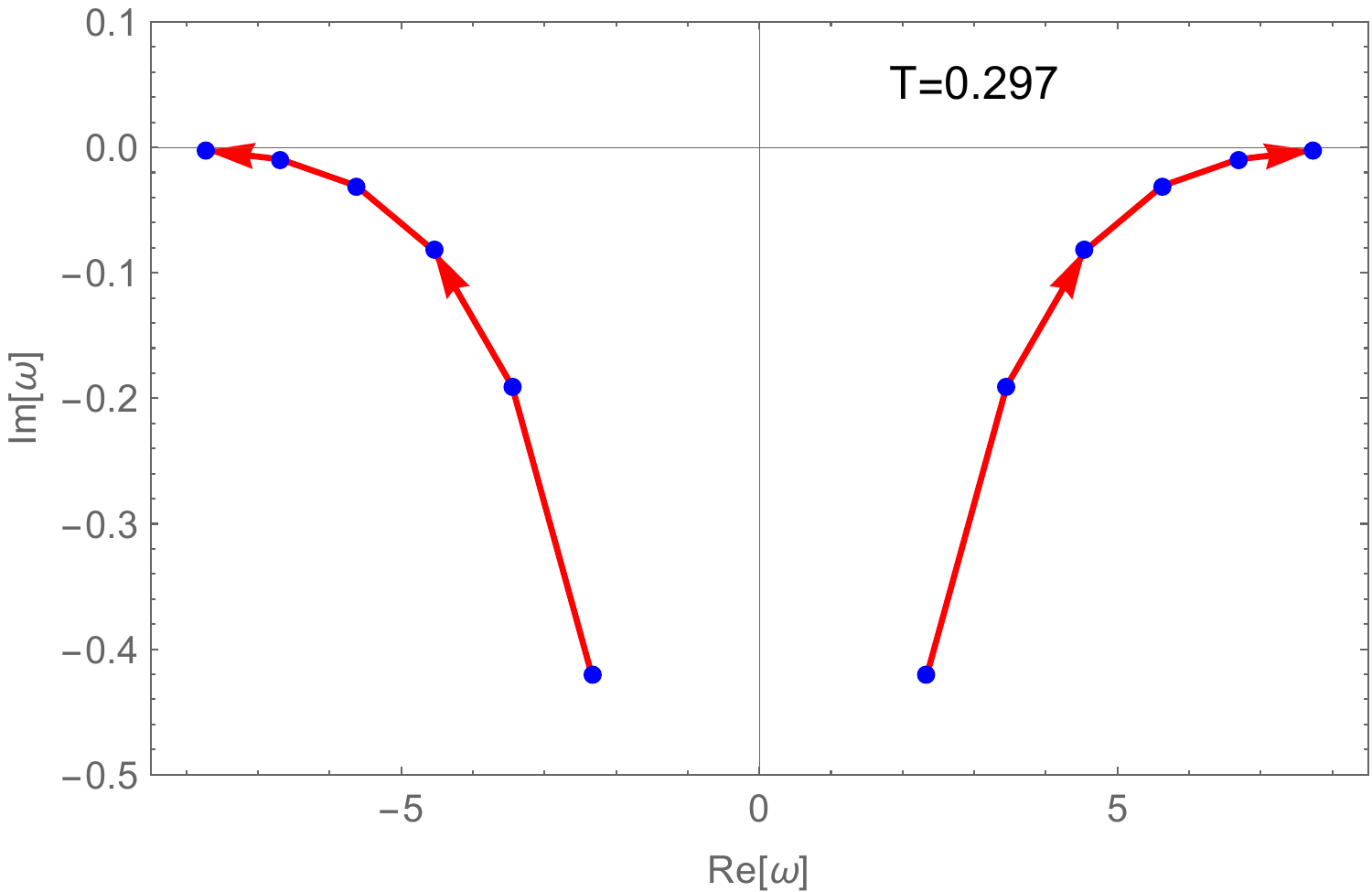}
\caption{The variation of the low-lying quasi-normal modes with angular quantum number $l$ at fixed temperature and chemical potential within normal-fluid state for large (top panels) and small (bottom panels) black holes. The leftmost panels are about scalar field channel ($q_1$ field and $q_2$ field), middle panels are about longitudinal gauge field channel ($a$ field and $c$ field), the rightmost panels are about transverse gauge field ($b$ field). $l$ is varied from 0 to 6 for leftmost panels, from 1 to 6 for the rest. \label{fig:normal fluids modes with L}}
\end{figure}

\begin{figure}
\includegraphics[scale=0.3]{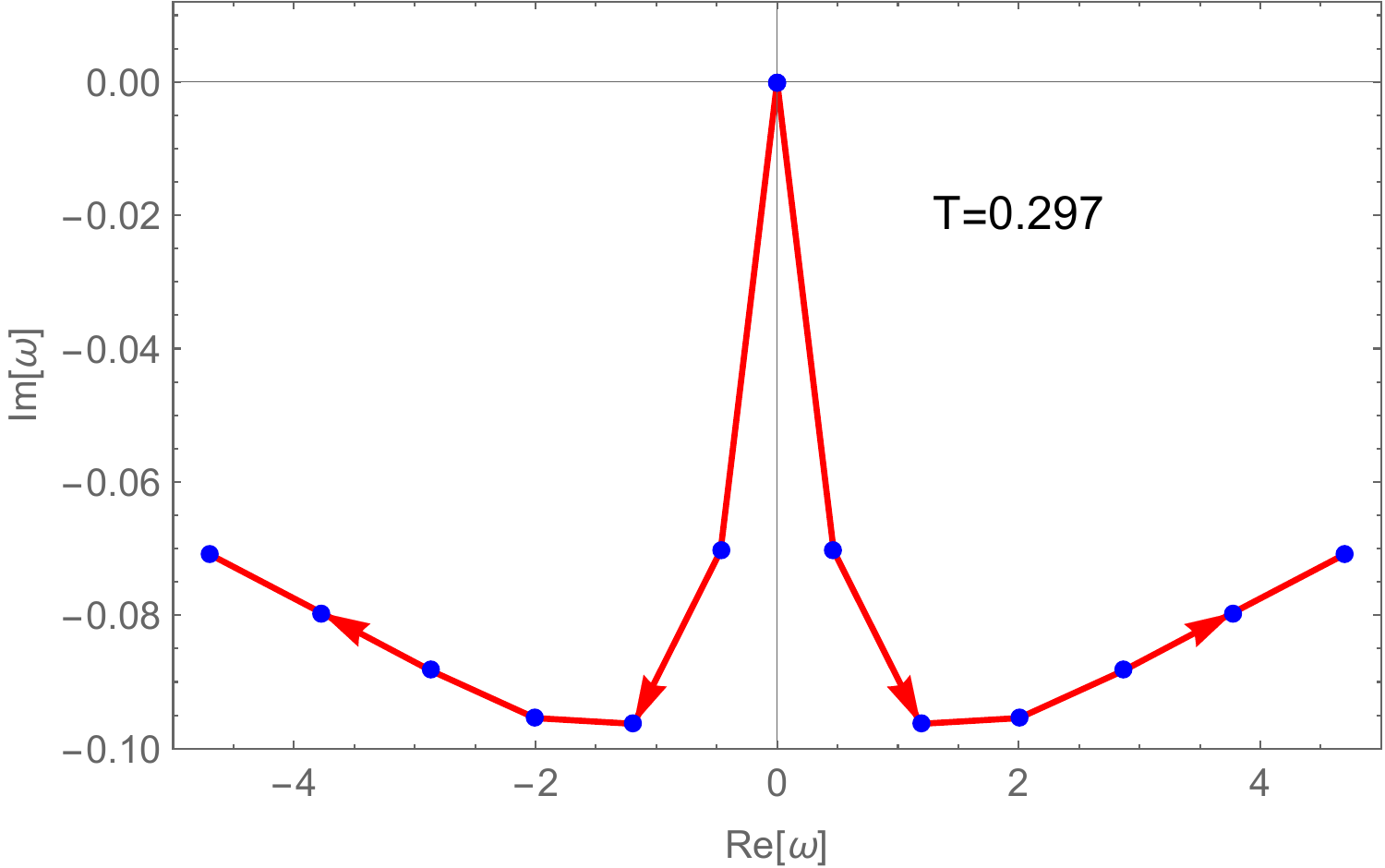}
\includegraphics[scale=0.3]{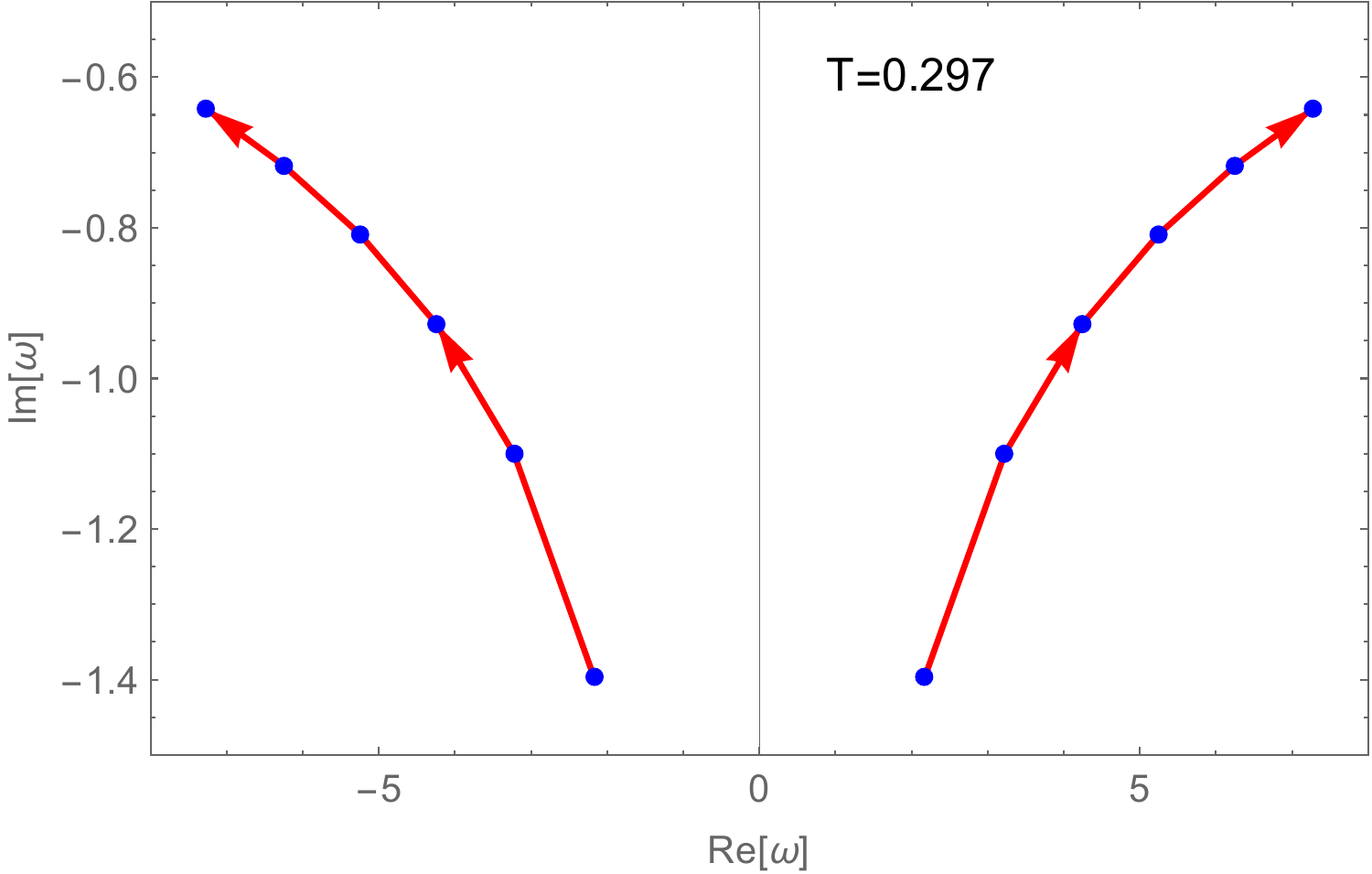}
\includegraphics[scale=0.3]{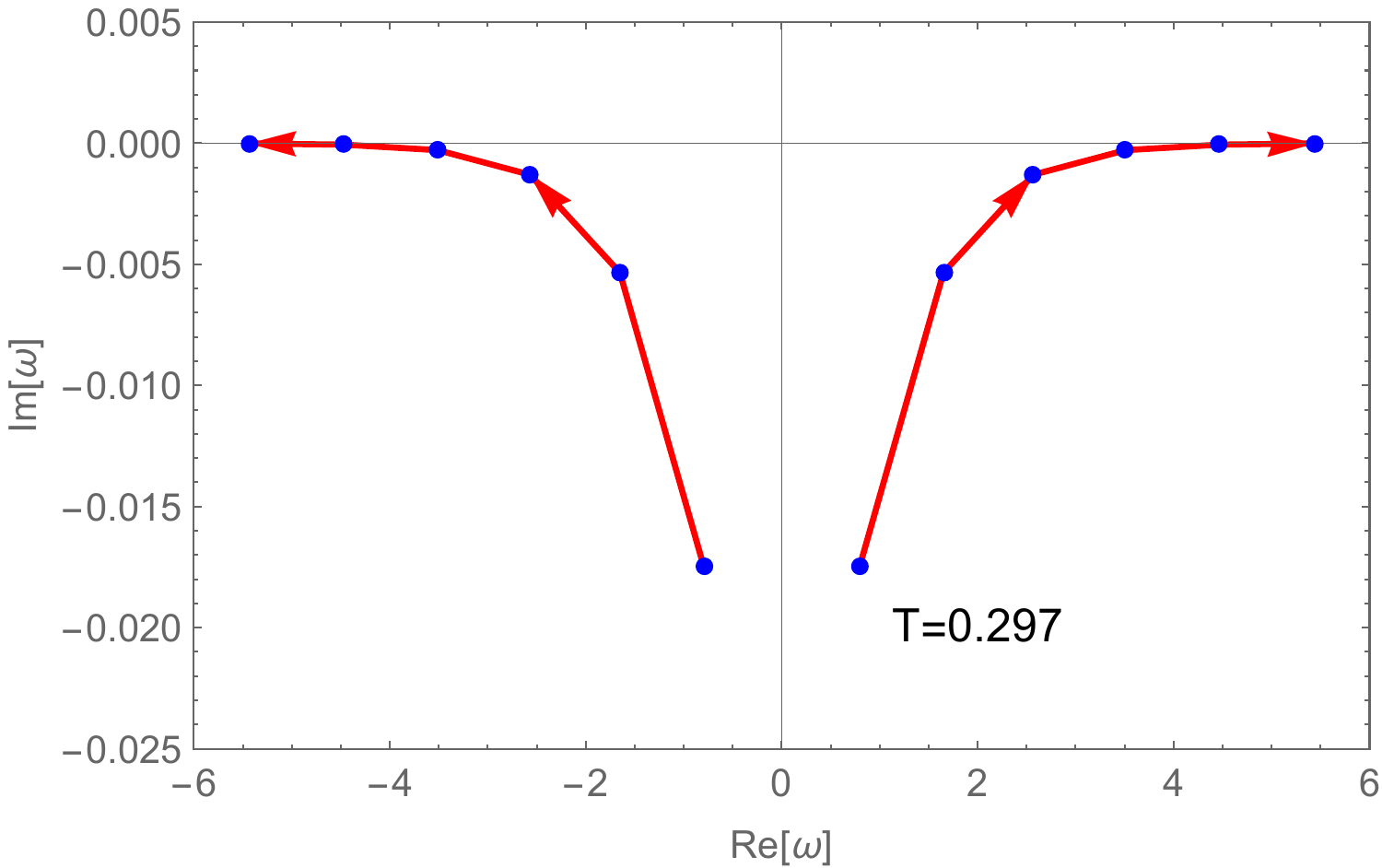}
\includegraphics[scale=0.3]{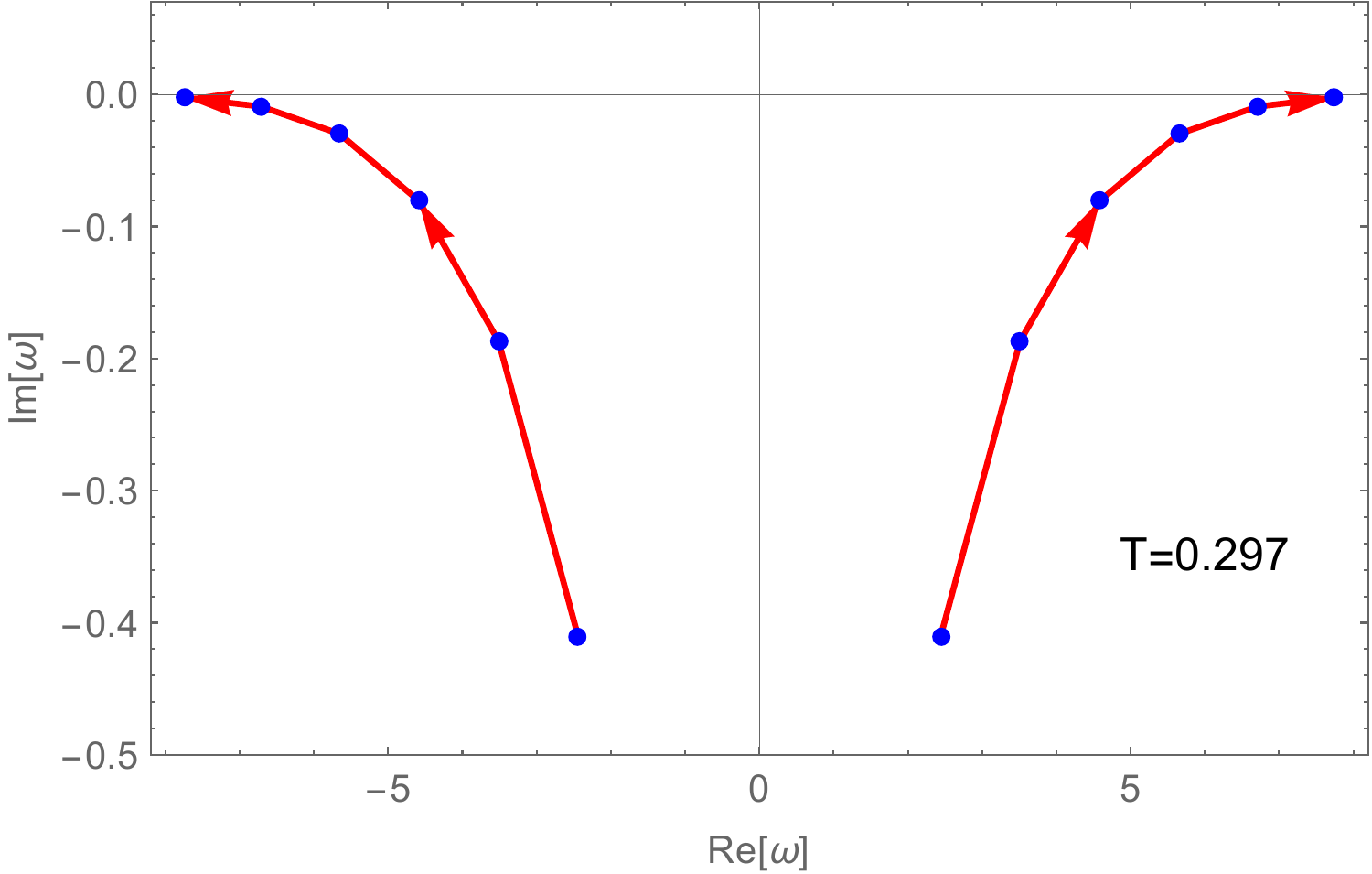}
\caption{The variation of the low-lying quasi-normal modes with angular quantum number $l$ at fixed temperature and chemical potential within superfluid state for large (top plots) and small (bottom plots) black holes. The left plots are from scalar field modes, right plots are from transverse gauge field. $l$ is varied from 0 to 6 for top left panel, from 1 to 6 for the rest.\label{fig:superfluids modes with L}}
\end{figure}

Furthermore, we investigate the impact of angular quantum numbers $l$ on collective modes. Figure \ref{fig:normal fluids modes with L} shows the dynamic behaviors of the collective modes with respect to the angular quantum number at fixed temperature and chemical potential for the normal-fluid state. Here, there exist three different channels of modes: the scalar field channel, the transverse gauge field channel, and the
longitudinal gauge field channel \cite{threechannels}. The three kinds of channels are decoupled when the system is in normal fluid state. In Figure \ref{fig:normal fluids modes with L}, the three top panels are based on a large black hole background, and bottom results are from a small black hole background. The left column depict
scalar field channel ($q_1$ and $q_2$ fields), middle column are from longitudinal gauge field channel (coupled $a$ and $c$ fields), and the right column are from transverse gauge field ($b$ field) channel. $l$ is varied from 0 to 6 for the scalar channel and from 1 to 6 for the gauge field channels. Since the fields of $b$ and $c$ are excitation modes of the angular components of the gauge fields, these excitation modes do not exist as $l=0$. Therefore, we calculate quasi-normal modes of $b$ field and $c$ field starting from $l=1$.
Moreover, $a,c$ fields are coupled as $l\neq 0$ for normal-fluid state. There is only one zero mode for $a$ field, which reflects charge diffusion. Since
$l=1$ on the sphere topology corresponds to the case of large wave number $k$ on the planar topology, the dynamic behaviors of the three channels for large black hole background in Figure \ref{fig:normal fluids modes with L} are similar to those in the planar case. However, we can identify that the dynamic behaviors of scalar modes in the backgrounds of the two types of black holes are completely different. Here, it also exhibits an unusual side for small black hole background. As such, compared to small black hole, the superfluid system with large black hole being background geometry in spherical topology is similar to the case that in planar topology.
Figure \ref{fig:superfluids modes with L} describes the superfluid case for different values of $l$. The left column is from scalar modes, the right column is from transverse gauge field. In this case, scalar field channel and longitudinal gauge field channel are coupled together, the transverse gauge field channel is always decoupled with other fields. Similarly, the dynamic behaviors of all modes under large black hole background in Figure \ref{fig:superfluids modes with L} are consistent with those of the planar case, but there are anomalies in the case of small black holes.

\begin{figure}
\includegraphics[scale=0.19]{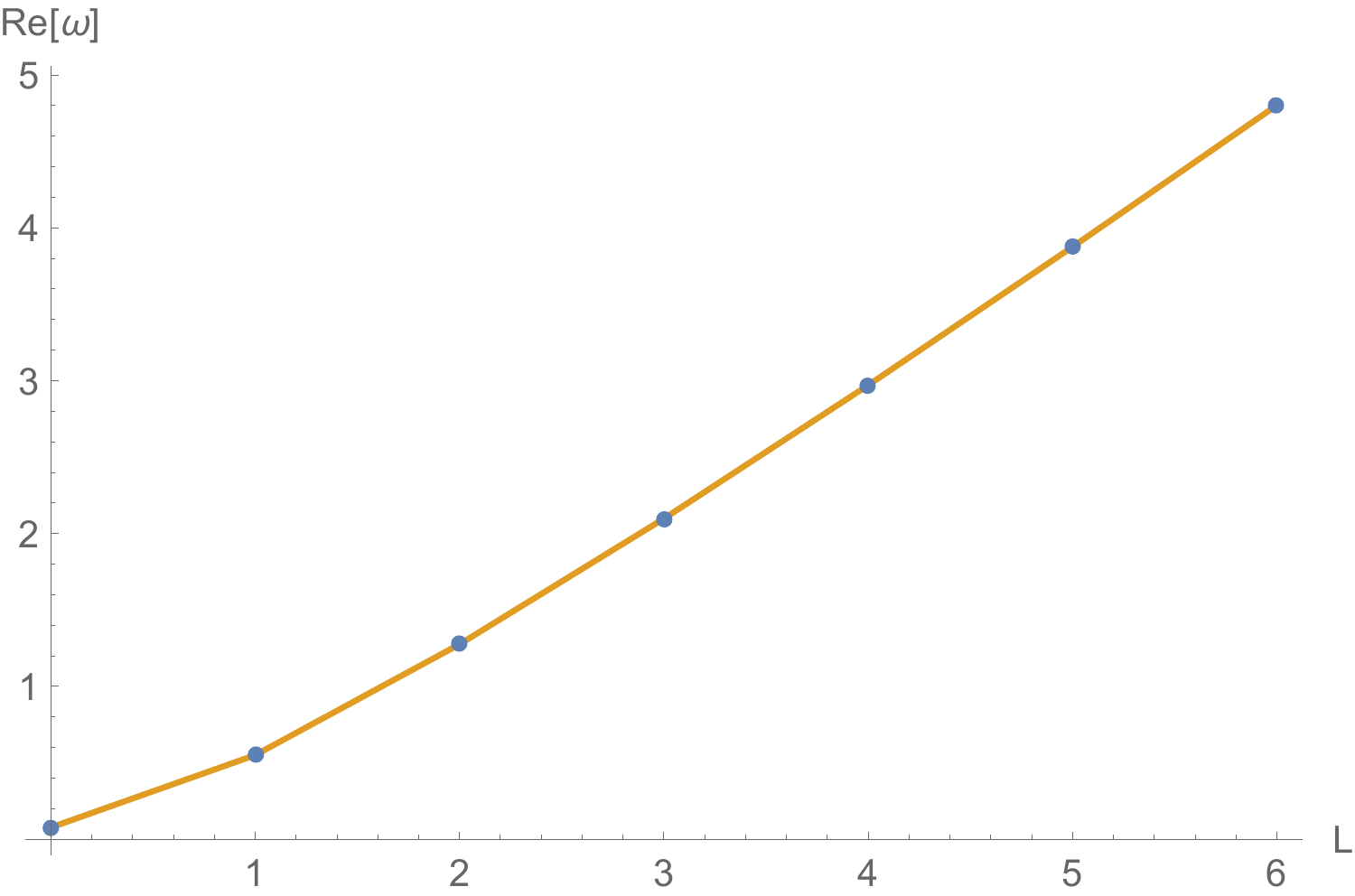}
\includegraphics[scale=0.19]{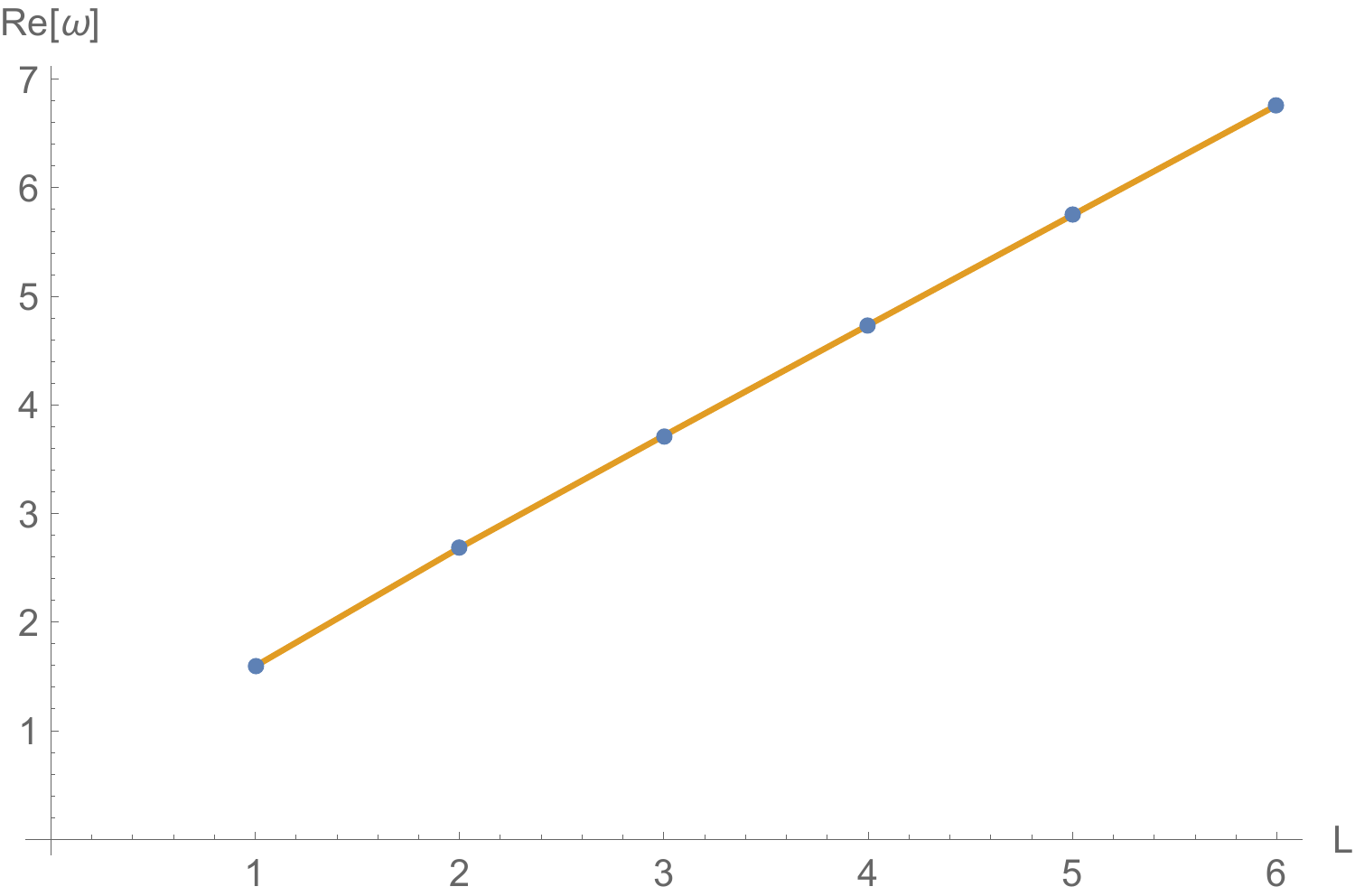}
\includegraphics[scale=0.19]{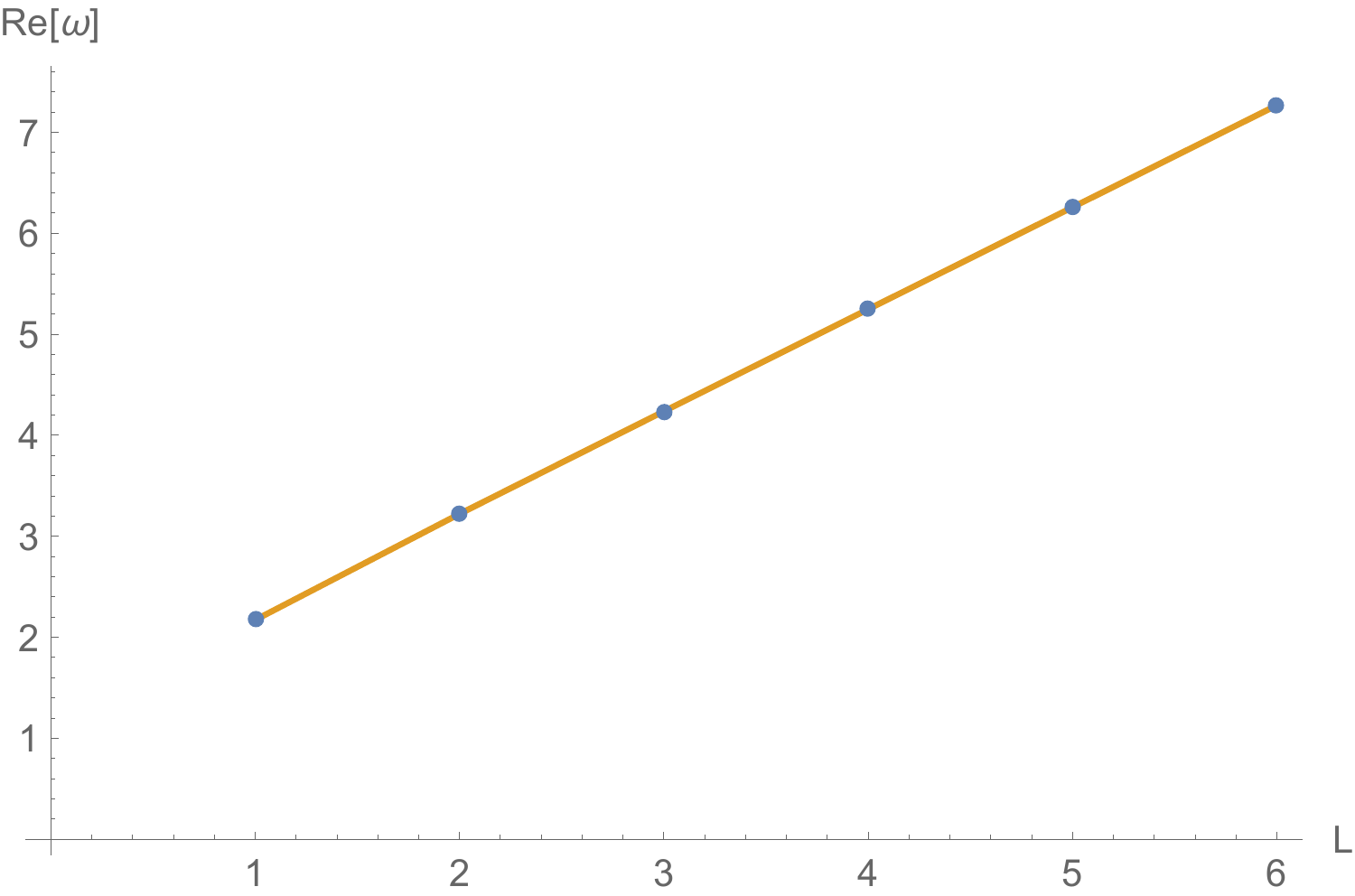}
\includegraphics[scale=0.19]{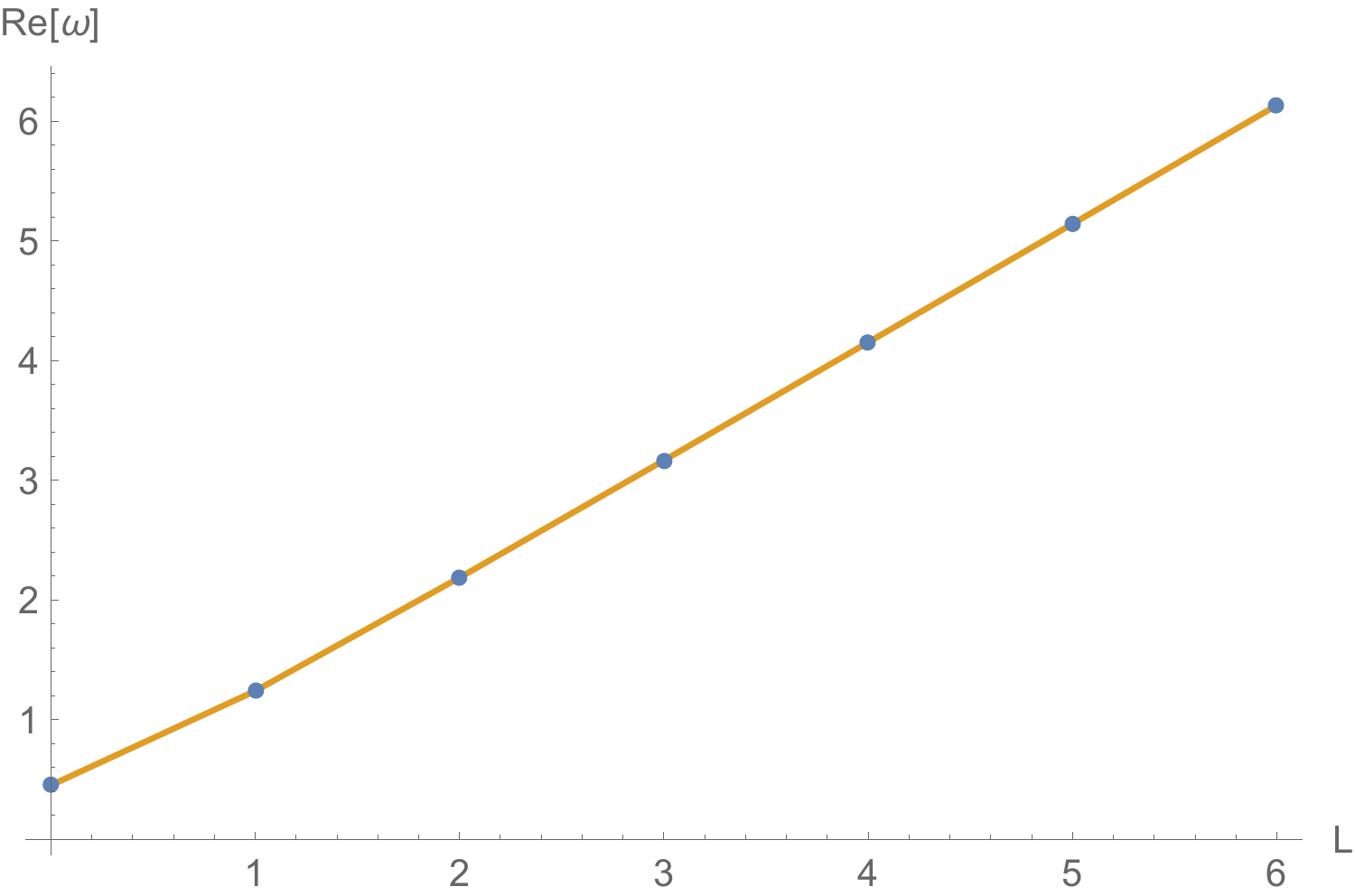}
\includegraphics[scale=0.19]{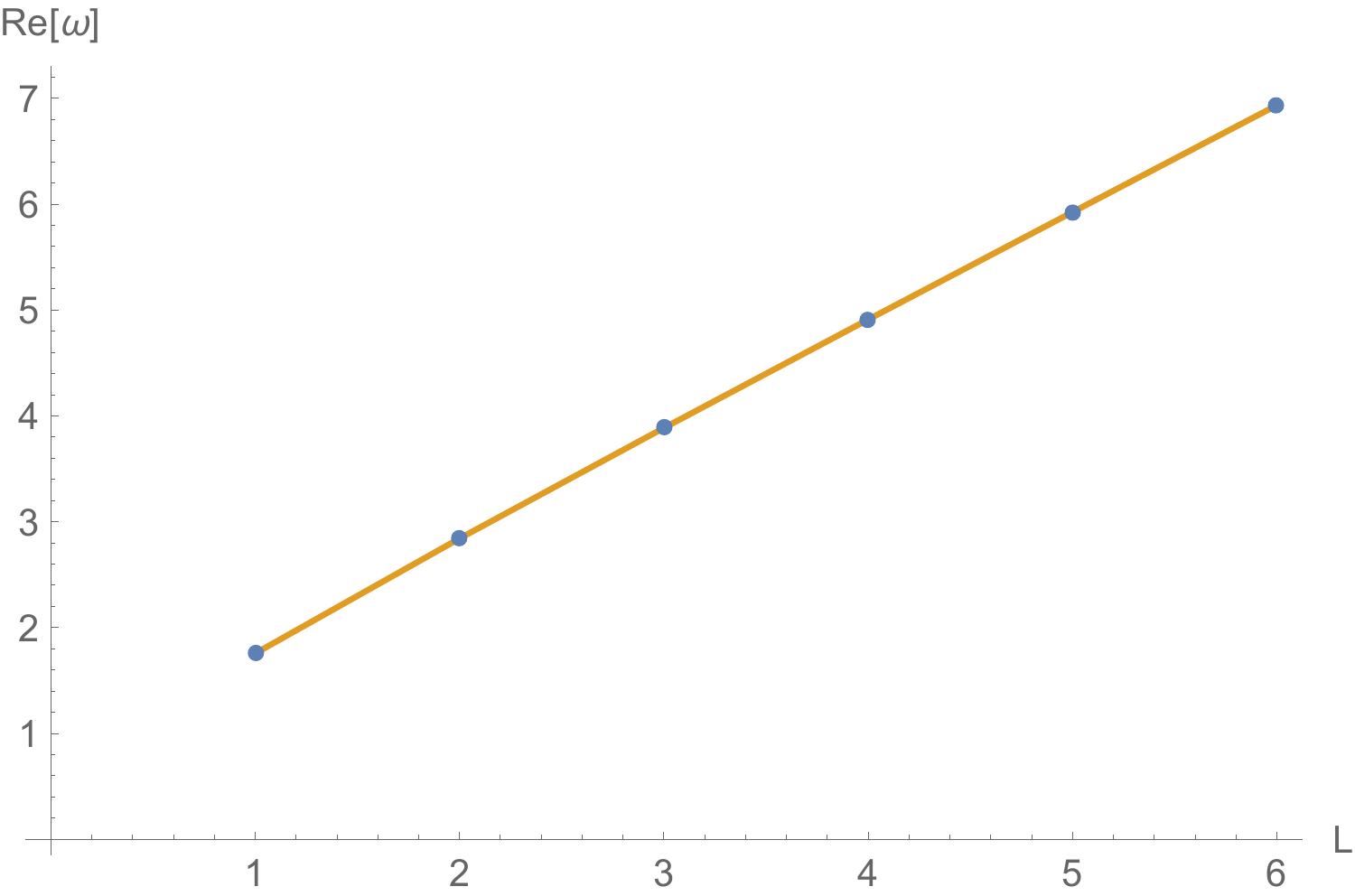}
\includegraphics[scale=0.19]{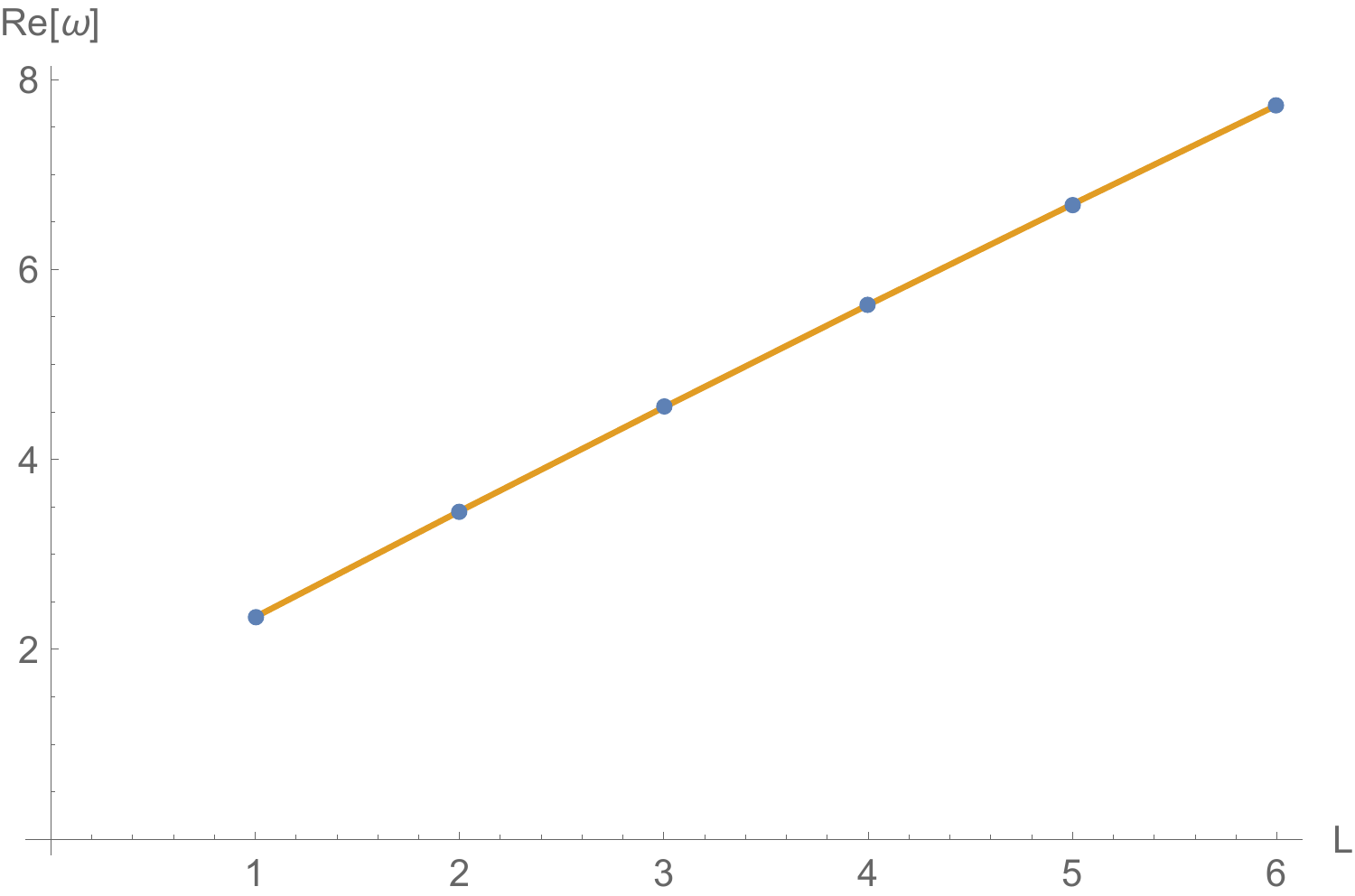}
\caption{The dispersion relation in normal fluid state for large (top plots) and small (bottom plots) black holes. The leftmost plots are from scalar field channel, middle plots are from longitudinal gauge field channel, rightmost plots are from transverse gauge field. \label{fig:Rew of normal fluids modes with L}}
\end{figure}

\begin{figure}
\includegraphics[scale=0.3]{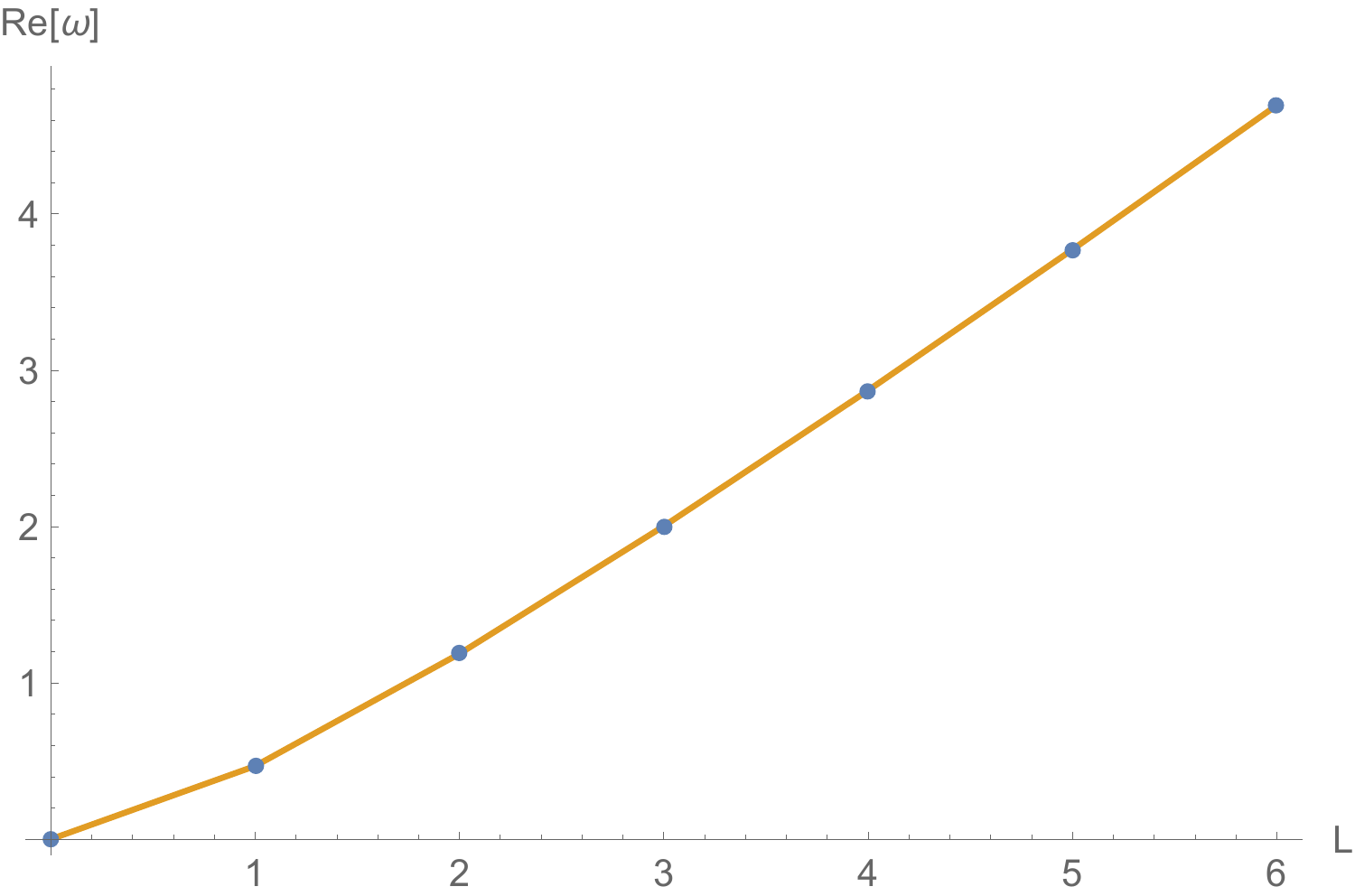}
\includegraphics[scale=0.3]{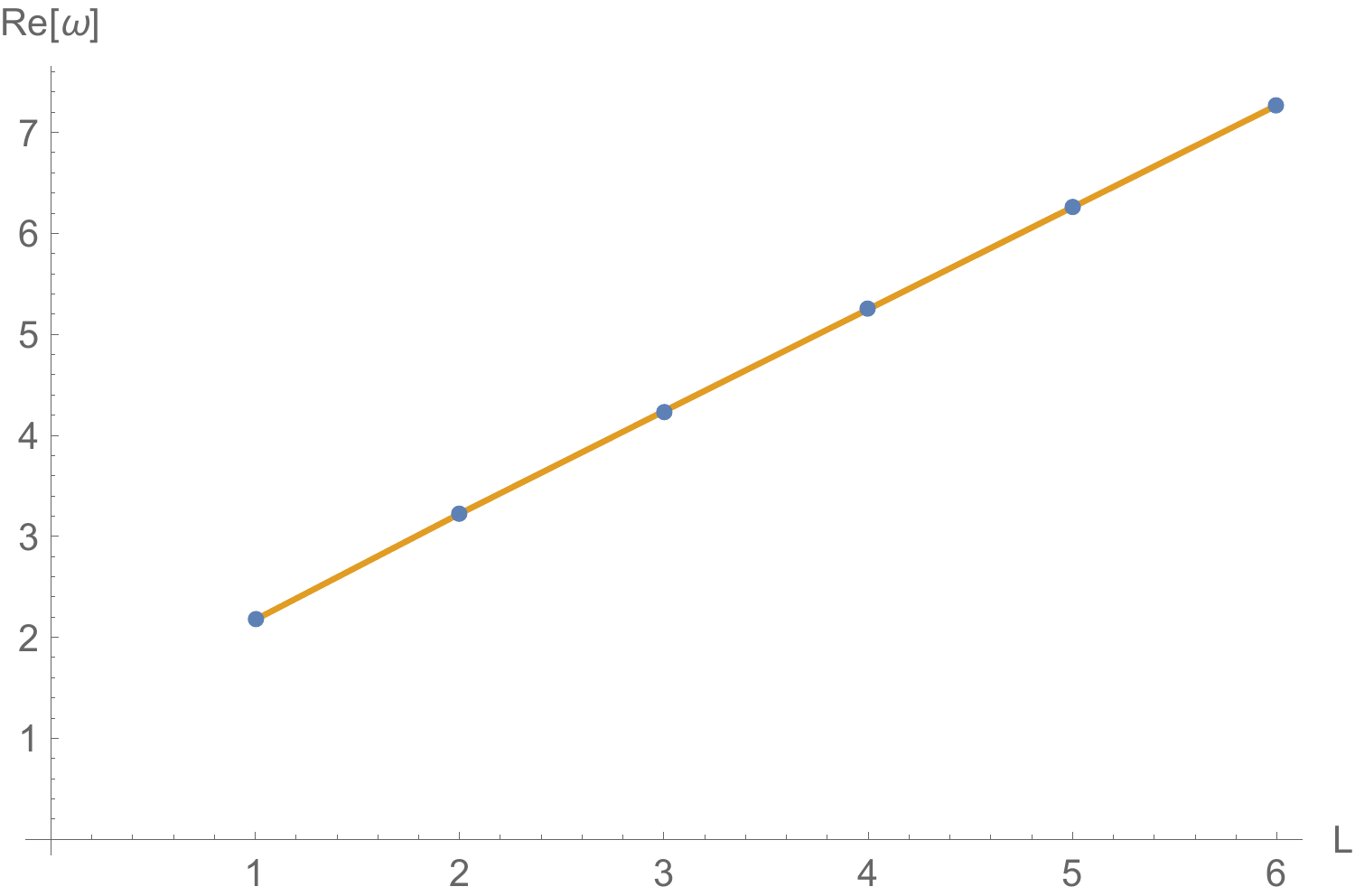}
\includegraphics[scale=0.3]{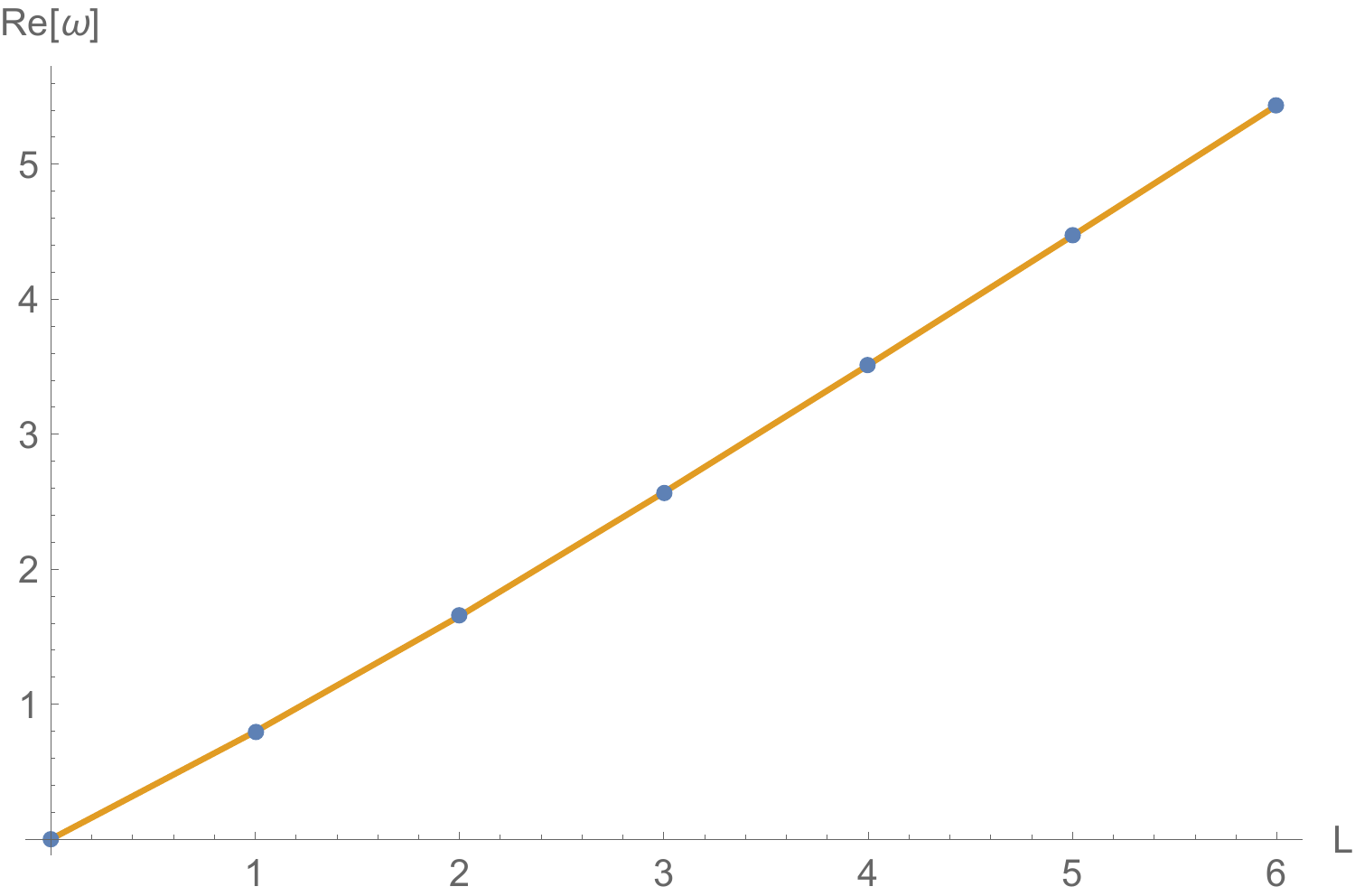}
\includegraphics[scale=0.3]{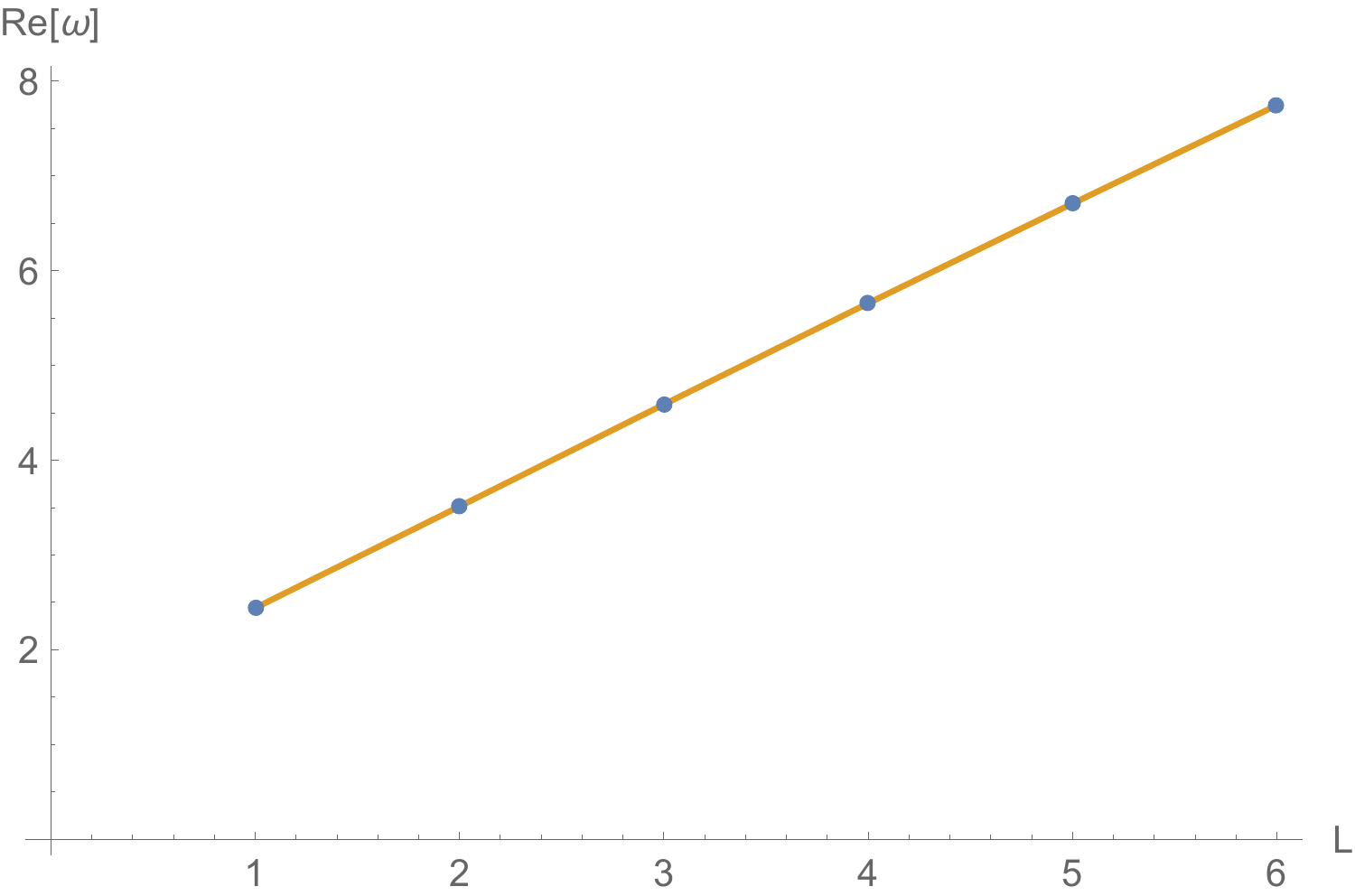}
\caption{The dispersion relation in superfluid state for large (top plots) and small (bottom plots) black holes. The left plots are from scalar mode, right plots are from transverse gauge field. \label{fig:Rew of superfluids modes with L}}
\end{figure}

\begin{figure}
\begin{center}
    \includegraphics[scale=0.5]{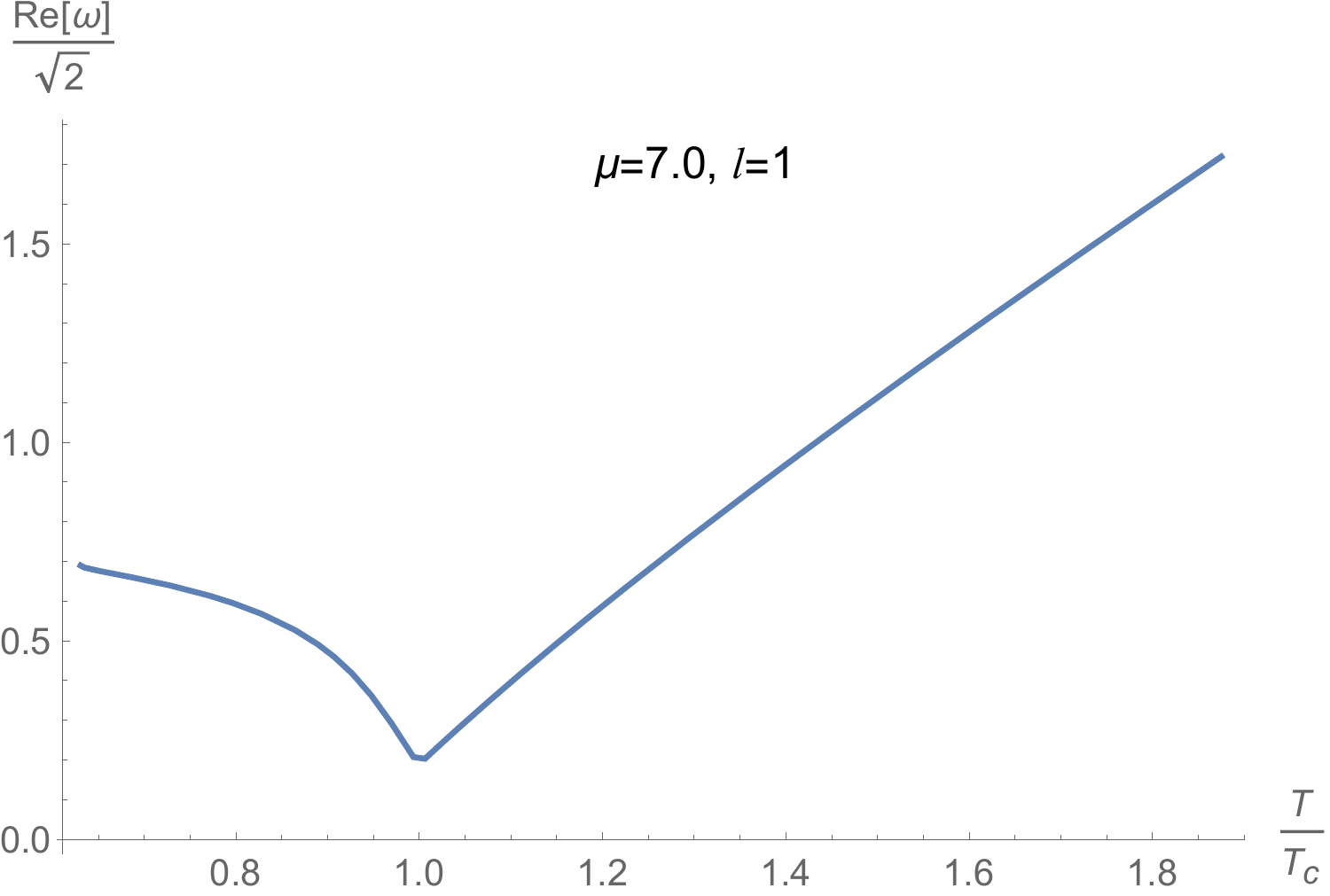}
\end{center}

\caption{The variation of the real part of low-lying mode with temperature based on large black hole background. This mode is exactly the ``first" hydrodynamic excitation within the spherical superfluid system \cite{Topologicalsuperfluid}. \label{fig:two mode with T}}
\end{figure}

To make the dispersion relation clearer under a spherical homogeneous superfluid, we depict the relationship between the real part of frequency and the angular quantum number $L$ for normal fluid state in Figure \ref{fig:Rew of normal fluids modes with L}. One can identify that, considering only the real part of the modes, the dispersion relations in the case of large black hole and small black hole are similar regardless of which of the three channels it is. For the superfluid state, it is presented in Figure \ref{fig:Rew of superfluids modes with L}. From the left column of panels in Figure \ref{fig:Rew of superfluids modes with L}, it can be seen that there exist gapless modes for both the two types of black holes, which is exactly the hydrodynamic mode. Whether in a superfluid system or normal fluid system, the modes for transverse gauge field are always gapped. For the gapless mode based on large black hole background, we observed an interesting dynamic behavior with increasing temperature, which is presented in Figure \ref{fig:two mode with T}. One can observe that this mode decreases initially and then increases as the temperature increases. The turning point coincides precisely with the phase transition point from superfluid to normal fluid with increasing temperature. This collective mode is so called the ``first" hydrodynamic excitation within a spherical superfluid system \cite{Topologicalsuperfluid}. 

\section{Summary and discussion}

This paper extends the study of holographic superfluids from planar to spherical topology, focusing on phase transitions and collective excitation modes within the spherical superfluid systems. The key technique is to expand the gauge field on the sphere using vector spherical harmonics and then perform linear analysis. As mentioned in the Introduction part, relevant experimental equipment has already been established in space, and studying Bose-Einstein condensation in a spherical system will be a promising research direction. The work presented in this paper will lay the foundation for using holographic gravity methods to study the spherical Bose-Einstein condensation, opening a window for subsequent research on spherical BECs, such as the previously mentioned issue about Bose-Einstein condensation of interacting bosons in the spherical system. Some of the research findings in this paper can be supported by previous studies \cite{BEC-sphere,Topologicalsuperfluid}, i.e. the critical temperature for the phase transition from normal fluid to the superfluid on the unit sphere is always larger than that on the plane, as well as the ``first" hydrodynamic excitation mode. 

In addition, we identify  a normal second-order phase transition with large black hole being the background geometry; however, the condensation associated with the small black hole as the background geometry presents an anomalous second-order phase
transition. For the four solutions in the context of two types of black hole backgrounds, we calculated their free energy, respectively. The results indicate that hairy large black hole have the greatest thermodynamic stability. Therefore, in future studies, we will no longer consider small black holes as the geometric background. Moreover, we analyzed the quasi-normal modes in a spherically uniform superfluid at the linear level. Although the small black hole is thermodynamically unstable, it is also dynamically stable at least at the linear level. In the study of excitation modes for $l\neq0$, we find interesting dynamic behaviors of the modes associated with temperature and chemical potential. As $l$ varies, it presents three different channels, which is similar to the case of the plane. 

The phase transition phenomena and dynamical behavior of collective modes discussed in this paper can await experimental verification. 
This work will build a foundation for the subsequent study of nonlinear structure, such as, soliton, vortex, as well as turbulence, on the unit sphere system with the help of AdS/CFT duality.

\begin{acknowledgments}
This work is partly supported by the National Key Research and Development Program of China with Grant
No. 2021YFC2203001 as well as the National Natural
Science Foundation of China with Grant Nos. 12075026, 12035016, 12361141825 and 12375058.
\end{acknowledgments}

\bibliographystyle{JHEP}
\bibliography{references}

\end{document}